\documentclass[12pt]{article}
\usepackage{amsmath}
\usepackage{amsfonts}
\usepackage{amssymb}
\usepackage{float}
\usepackage{ulem}
\usepackage{graphicx}

\usepackage{color}
\restylefloat{table}

\newlength{\abstractwidth}

\usepackage{epsf}
\usepackage{cancel}

\renewcommand{\thanks}[1]{\footnote{#1}}

\renewcommand{\theequation}{\thesection.\arabic{equation}}
\newcommand{\bea}{\begin{eqnarray}}
\newcommand{\eea}{\end{eqnarray}}
\newcommand{\ee}{\end{equation}}
\newcommand{\be}{\begin{equation}}

\newcommand{\ea}{\end{array}}
\newcommand{\bac}{\begin{array}{c}}
\newcommand{\bacc}{\begin{array}{cc}}
\newcommand{\barcl}{\begin{array}{r@{}c@{}l}}
\newcommand{\brcl}{\begin{array}{rcl}}
\newcommand{\bdm}{\begin{displaymath}}
\newcommand{\edm}{\end{displaymath}}
\newcommand{\nn}{\nonumber}
\newcommand{\eps}{\epsilon}

\def\cF{{\cal F}}
\def\cG{{\cal G}}
\def\cH{{\cal H}}

\def\cL{{\cal L}}
\def\cM{{\cal M}}
\def\cN{{\cal N}}

\def\cP{{\cal P}}
\def\cQ{{\cal Q}}

\def\cV{{\cal V}}

\def\Re{{\rm Re}}
\def\Im{{\rm Im}}

\def\eps{\epsilon}

\def\hi{{\hat \imath}}
\def\hj{{\hat \jmath}}
\def\hk{{\hat k}}

\def\no{\nonumber}

\def\eqn#1{eq.~(\ref{#1})}

\def\sec#1{section~\ref{#1}}
\def\app#1{appendix~\ref{#1}}
\def\tf{\tilde{f}}

\def\fig#1{figure~{\ref{#1}}}
\def\tab#1{table~{\ref{#1}}}

\def\nvec5{{\tilde n}}
\def\vs{{V_s}}
\def\Kun{{\tilde K}}
\def\cchi{x^2}
\def\hh{y^2}
\def\vym{V}

\def\si{a}
\def\sj{b}
\def\Tr{{\rm Tr}}
\def\id{\protect{{1 \kern-.28em {\rm l}}}}

\def\spa#1.#2{\left\langle#1\,#2\right\rangle}
\def\spb#1.#2{\left[#1\,#2\right]}

\newcommand{\ha}{{\hat a}}
\newcommand{\hb}{{\hat b}}
\newcommand{\hc}{{\hat c}}
\newcommand{\hd}{{\hat d}}
\newcommand{\he}{{\hat e}}
\newcommand{\haa}{{\hat \alpha}}
\newcommand{\hbb}{{\hat \beta}}
\newcommand{\hgg}{{\hat \gamma}}
\newcommand{\hdd}{{\hat \delta}}
\newcommand{\hee}{{\hat \epsilon}}

\newcommand{\ffm}{\ensuremath{m}}
\newcommand{\ffn}{\ensuremath{n}}
\newcommand{\ffp}{\ensuremath{o}}

\newcommand{\mscr}[1]{\mbox{\scriptsize #1}}

\newcommand{\tI}{{\tilde I}}
\newcommand{\tJ}{{\tilde J}}
\newcommand{\tK}{{\tilde K}}

\newcommand{\la}{\lambda}
\newcommand{\prt}{\partial}
\newcommand{\nnu}{\nonumber}
\newcommand{\Dsl}{D\negthickspace\negthickspace\negthinspace /}
\newcommand{\prtsl}{\partial\negthickspace\negthickspace /}

\newcommand{\lra}{\longrightarrow}
\newcommand{\gs}{g_{\scriptscriptstyle S}}
\newcommand{\fD}{\mathfrak{D}}
\newcommand{\uS}{{\scriptscriptstyle (S)}}

\usepackage{color}
\def\rep{U}
\def\repbar{\overline U}

\def\brk#1{{\cal #1}}

\renewcommand{\theequation}{2.\arabic{equation}}
\def\LCivita{\epsilon}
\def\Pvec{\varepsilon}

\begin{document}
 
\textwidth 170mm
\textheight 230mm
\topmargin -1cm
\oddsidemargin-1cm \evensidemargin -1cm
\topskip 9mm
\headsep9pt

\overfullrule=0pt
\parskip=2pt
\parindent=12pt
\headheight=0in \headsep=0in \topmargin=0in \oddsidemargin=0in

\vspace{ -3cm} \thispagestyle{empty} \vspace{-1cm}

\begin{center}
UUITP-18/15;
NORDITA-2015-119;
CERN-PH-TH-2015-235;
NSF-KITP-15-136
\end{center}


\vspace{ -0.1cm}

\begin{center}

{\Large \bf Spontaneously Broken Yang-Mills-Einstein Supergravities as Double Copies}

\bigskip

{\large Marco Chiodaroli,${}^{a}$ Murat G\"{u}naydin,${}^{b}$ Henrik Johansson,${}^{c,d,e}$ \\ and Radu Roiban${}^{b,f}$}

\medskip

\small 

${}^a${ Max-Planck-Institut f\"ur Gravitationsphysik \\ 
             Albert-Einstein-Institut, Am M\"uhlenberg 1, 14476 Potsdam, Germany\\
}
\smallskip
${}^b${ Institute for Gravitation and the Cosmos \\
 The Pennsylvania State University, University Park PA 16802, USA \\ 
}
\smallskip
${}^c${ Theory Division,
Physics Department, CERN \\
CH-1211 Geneva 23, Switzerland\\
}
\smallskip
${}^d${ Department of Physics and Astronomy, Uppsala University,
 SE-75108 Uppsala, Sweden\\
 }
 \smallskip
${}^e${ Nordita, KTH Royal Institute of Technology and Stockholm University, \\
Roslagstullsbacken 23, SE-10691 Stockholm, Sweden\\
}
 \smallskip
${}^f${ Kavli Institute for Theoretical Physics, University of California\\
Santa Barbara, CA 93106-4030, USA \\
}
\bigskip

\end{center}

\begin{abstract}

Color/kinematics duality and the double-copy construction have proved to be systematic tools for gaining new insight into gravitational theories. 
Extending our earlier work, in this paper we introduce new double-copy constructions for large classes of spontaneously-broken 
Yang-Mills-Einstein theories with adjoint Higgs fields.
One gauge-theory copy entering the construction is a spontaneously-broken (super-)Yang-Mills theory,
while the other copy is a bosonic Yang-Mills-scalar theory with trilinear 
scalar interactions that display an explicitly-broken global symmetry. We show that the kinematic numerators of these gauge theories can be made to obey color/kinematics duality by exhibiting particular additional Lie-algebraic relations.
We discuss in detail explicit examples with ${\cal N}=2$ supersymmetry,
focusing on Yang-Mills-Einstein supergravity theories belonging to the generic Jordan family in four and five dimensions, and
identify the map between the supergravity and double-copy fields and parameters.
We also briefly discuss the application of our results to $\cN=4$ supergravity theories.
The constructions are illustrated by explicit examples of tree-level and one-loop scattering amplitudes.

\end{abstract}

\baselineskip=16pt
\setcounter{equation}{0}
\setcounter{footnote}{0}

\newpage

\tableofcontents

\newpage

\section{Introduction}
\renewcommand{\theequation}{1.\arabic{equation}}
\setcounter{equation}{0}

Einstein's theory of gravity and spontaneously-broken gauge theory 
are two of the pillars of our current understanding of the known fundamental interactions of Nature.
While supersymmetric field theories that combine gravitational interactions and spontaneous symmetry breaking have been studied extensively 
at the Lagrangian level, the perturbative 
$S$ matrices of these theories have largely been unexplored.  

Modern work on scattering amplitudes in matter-coupled gravitational theories has been 
largely focused on pure supergravities and on cases in which additional matter consists of abelian vectors (i.e. Maxwell-Einstein supergravities) or fermion/scalar fields. A key tool has been the double-copy construction \cite{Bern:2008qj, Bern:2010ue}, 
which has led to a dramatic simplification of perturbative calculations. For example, explicit expressions of one-, two-, three- and four-loop 
amplitudes have been obtained for ${\cal N}= 4$, ${\cal N}=5$ and ${\cal N}=8$ supergravities 
in refs. \cite{Bern:2010ue, Carrasco:2011mn,Bern:2011rj,BoucherVeronneau:2011qv,Bern:2012uf, Bern:2012cd, 
Bern:2012gh, Bern:2013qca,Bjerrum-Bohr:2013iza,Bern:2014sna,Bern:2014lha,Mafra:2015mja,He:2015wgf}. For the case of ${\cal N}\le 4$,
one-loop four-point superamplitudes have been obtained for the generic Jordan family of $\cN=2$ Maxwell-Einstein supergravity (MESG) 
theories~\cite{Carrasco:2012ca,Ochirov:2013xba,Johansson:2014zca}, for pure supergravities with ${\cal N}\le 4$~\cite{Carrasco:2012ca,Nohle:2013bfa,Johansson:2014zca}, 
and for orbifolds thereof~\cite{Carrasco:2012ca,Chiodaroli:2013upa}. 

The double-copy construction assumes the existence of presentations of gauge-theory scattering
amplitudes that exhibit color/kinematics duality. The duality states that, in an amplitude's Feynman-like diagrammatic expansion, 
one can find numerator factors that obey Lie-algebraic kinematic relations mirroring the relations satisfied by the corresponding gauge-group color factors. 
Once found, the numerators may play the role of these color factors in any gauge theory amplitude, and upon substitution one obtains valid gravitational amplitudes.
There is by now extensive evidence for the duality and for the double-copy 
construction in wide classes of Yang-Mills (YM) theories  and in the associated 
(super)gravity theories. Examples where color/kinematics 
duality has been demonstrated include: pure super-Yang-Mills 
(SYM) theories~\cite{Bern:2008qj,Bern:2010ue,BjerrumBohr:2009rd,Stieberger:2009hq,Mafra:2011kj}, SYM theories with adjoint
matter~\cite{Carrasco:2012ca,Nohle:2013bfa,Johansson:2014zca}, self-dual Yang-Mills theory~\cite{Monteiro:2011pc,Boels:2013bi}, 
QCD and super-QCD~\cite{Johansson:2015oia, delaCruz:2015dpa}, YM coupled to $\phi^3$ theory~\cite{Chiodaroli:2014xia}, 
and YM theory extended by a higher-dimensional operator~\cite{Broedel:2012rc}. 
It has also been observed that the duality is not limited to YM gauge theories, but it also applies 
to certain Chern-Simons-matter theories~\cite{Bargheer:2012gv, Huang:2012wr, Huang:2013kca}, as well as to 
the non-linear sigma model/chiral Lagrangian~\cite{Chen:2013fya} and to the closed (heterotic) 
sector of string theory~\cite{Stieberger:2014hba}.  

Amplitudes in Maxwell-Einstein supergravities are obtained by a double-copy construction 
of the form
(pure SYM)$\otimes$(YM coupled to scalars). 
Subgroups of the global symmetries of Maxwell-Einstein 
supergravities can be gauged.\footnote{While gauging part of the R-symmetry group is very interesting, here we will focus on gaugings that only affect 
the other global symmetries.}  In the resulting theories some of  the vector fields become gauge fields of the chosen gauge group  and transform in its adjoint 
representation. Therefore, the only subgroups of the global symmetry that can be gauged are those whose adjoint representation
is smaller than the number of  vector fields that transform non-trivially under the global symmetry group. In five dimensions, gauging 
only a subgroup of the  global symmetry group in $\cN=2$ Maxwell-Einstein supergravity theories does not introduce a potential for the scalar 
fields and hence the resulting theory is guaranteed to have a Minkowski vacuum state. 

The double-copy construction of a wide class of Yang-Mills-Einstein supergravity (YMESG) theories was given in \cite{Chiodaroli:2014xia}, where it was shown that one of the two gauge-theory factors is a pure SYM theory, and the other is a bosonic YM theory coupled to scalars that transforms in the adjoint representation of both the gauge group and a global symmetry group. 
The latter theory has trilinear $\phi^3$ couplings, and hence we refer to it 
as YM~+~$\phi^3$ theory. Through the double-copy construction, the global symmetry of the
non-supersymmetric gauge-theory factor becomes a local symmetry, and the trilinear scalar couplings generate the minimal couplings of the corresponding gauge fields. The gravitational supersymmetry is directly inherited from the SYM theory, thus accommodating ${\cal N}=1,2,4$ YMESG theories and ${\cal N}=0$ Yang-Mills-Einstein (YME) theories. 
Earlier work~\cite{Bern:1999bx} introduced the same type of construction for single-trace tree-level YME amplitudes. 
Recent work on YME amplitudes takes several different approaches, see refs.~\cite{Stieberger:2014cea,Cachazo:2014nsa,Stieberger:2015qja, Casali:2015vta,Adamo:2015gia, Stieberger:2015kia,Stieberger:2015vya}.

It is essential to explore the validity of the double-copy construction away from the origin of the moduli space.
In particular, a natural and physically-motivated extension is to consider cases in which the supergravity gauge symmetry is spontaneously broken 
through the Brout-Englert-Higgs mechanism.  We will present such an extension in the present paper. 
As a key result, we find that one of the two gauge-theory factors is the 
spontaneously-broken pure SYM theory 
(or, alternatively stated, the Coulomb branch of pure SYM theory), while the other is a particular massive deformation 
that explicitly breaks the global symmetries of the YM~+~$\phi^3$ theory. 

Identifying the relation between asymptotic states of the supergravity theory and the corresponding states of the gauge-theory factors 
is an important aspect of the double-copy construction. 
For gauge theories with only adjoint fields, the double copy gives a supergravity 
state for every tensor product of gauge-theory states (not counting the degeneracy of the representation). 
In cases in which the gauge-theory matter transforms in non-adjoint representations of the gauge group,
the double-copy construction allows for better tuning of the matter content of the gravitational theory, 
since only certain tensor products of the gauge-theory matter are allowed. 

In ref.~\cite{Chiodaroli:2013upa} color/kinematics duality was extended to non-adjoint representations 
in the context of orbifolds of ${\cal N}=4$ SYM, and the associated double copies were 
found to be matter-coupled supergravity theories. 
The construction required that: (1) the gauge groups of the two gauge theories should be identified, and 
(2) supergravity states correspond to  gauge-invariant bilinears that can be formed out of the gauge-theory states. 
This construction correlates gauge- and global-group representations appearing in the resulting gauge theories.

In ref.~\cite{Johansson:2014zca}  color/kinematics duality was extended to theories
with fields in the fundamental representation 
and used to construct pure ${\cal N} \le 4$ supergravity theories 
as well as matter-coupled theories. 
In this construction, the necessary condition for the double copy to be valid is that the kinematic matter-dependent numerators obey 
the same relations as the corresponding color factors with fundamental representations. Upon replacing the
color factors with kinematic numerator factors one similarly obtains a double copy that correlates the representations of the states of the two gauge-theory sides. 

For the double-copy constructions of supergravity theories with spontaneously-broken gauge symmetry, 
the identification of the asymptotic states will follow closely the non-adjoint or fundamental cases. 
However, the details of the kinematic algebra obeyed by the numerators will differ substantially compared 
to previous situations. The kinematic Jacobi identities 
and commutation relations will be extended by additional identities
which are inherited from the Jacobi relations of the theory with unbroken gauge symmetry. 
We stress that our construction works well with -- but does not require -- supersymmetry, 
and similarly works in all dimensions in which the theories are defined, as it is expected for color/kinematics duality.

The paper is organized as follows. In \sec{sec.ck2copy} we review color/kinematics duality, and identify matter-coupled gauge theories 
with fields in several different representations
of the gauge group and specific cubic and quartic couplings which obey the duality. 
We extend color/kinematics duality and the  double-copy construction 
to massive field theories, as well as to field
theories with spontaneously-broken gauge symmetry, 
paying close attention to the construction of asymptotic states. 
In particular subsection \ref{2copySec} discusses extensions of the double-copy construction and contains our main results of this generalization.  

In section~3 we review, from the Lagrangian perspective, the Higgs mechanism in four- and five-dimensional ${\cal N}=2$ 
Yang-Mills-Einstein supergravities. Such theories are uniquely specified by their cubic interactions and   
provide simple examples of our construction. 
In particular, we identify the four-dimensional 
symplectic frame  in which the amplitudes from the spontaneously-broken 
Yang-Mills-Einstein supergravity Lagrangian reproduce the ones from the double-copy construction.

In section~4, we compute tree-level scattering amplitudes in the gauge theories discussed in section~2 and in the 
supergravity theories discussed in section~3. We find the constraints imposed by color/kinematics duality on the 
cubic and quartic couplings of the gauge theories, identify the precise map between supergravity states and gauge-invariant 
billinears of gauge-theory states, and give the relation between the gauge-theory and supergravity parameters. 

In section~5,  we discuss  loop-level calculations in theories  formulated in the earlier sections.  
Section~6,  discusses briefly 
spontaneously-broken $\cN=4$ Yang-Mills-Einstein supergravity theories. 
We review the bosonic part of their Lagrangians in five dimensions and discuss how their amplitudes can be obtained through the double-copy construction with a straightforward extension 
of the results obtained for $\cN=2$ theories.

\section{Color/kinematics duality and double copy \label{sec.ck2copy}}
\renewcommand{\theequation}{2.\arabic{equation}}
\setcounter{equation}{0}

In this section, we review the color/kinematics duality applied to gauge theories that have fields in 
complex representations of the gauge group. Giving concrete examples, we write down Lagrangians of several gauge theories
where the duality should be present. We then spontaneously (and explicitly) break the symmetries of these theories, 
and in the process generalize color/kinematics duality to such situations. Finally, we give the double-copy prescription for spontaneously- and explicitly-broken theories. 

\subsection{Review: color/kinematics duality for complex representations \label{sec.ckdu}}

The scattering amplitudes in a gauge theory with fields in both the adjoint representation and some generic complex 
representation\footnote{By generic complex representation, we mean a representation that only has quadratic and 
cubic invariants $\overline{U} U$, and $\overline{U}\,{\rm (Adj)}\,U $, respectively. A canonical example of such an $U$ is the fundamental representation.} 
$\rep$ of a Lie group
can be organized in terms of cubic graphs.\footnote{Quartic and higher-degree interactions are absorbed into the numerators of
the cubic graphs. This corresponds to having introduced suitably-chosen auxiliary fields to make the Lagrangian cubic.} At $L$ loops and in $D$ dimensions,
such amplitude has the following form\footnote{We use a different numerator normalization compared to ref.~\cite{Bern:2010ue}. 
Relative to that work, we absorb one factor of $i$ into the numerator, giving a uniform overall $i^{L-1}$ to the gauge and gravity amplitudes.}
\be
   {\cal A}^{(L)}_{n} = i^{L-1} g^{n-2+2L} \sum_{i \in \text{cubic}}\, \int \frac{d^{LD}\ell}{(2\pi)^{LD}} \frac{1}{S_i} \frac{c_{i} n_i}{D_i} \,,
\label{BCJformYM}
\ee
where $c_{i}$ are color factors, $n_i$ are kinematic numerators and $D_i$ are denominators encoding the propagator structure of the cubic graphs.
The denominators may contain masses, corresponding to massive fields in the representation $\rep$. The $S_i$ are standard symmetry
factors that also appear in Feynman loop diagrams.

\begin{figure}[t]
      \centering
      \includegraphics[scale=0.80]{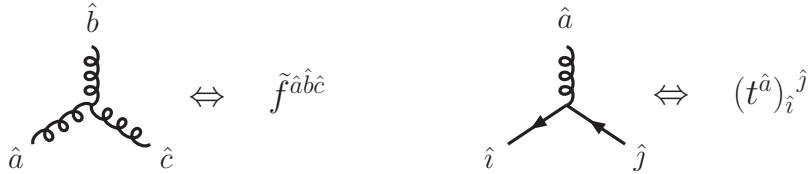}
      \caption[a]{\small The two cubic types of interactions for fields in adjoint representation and a generic complex representation. We organize the amplitudes around cubic graphs with these two types of vertices, and the corresponding color factors are contractions of the structure constants and the generators.   
      \label{BasicVertexFigure} }
\end{figure}

The cubic form~\eqref{BCJformYM} directly follows the organization of the color factors $c_{i}$, which are constructed from two cubic building blocks. 
These are the structure constants $\tf^{\ha \hb \hc}$ for vertices linking three adjoint fields and the generators $(t^{\ha})_{\hi}^{\ \hj}$ for the 
$\rep$-$\repbar$-adjoint vertices, as shown in \fig{BasicVertexFigure}.
When isolating color from kinematics, the crossing symmetry of a vertex only holds up to 
signs dependent on the signature of the permutation. These signs are apparent in the total antisymmetry of $\tf^{\ha \hb \hc}$ and 
may be made uniform by defining the generators in the representation $\rep$
to have a similar antisymmetry:
\be
( t^{\ha})^{\hj}_{\ \hi} \equiv  - (t^{ \ha})_{\hi}^{ \  \hj} ~~~ \Leftrightarrow~~~\tf^{\hc \ha \hb} =
- \tf^{\hb \ha \hc} \  .
\label{signflip}
\ee
The effect of such a relabeling is that any color factor picks a minus sign, $c_i \rightarrow -c_i$, under the permutation of any two graph 
edges meeting at a vertex.

\begin{figure}[t]
      \centering
      \includegraphics[scale=0.93]{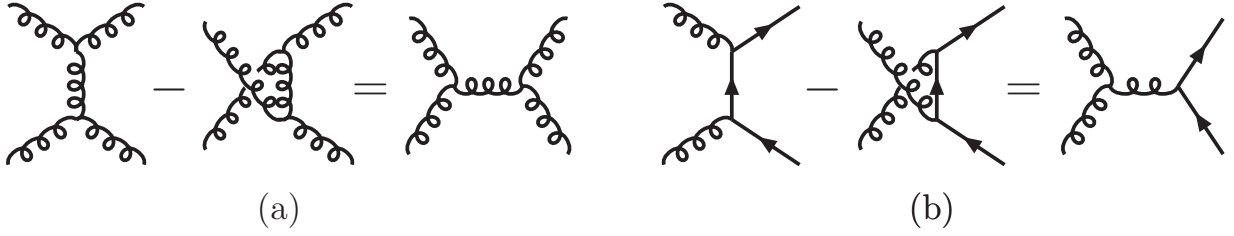}
      \caption[a]{\small Pictorial form of the basic color and kinematic Lie-algebraic relations: (a) the Jacobi relations for fields in the adjoint representation, and (b) the commutation relation for fields in a generic complex representation.  
      \label{BasicJacobiFigure} }
\end{figure}

The color factors obey simple linear relations arising from the Jacobi identities and commutation relations of the gauge group,
\be
 \left. \begin{array}{r}
\tf^{\hd \ha \hc} \tf^{\hc \hb \he} - \tf^{\hd \hb \hc} \tf^{\hc \ha \he}  = \tf^{\ha \hb \hc} \tf^{\hd \hc \he}  \\
   ( t^{\ha})_{\hi}^{\ \hk} \, ( t^{\hb})_{\hk}^{ \ \hj}   -
   ( t^{\hb})_{\hi}^{\  \hk} \, ( t^{\ha})_{\hk}^{ \ \hj}  = \tf^{\ha \hb \hc} \, 
   ( t^{\hc})_{\hi}^{ \ \hj}  \,
\end{array} \right\} ~\Rightarrow ~~  c_{i}-c_{j}=c_{k}\ ;
\label{coloralgebra}
\ee
these relations are shown diagrammatically in \fig{BasicJacobiFigure}. The identity $ c_{i}-c_{j}=c_{k}$ is understood to hold for triplets of diagrams $(i,j,k)$
that differ only by the subgraphs in \fig{BasicJacobiFigure} and otherwise have common graph structure. The linear relations among the color factors~$c_{i}$ imply that the corresponding kinematic parts of the graphs, $n_i/D_i$, are in general not unique. This should be expected, given that individual (Feynman) diagrams are gauge-dependent quantities.

It was observed by Bern, Carrasco and one of the current authors (BCJ)~\cite{Bern:2008qj,Bern:2010ue} that, within the gauge freedom of individual graphs, there exist particularly nice amplitude presentations that make the kinematic numerator factors $n_i$ obey the same general algebraic identities as the color factors $c_{i}$.
In the present context, this implies that there is a numerator relation for every color Jacobi or commutation relation~\eqref{coloralgebra} and a numerator sign flip for every color factor sign 
flip~\eqref{signflip}:
\bea
      n_i - n_j = n_k ~~~&\Leftrightarrow&~~~ c_i - c_j=c_k \,, \nn \\
      n_i \rightarrow -n_i  ~~~&\Leftrightarrow&~~~ c_i \rightarrow -c_i \,.
\label{duality}
\eea
In a more general context, there could exist color identities beyond the Jacobi or commutation relation,
which would justify the introduction of corresponding kinematic numerator identities. 
Indeed, we will encounter this in \sec{Higgs} after introducing additional (bi-fundamental) complex representations of the gauge group. 

Amplitudes built out of numerators that satisfy
the same general identities as the color factors are said to exhibit color/kinematics duality manifestly. 
Theories whose amplitudes can be presented in a form that exhibits this property are said to obey the color/kinematics duality.

It is interesting to note that \eqn{duality} defines a kinematic algebra in terms of the numerators,
which suggests the existence of an underlying Lie algebra.  While not much is known about this kinematic Lie algebra, 
it should be infinite-dimensional due to the momentum-dependence of the numerators.  In the restricted case of self-dual YM theory the kinematic 
algebra has been  shown to be isomorphic to that of the area-preserving diffeomorphisms~\cite{Monteiro:2011pc} (see also ref.~\cite{Monteiro:2013rya}).

A central aspect of the color/kinematics duality is that, once numerators have been found to obey the duality, they can replace the color factors in \eqn{BCJformYM}. This gives a double-copy construction for amplitudes of the form
\be
   {\cal M}^{(L)}_{n} = i^{L-1}\;\!\Big(\frac{\kappa}{2}\Big)^{n-2+2L} \sum_{i \in \text{cubic}}\, \int \frac{d^{LD}\ell}{(2\pi)^{LD}} \frac{1}{S_i} \frac{n_i \tilde{n}_i}{D_i} \,,
\label{BCJformGravity}
\ee
which describe scattering in a gravitational theory.\footnote{If vector contributions are absent in either $n_i$ or $\tilde{n}_i$, 
then \eqn{BCJformGravity} describes a non-gravitational sector.} 
The tilde notation is necessary since the two copies of numerators may not be identical. 
The two sets of numerators entering the double-copy construction may belong to different gauge theories, 
and at most one set is required to manifestly obey the duality~\cite{Bern:2008qj,Bern:2010ue}.

While the double copy discussed here strictly applies to the construction of a gravitational amplitude using the 
scattering amplitudes of two gauge theories as building blocks, it is often convenient the shorten 
the description using the notation  gravity =  ${\rm gauge}\otimes \widetilde{\rm gauge}$. 
This emphasizes the tensor structure of the asymptotic
states of the double copy, and at the same time gives essential information about 
the theories that enter the construction. 
The notation is also motivated by the observations that the double copy appears to have 
extensions beyond perturbation theory~\cite{Monteiro:2011pc,Saotome:2012vy,Monteiro:2014cda,Luna:2015paa}. 

As examples of double copies, we note that pure Yang-Mills theory ``squares'' to gravity coupled to a dilaton and a two-index anti-symmetric tensor: 
GR + $\phi$ + $B^{\mu\nu}$ = ${\rm YM}\otimes {\rm YM}$~\cite{Kawai:1985xq,KLT2}.
Pure Einstein gravity may be obtained by removing these extra particles via a ghost-like double-copy prescription for
massless quarks~\cite{Johansson:2014zca}. An asymmetrical double copy, YM $\otimes$ (YM~+~$\phi^3$), is needed for the amplitudes that 
couple Yang-Mills theory to gravity~\cite{Chiodaroli:2014xia}.
For the double copies of 
YM theories with matter in a complex representation $U$, 
such as described in \eqn{BCJformGravity}, one obtains amplitudes that involve gravitons, dilatons, 
two-index antisymmetric tensors and matter fields~\cite{Johansson:2014zca}. 
In supersymmetric extensions of these theories, superamplitudes are labeled by the corresponding 
supermultiplets; the tensor product of two supermultiplets is typically reducible to a sum of smaller multiplets of the resulting supersymmetry algebra.

While the color/kinematics duality has a conjectural status at loop level, amplitudes up to four loops for diverse theories 
(with and without additional matter) have been explicitly constructed in forms consistent with the duality 
and the double copy~\cite{Bern:2010ue,Carrasco:2011mn,Bern:2011rj, Bern:2012uf,Carrasco:2012ca,Boels:2013bi,Bjerrum-Bohr:2013iza, Bern:2013yya, Nohle:2013bfa,Bern:2013uka,Chiodaroli:2013upa, Johansson:2014zca,Bern:2014sna,Bern:2015ooa, Badger:2015lda, Bern:2015xsa}.

At tree level, the double-copy construction restricted to fields in the adjoint representation 
is known~\cite{Bern:2008qj,Bern:2010yg} to be equivalent to the field-theory limit of the Kawai-Lewellen-Tye (KLT) relations~\cite{Kawai:1985xq,KLT2} 
between open- and closed-string amplitudes. 
Color/kinematics duality has been used to derive a number of impressive results for string-theory 
amplitudes~\cite{BjerrumBohr:2009rd, Stieberger:2009hq, Mafra:2011nv, Mafra:2011nw, Mafra:2012kh, Broedel:2013tta, Stieberger:2014hba};
more generally, the duality combined with string-theory methods provides powerful new tools for field theory~\cite{Mafra:2011kj, Ochirov:2013xba, Mafra:2014oia, Mafra:2014gja, Mafra:2015gia, Mafra:2015mja, He:2015wgf, Lee:2015upy, Mafra:2015vca}. 
Recently, the double-copy construction has been extended to express certain Kerr-Schild-type solutions of general relativity in terms of classical 
solutions of the Yang-Mills equations of motion~\cite{Monteiro:2014cda,Luna:2015paa}.
The duality implies the BCJ amplitude relations~\cite{Bern:2008qj} that limit the number of independent tree amplitudes to $(n-3)!$ in the purely adjoint case,
and otherwise to $(n-3)!(2k-2)/k!$ when $k>1$ fundamental-antifundamental pairs are present~\cite{Johansson:2015oia}.
The BCJ amplitude relations have a close connection to the scattering equations and to the associated string-like formulae 
for gauge and gravity tree amplitudes~\cite{Cachazo:2013iaa, Cachazo:2013gna, Cachazo:2013hca, Cachazo:2013iea,Litsey:2013jfa,Naculich:2014naa,Weinzierl:2014ava,Naculich:2015zha, Geyer:2015bja, delaCruz:2015raa}. Finally, a formulation of the double copy at the level of off-shell linearized supermultiplets was obtained in \cite{Borsten:2013bp,Anastasiou:2013hba,Anastasiou:2014qba,Anastasiou:2015vba}.

\subsection{Scalar $\phi^3$ theories \label{scalar_ths}}

As a warm-up exercise, consider a simple scalar model that exhibits the properties described in the previous section 
where all fields transform either in the adjoint of a group $G_c$ or in a generic complex representation $U$ of this group (and corresponding conjugate $\overline{U}$). 

Suppressing all $G_c$ indices, assume we have a family of real massless scalars transforming in the adjoint representation, labeled as $\phi^a$.  And, similarly, a family of identical-mass complex scalars transforming in the $U$ ($\overline{U}$) representation, labeled as $\varphi_i$ ($\overline{\varphi}^{i}$). 
For a scalar theory with at most cubic interactions the Lagrangian is then\footnote{Scalar and gauge-theory Lagrangians are written in mostly-minus spacetime signature, whereas gravity Lagrangians use mostly-plus signature.}
\be
g^2 \, {\cal L}_{\rm scalar} = {\rm Tr}\Big(\frac{1}{2}\partial_\mu\phi^a \partial^\mu\phi^a + \frac{i}{3!}\lambda \, F^{abc}[\phi^a,\phi^b]\phi^c\Big) 
+ \partial_\mu\overline{\varphi}^{i}  \partial^\mu \varphi_i -
m^2 \overline{\varphi}^{ i} \varphi_i + \lambda \, T^{a\  j}_{\ i} \, (\overline{\varphi}^i \phi^a \varphi_j) \, .
\label{Lagrangian_ini}
\ee
Note that the indices $a,b,c,\ldots$ and $i,j,\ldots$ are not $G_c$ indices, but rather labels that distinguish fields in the same representation (see \app{appendixA} for a summary of notation). The coefficients $F^{abc}$ and $ T^{a \  i}_{\ j}$ are arbitrary couplings between these fields, and $\lambda$ is a dimension-one constant (in four dimensions) such that all terms in ${\cal L}_{\rm scalar}$ have uniform dimension. For later convenience we have also introduced a dimensionless coupling $g$.

Denoting by $(t^{\ha})_{\hi}^{\ \hj}$ the generators\footnote{We normalize the generators as ${\rm Tr}( t^{\ha} t^{\hb})={1 \over 2}\delta^{\ha \hb}$.} 
of $G_c$ in the representation $U$ and expressing the adjoint fields as $\phi^a = t^{\ha} \, \phi^{a \ha}$, a more explicit form of the Lagrangian can be obtained,
\be
{\cal L}_{\rm scalar} = \frac{1}{2}\partial_\mu\phi^{a \ha} \partial^\mu\phi^{a \ha} +  
\frac{1}{3!}g \lambda  F^{abc} f^{\ha \hb \hc} \phi^{a \ha}\phi^{b \hb}\phi^{c \hc} + 
\partial_\mu\overline{\varphi}^{i}  \partial^\mu\varphi_{i}
-\overline{\varphi}^i m^2 \varphi_i  +  g \lambda T^{a \ j}_{\ i} \phi^{a \ha} \ \overline{\varphi}^{i} t^{\ha}  \varphi_{j}
\label{Ls} \ . \ee
Here $f^{\ha \hb \hc}= - i {\rm Tr}([t^{\ha},t^{\hb}] t^{\hc})$ are 
the structure constants of the group~$G_c$, the coupling constant 
$g$ has been moved to the cubic interactions via the redefinition 
$\phi \rightarrow g\phi$, $\varphi \rightarrow g\varphi $, and the indices of the complex representation $U$ remain suppressed. 

The symmetry $G_c$ can be gauged, as we will do in the next section. Even before gauging, scattering amplitudes from ${\cal L}_{\rm scalar}$ have the same form as \eqn{BCJformYM}, with the coefficients $c_i$ given in terms of the generators and structure constants of $G_c$. Anticipating the gauging of $G_c$ we can constrain the Lagrangian~\eqref{Ls} such that amplitudes expressed in this form have numerators $n_i$ that obey the duality \eqref{duality}, in one-to-one correspondence with those obeyed by the group-theoretic factors $c_i$. 
This simple theory has no derivative couplings, and therefore the numerator factors $n_i$ have no momentum dependence, they are only built out of the couplings $F^{abc}$ and $T^{a\  j}_{\ i}$. An inspection of the Lagrangian shows that the duality holds if the couplings are in one-to-one correspondence with the structure constants and generators of $G_c$,
\be
 F^{abc}~ \Leftrightarrow ~ f^{\ha \hb \hc}~~~~~~\text{and} ~~~~~~ T^{a \ j}_{\ i}~ \Leftrightarrow ~ (t^{\ha})^{\ \hj}_{\hi}\,,
\ee
in the sense that the pair $(F^{abc}, T^{a})$ obeys the same general algebraic relations as $(f^{\ha \hb \hc}, t^{\ha})$.

This implies that 
\begin{enumerate}
\item
  $(T^{a})^{\ j}_{ i} \equiv  T^{a \ j}_{\ i}$ are the generators of a generic complex representation $U'$  
  of a ``kinematic'' Lie algebra\footnote{This name is convenient because, once the $G_c$ symmetry is gauged, the Lie algebra of $G_k$ becomes a subalgebra of the full kinematic algebra obeyed by the numerator factors.} of some group $G_k$. They can be taken to be normalized as ${\rm Tr}(T^aT^b)={1 \over 2}\delta^{ab}$.
\item
 $F^{abc}$ are the structure constants of that algebra given by $F^{abc}= - 2 i {\rm Tr}([T^a,T^b] T^c)$. 
\item
 The ranges of indices $a, b, c,\dots$ and $i, j, k, \dots$ are the dimensions of the adjoint representation of 
 $G_k$ and  its representation $U'$, respectively.
\end{enumerate}
The resulting theory describes a  $G_c \otimes G_k$ invariant scalar field theory, with massless scalars $\phi^{a{\hat a}}$ 
in the ``bi-adjoint'' representation and  massive complex scalar fields $\varphi^{i{\hat\imath}}$ in the representation $U\otimes U'$. 
This is one of the simplest realizations of a theory that exhibits a duality of the type described in \sec{sec.ckdu} which is manifest in the Lagrangian. 

Note that it is straightforward to modify the mass spectrum of the theory while preserving 
the duality. If the $G_c$ representation $U$ and/or 
the $G_k$ representation $U'$ are reducible, the mass $m$ in \eqn{Ls} can carry labels identifying the irreducible components of $U$ and $U'$. Hence,
the $U\otimes U'$ representations can be decomposed into irreps of $G_c \otimes G_k$, each with a different mass term in the Lagrangian.

As a concrete example of this generalization, take the kinematic algebra to be $G_k=SU(N_k)$, and let the representation $U'$ be $N_f$ copies of the fundamental 
representation; these copies are labeled by the {\it flavor} indices $m,n=1,\dots,N_f$. Next take $G_c=SU(N_c)$, and let the representation $U$ be its fundamental 
representation. With these choices, the scalar theory takes the form
\bea
{\cal L}_{\rm scalar}' &=& \frac{1}{2}\partial_\mu\phi^{a \ha} \partial^\mu\phi^{a \ha} 
+\frac{1}{3!}g \lambda  F^{abc} f^{\ha \hb \hc} \phi^{a \ha}\phi^{b \hb}\phi^{c \hc} \nn \\
&& \null + \partial_\mu\overline{\varphi}^{i \ffm}  \partial^\mu\varphi_{i \ffm} 
- (m^2)_{\ffm}^{\ \ffn} \overline{\varphi}^{i \ffm}  \varphi_{i \ffn} + 
g \lambda  T^{a \ j}_{\ i } \phi^{a \ha} \overline{\varphi}^{i \ffm} t^{\ha}  \varphi_{ j \ffm} \ ,
\label{chooseReps}
\eea
where $t^{\ha}$ and $T^{a}$ are generators in the fundamental representation of respective group. 
The fundamental S$U(N_c)$ indices $\hi, \hj$ are not shown explicitly. The mass matrix is assumed to be
diagonalized, $m_{\ffm}^{\ \ffn}= \delta_{\ffm}^{\ \ffn} \, m_\ffn $ (no sum), corresponding 
to the mass eigenstates: $\varphi_{i \hi \ffn}$ and $\overline{\varphi}^{i \hi \ffn}$.
In the limit that $m_{\ffn} \rightarrow 0$ (or $m_{\ffn} \rightarrow m$) 
this theory has ${SU}(N_c)\times {SU}(N_k) \times {SU}(N_f)$ symmetry, 
where ${SU}(N_f)$ is the flavor group.  For generic $m_{\ffn}$ the flavor group is explicitly broken to ${SU}(N_f)\rightarrow {U}(1)^{N_f}$.
The case $N_f = 0$ is that of the pure bi-adjoint $\phi^3$ theory,
\be
{\cal L}_{\rm \phi^3} = \frac{1}{2}\partial_\mu\phi^{a \ha} \partial^\mu\phi^{a \ha} +  \frac{1}{3!}g \lambda F^{abc} f^{\ha \hb \hc} 
\phi^{a \ha}\phi^{b \hb}\phi^{c \hc}
\label{Lphi3} \ ,
\ee
which was identified in refs. \cite{Bern:1999bx,Chiodaroli:2014xia} to be useful for obtaining amplitudes in gravity
theories coupled to non-abelian gauge fields with ${SU}(N_k)$ symmetry.\footnote{Compared to the notation used in ref.~\cite{Chiodaroli:2014xia}, 
we have renamed the two couplings: ${\rm g} \rightarrow g$, $g' \rightarrow \lambda$.} See also refs.~\cite{Du:2011js, BjerrumBohr:2012mg} 
for other applications of this theory in the context of color/kinematics duality.

\subsection{Yang-Mills-scalar theories\label{YMscalar1}: gauging $G_c$ \label{YMscalar_unbroken}}

Let us now gauge the symmetry group $G_c$ and include the self-interactions of the corresponding non-abelian gauge fields. In \eqn{Lagrangian_ini} we may replace all derivatives by covariant derivatives in the representation $U$, $\partial_\mu\rightarrow D_\mu$, 
and add the standard pure-Yang-Mills Lagrangian with gauge group $G_c$.

Gauging the $G_c$ symmetry is not sufficient for the resulting theory to obey color/kinematics duality; indeed, it is known from the $N_f =0$ case~\cite{Chiodaroli:2014xia,Johansson:2013nsa} as well as from the case of fundamental and orbifold field theories \cite{Chiodaroli:2013upa,Johansson:2014zca} that 
quartic scalar terms like $\phi^4$, $\phi^2 \overline{\varphi} \varphi$ and $(\overline{\varphi} \varphi)^2$ are required. For the particular theories discussed in this subsection, color/kinematics duality will uniquely dictate the $\phi^4$ and $\phi^2 \overline{\varphi} \varphi$ terms, whereas all terms of $(\overline{\varphi} \varphi)^2$ type will be unconstrained. However,
if $\overline{\varphi}$ and $\varphi$ are in special complex representations for which the color factors obey extra 
identities, then the $(\overline{\varphi} \varphi)^2$ terms may be constrained by color/kinematics duality. We will see that these special representations include the 
ones arising from the spontaneous symmetry breaking of a larger gauge group.

In ref.~\cite{Chiodaroli:2014xia} we showed that the specific $\phi^4$ term that is consistent with color/kinematics duality is
\be
{\cal L}_ {\phi^4} =  -\frac{g^2}{4} f^{\ha \hb \he} f^{\he \hc \hd}\phi^{a \ha}\phi^{b \hb}\phi^{a \hc}\phi^{b \hd} \ .
\label{adjointContact} 
\ee
In \sec{secamp} we will compute four-point amplitudes in the YM-scalar theories and see that they obey color/kinematics duality
only if the Lagrangian also contains the term
\be
{\cal L}_ {\phi^2 \overline{\varphi} \varphi } = - 
g^2  \phi^{a \ha} \phi^{a \hb} \overline{\varphi}^{i} t^{\ha} t^{ \hb} \varphi_{i} \ .
\ee
There are several terms involving four fields in complex representations
that can in principle be freely added; we find that the combination 
\be
{\cal L}_ {(\overline{\varphi} \varphi)^2} = - g^2    \overline{\varphi}^{i } 
t^{\ha} \varphi_{j } \, \overline{\varphi}^{j} t^{\ha} 
\varphi_{i }+ \frac{g^2}{2}   
\overline{\varphi}^{i} 
t^{\ha} \varphi_{i} \, \overline{\varphi}^{j} t^{\ha} 
\varphi_{j}\,
\ee
is particularly natural as it is in a certain sense (discussed in \sec{YMBreakScalarEx}) the 
complex generalization of the adjoint contact term~(\ref{adjointContact}). 

Thus, the Lagrangian with local symmetry $G_c$ and global symmetry $G_k$, giving Yang-Mills theory coupled to scalar fields, takes the following form:
\be
{\cal L}_{\rm YM+scalar} = -\frac{1}{4}F_{\mu\nu}^{\ha}F^{\mu\nu \ha} + {\cal L}_{\rm scalar}\Big|_{\partial \rightarrow D} +{\cal L}_ {\phi^4}+ {\cal L}_ {\phi^2 \overline{\varphi} \varphi }+{\cal L}_ {(\overline{\varphi} \varphi)^2} \,.
\label{YMscalar_general}
\ee
For the particular choices of groups and representations that led to the theory \eqref{chooseReps}, the Lagrangian is
\bea
{\cal L}'_{\rm YM+scalar} &=&-\frac{1}{4}F_{\mu\nu}^{\ha}F^{\mu\nu \ha}+\frac{1}{2}(D_\mu\phi^{a})^{\ha} (D^\mu\phi^{a})^{\ha} +
\frac{1}{3!}g \lambda  F^{abc} f^{\ha \hb \hc} \phi^{a \ha}\phi^{b \hb}\phi^{c \hc} \nn \\
&& \null +  \overline{D_\mu \varphi^{ i \ffm}}  D^\mu \varphi_{i \ffm} 
- (m^2)_{\ffm}^{\ \ffn}\, \overline{\varphi}^{ i \ffm}  \varphi_{i \ffn} +
g \lambda  T^{a \ j}_{\ i} \phi^{a \ha} \overline{\varphi}^{ i \ffm} t^{\ha} \varphi_{j \ffm} \nn \\
&& \null +\frac{g^2}{4} f^{\ha \hb \he} f^{\he \hc \hd}\phi^{a \ha}\phi^{b \hb}\phi^{a \hc}\phi^{b \hd} - g^2  \phi^{a \ha} \phi^{a \hb} \overline{\varphi}^{i \ffm} t^{\ha} t^{ \hb} \varphi_{i \ffm} \nn \\
&& \null - g^2    \overline{\varphi}^{i \ffm} 
t^{\ha} \varphi_{j \ffn} \, \overline{\varphi}^{j \ffn} t^{\ha} 
\varphi_{i \ffm}+ \frac{g^2}{2}   
\overline{\varphi}^{i \ffm} 
t^{\ha} \varphi_{i \ffm} \, \overline{\varphi}^{j \ffn} t^{\ha} 
\varphi_{j \ffn}\,.
\label{YMscalar}
\eea
This theory has a local symmetry $SU(N_c)$, a global symmetry $SU(N_k)$, and a broken flavor symmetry ${SU}(N_f)\rightarrow {U}(1)^{N_f}$ generically (for special choices of mass matrix, it is broken to some subgroup ${SU}(N_f)$). 
We will derive the Lagrangian \eqref{YMscalar} in \sec{YMBreakScalarEx} as a particular truncation of a gauge theory with broken global symmetry. We expect that it obeys color/kinematics duality, at least at tree level, as it should inherit this property from the broken theory considered in \sec{YMBreakScalar}. The corresponding BCJ relations for tree-level amplitudes in the theories \eqref{YMscalar_general} and \eqref{YMscalar} should be the same as those of QCD~\cite{Johansson:2015oia}.  Note that  theories \eqref{YMscalar_general} and \eqref{YMscalar} do not admit obvious supersymmetric extensions unless $\lambda=0$. 

In the next two sections we consider spontaneous symmetry breaking for dimensionally-reduced YM theories (obtained by setting $\lambda=0$ and $N_f=0$) including supersymmetric extensions, and, similarly, explicit symmetry breaking in a YM + $\phi^3$ theory (obtained by setting $N_f=0$).

\subsection{Adjoint Higgs mechanism: breaking $G_c$ \label{Higgs}}

Here we briefly review Yang-Mills theories for which the gauge symmetry is spontaneously broken by an adjoint Higgs field, and introduce the color/kinematics duality in this setting. This a necessary ingredient in the double-copy construction of Yang-Mills-Einstein supergravity theories with spontaneously-broken gauge symmetry.  While supersymmetry is not required by the construction, its presence facilitates the identification of gravitational theories generated by the double-copy prescription. 

Consider a YM-scalar theory that is the dimensional reduction of pure YM theory,
\be 
{\cal L}_{\rm YM_{DR}} =  -{1\over 4} {\cal F}^{\hat A}_{\mu \nu} {\cal F}^{\mu \nu \hat A} + {1 \over 2} ({\cal D}_{\mu} \phi^{\si})^{\hat A} ({\cal D}^{\mu} \phi^{\si})^{\hat A}  
- {g^2 \over 4} f^{\hat A \hat B \hat E} f^{\hat C \hat D \hat E} \phi^{ \si \hat A} \phi^{ \sj \hat B} \phi^{ \si \hat C} \phi^{\sj \hat D}\,,
\label{unbroken}
\ee
where $\hat A, \hat B,\ldots$ are adjoint gauge indices, and $a, b,\ldots$ are global symmetry indices. 
The indices $\si, \sj = 0,1, \ldots N'_\phi-1$ run over the 
different real scalar fields in the theory. 
For example, considering the particular cases $N'_\phi=2$ or $N'_\phi=6$ in $D=4$ dimensions, we obtain the bosonic part of the $\cN=2$ or $\cN=4$ SYM 
Lagrangians, respectively. 
In the $\cN=4$ case, the scalars transform in the anti-symmetric tensor representation of the $R$-symmetry group $SU(4)$, and  
in $\cN=2$ theories the scalars carry a charge only under the $U(1)$ part of the full $R$-symmetry group $SU(2)\times U(1)$.

It is well-known that the Coulomb-branch vacua of this theory are described by constant scalar fields solving
\be
[\phi^a, \phi^b] = 0 \ ,~~~ \phi^a\equiv  \phi^{{\hat A} a} t^{\hat A} \ , 
\ee
where $t^{\hat A}$ are the generators of the gauge group.
We choose a vacuum with scale $V$ such that the vacuum expectation value (VEV) of the field $\phi^0$ is proportional  to a single gauge group generator $t^0$,
\be
\langle \phi^a\rangle = V t^0 \, \delta^{a0} \ .
\label{scalarVEV}
\ee
With this choice, we can interpret the theory with $N'_\phi=2$ as the dimensional reduction of a spontaneously-broken half-maximal SYM theory in five dimensions where $\phi^0$ is the scalar of the vector multiplet.
The fact that our construction uplifts to $D=5$ dimensions will be useful when identifying 
the corresponding supergravity Lagrangian obtained by the double-copy construction. Similarly, for $N'_\phi=6$, the theory can be uplifted  to the spontaneously-broken maximally-supersymmetric YM theory in $D\le9$. For convenience of presentation, in the following we will ignore terms containing fermions in the supersymmetric Lagrangians. 

\begin{figure}[t]
      \centering
      \includegraphics[scale=0.84]{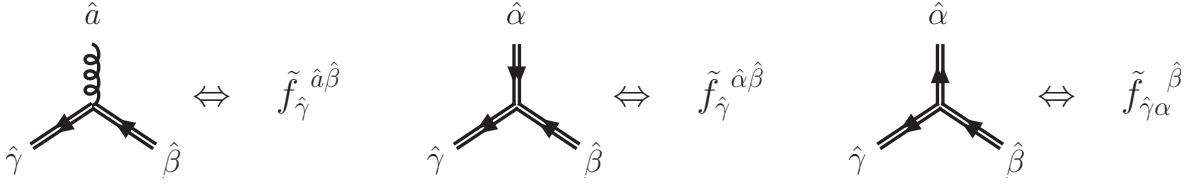}
      \caption[a]{\small Additional types of cubic interactions that are obtained after the gauge symmetry is spontaneously-broken in a purely adjoint theory.  The resulting amplitudes are organized around cubic graphs where these vertices are included. The corresponding color factors are contractions of the various types of structure constants.  
       \label{BrokenVertexFigure} }
\end{figure}

The Higgsed Lagrangian corresponding to (\ref{unbroken}) is obtained by splitting the scalar and vector fields as
\bea 
A^{\hat  A}_\mu = \big( A^{\hat  a}_\mu , W_{\mu \hat  \alpha}, \overline{W}^{ \hat { \alpha}}_\mu\, \big) \ , \qquad 
\phi^{ \hat  A \si} = \big( \phi^{ \si \hat  a} , \varphi^{\si}_{ \ \hat  \alpha}, \overline{\varphi}^{  \si \hat { \alpha}} \big) \ , \eea
so that the index 
$\ha$ runs over the adjoint representation of the unbroken part of the gauge group and the index $\haa$ runs over the matter-like non-adjoint complex representations. 
Under this split, the only non-zero entries of the structure constants $f^{\hat A \hat B \hat C}$ are\footnote{Note that we may freely cyclicly permute the indices, e.g. $f^{\haa \  \hgg}_{ \ \hbb}= f^{ \  \hgg \haa}_{\hbb}=f^{\hgg \haa}_{\ \ \, \hbb}$. }
\be 
f^{\ha \hb \hc}=  - i {\rm Tr}([t^\ha, t^\hb] t^\hc) \,,~~~f^{\ha \ \haa}_{ \ \hbb}= - i {\rm Tr}([t^\ha, (t^\hbb)^\dagger] t^\haa)  \,,~~~f^{\haa \  \hgg}_{ \ \hbb}=(f_{\hgg \ \haa}^{\ \hbb})^\dagger= - i {\rm Tr}([t^\haa, (t^\hbb)^\dagger] t^\hgg) \,, \label{fsplitgauge}
\ee
which are the structure constants of the unbroken part of the gauge group, the generators and Clebsch-Gordan coefficients for the matter-like representations.
Since the scalar VEV has been taken along the gauge group generator $t^0$,  the mass matrix has the expression 
\be 
m_{\haa}^{\ \hbb} = {i g \vym} f^{0 \ \hbb}_{\ \hat{\alpha}} \label{massM} \ .
\ee

Expanding the original covariant derivative $(D_{\mu} \phi^{\si})^{\hat A} = \partial_\mu \phi^{\hat A \si} + g f^{\hat A \hat B \hat C} A^{\hat B}_\mu \phi^{\hat C \si}$ 
and the covariant field strengths around the scalar VEV, and decomposing these objects 
in representations of the unbroken part of the gauge group leads to\footnote{
We use the shorthand notation
\bea
\overline{V}^{\hbb}  f^{\ha \ \hgg}_{\ \hbb} U_\hgg  \rightarrow  \overline{V} f^{\ha} U\,,~~~\overline{V}^{\hbb}  f^{\haa \ \hgg}_{\ \hbb} U_\hgg  \rightarrow  \overline{V} f^{\haa} U\,,~~~V_{\hbb}  f^{\hbb \ \hgg}_{\ \haa} U_\hgg \rightarrow V f_{\haa} U\,,~~~\overline{V}^{\hbb}  f_{\hbb \ \hgg}^{\ \haa} \overline{U}^\hgg  \rightarrow \overline{V} f^{\haa} \overline{U}\,.
\nonumber
 \eea
 }
\bea
{(\brk{D}_\mu \phi^{\si})\!}^{\,\hat A }
  \!  \!\!\!  &=& \!\!\! \! \! \left( \!\! \begin{array}{c}   (D_\mu \phi^{\si})^{\ha } + g 
     \overline{W}_\mu f^{\ha} \varphi^\si - g {\overline{ \varphi}}^\si f^{\ha} W_\mu  \\[5pt]
    (D_\mu \varphi^{\si})_{\haa} - {i} \delta^{\si 0} (m {W}_{\mu})_{\haa}  +
    g {\phi}^{ \si \ha} (f^{\ha} W_\mu)_{\haa} +
    g  \overline{W}_\mu  f_{\haa} \varphi^\si  -
    g  \overline{\varphi}^\si f_{\haa} W_\mu  +
    g{\varphi}^\si f_{\haa} W_\mu
      \\[5pt]
    (\overline{D_\mu{\varphi}^{\si}})^{\haa} \! +  {i} \delta^{\si 0} 
    { (\overline{W}_\mu m)}^{\haa} \! -g \phi^{\si \ha} ( {\overline{W}}_\mu f^{\ha})^{\haa}
    \! +g \overline{W}_\mu  f^{\haa}  \varphi^{\si} \! -g
    \overline{\varphi}^{\si}   f^{\haa}  W_\mu \! -g
      \overline{W}_\mu f^{\haa} \overline{\varphi}^{\si}
     \end{array} \!\! \right), \no \\ \eea \bea
   %
  %
\brk{F}^{\hat A}_{\mu \nu}
 & \!\!\! \!\!\!  =& \!\!\! \!\!\! \left( \begin{array}{c}   F^{\ha}_{\mu \nu} +2 g \overline{W}_{[\mu}f^{\ha} W_{\nu]}  \\[5pt]
    2 (D_{[\mu}{ W}_{\nu]})_{\haa} + 2  g  \overline{W}_{[\mu} f_{\haa} {W}_{\nu]}  -  g
    W_{\mu} f_{\haa} W_\nu \\[5pt]
    2 (\overline{D_{[\mu} W_{\nu]}})^{\haa} + 2 g  \overline{W}_{[\mu} f^{\haa} {W}_{\nu]}  
    - g\overline{W}_{\mu}f^{\haa}  \overline{W}_\nu
     \end{array} \right) \ .
\label{FAmn} 
\eea
In general, the matrix $m_{\haa}^{\ \hbb}$ is block diagonal, with each block corresponding 
to different irreducible representations. As usual, the mass of the scalar fields in the matter-like representation corresponding to the generator $t^0$ ($\varphi^{0{\hat\alpha}}$) depends 
on the choice of gauge. In the unitary gauge its mass is infinite and this field decouples.
In this gauge the Lagrangian is
\bea
{\cal L}_{\cancel{\rm YM}_{\rm DR}} &=&  -{1\over 4}  \brk{F}^{\hat A}_{\mu \nu} \, \brk{F}^{\mu \nu \hat A} + {1 \over 2}\,(\brk{D}_\mu \phi^{\si})^{\,\hat A } \,{(\brk{D}^\mu \phi^{\si})\!}^{\,\hat A }
- {g^2 \over 4} f^{\ha \hb \he} f^{\hc \hd \he} \phi^{ \si \ha} \phi^{ \sj \hb} \phi^{ \si \hc} \phi^{\sj \hd} - \overline{\varphi}^{a \haa} (m^2)_{\haa}^{\ \hbb} \varphi^{a}_{\ \hbb} \nn \\
&&\null 
- 2 i g f^{\hat a \ \hgg}_{\ \haa} m_{\hgg}^{\ \hbb} \phi^{0 \ha} \overline{\varphi}^{a \haa} \varphi^a_{\hbb} + V_{4}\big( \phi, \varphi \big) \ ,
\label{brokenGeneral}
\eea
where we have written explicitly the cubic term in the scalar potential of \eqn{unbroken}. 


\begin{figure}[t]
      \centering
      \includegraphics[scale=0.92]{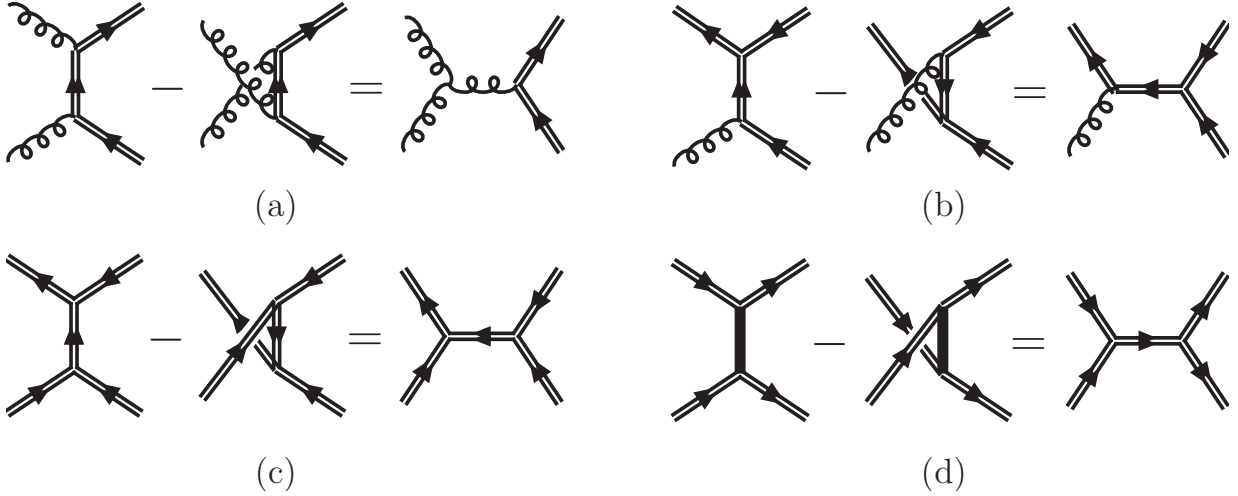}
      \caption[a]{\small Pictorial representation of additional color Lie-algebra relations that are obtained after the gauge symmetry spontaneously-broken in a purely adjoint theory. These are also pictorial representations of the kinematic algebra that should be imposed on diagram numerators in the context of color/kinematics duality. The relations are generalizations of the Jacobi identity. Curly lines represent unbroken adjoint states (massless fields) and double lines represent broken non-hermitian states (massive fields). Solid fat lines in (d) represent sums over all three types of states (the massless and two conjugates of the massive ones), giving seven terms in the (d) identity. \label{BrokenJacobiFigure} }
\end{figure}

Focusing on the $N'_\phi=2$ case (corresponding to the bosonic part of $\cN=2$ SYM) in the unitary gauge, 
we have only one family of massive real scalars $\varphi_\haa \equiv  \varphi^{1}_{\ \haa}$; 
the other 
family, $\varphi^{0}_{\ \haa}$, becomes the longitudinal component of the $W$ bosons.

The structure constants, generators and Clebsch-Gordan coefficients obey relations inherited from the Jacobi relations of the original gauge group. A first set of relations is
\bea
 f^{\hd \ha \hc} f^{\hc \hb \he} - f^{\hd \hb \hc} f^{\hc \ha \he}  &=& f^{\ha \hb \hc} f^{\hd \hc \he} \,, \nn \\
 f^{\ha \ \hbb}_{\ \hgg } f^{\hb \ \hgg }_{\ \haa}-  f^{\hb \ \hbb}_{\ \hgg }f^{\ha \ \hgg }_{\ \haa}  &=& f^{\ha \hb \hc} f^{\hc \ \hbb }_{\ \haa} \, , \no \\
f^{\ha \ \hgg}_{\ \hee} f^{\hee \ \hbb}_{\ \hdd} -
f^{\ha \ \hbb}_{\ \hee} f^{\hee \ \hgg }_{\ \hdd} &=& 
f^{\ha \ \hee}_{\ \hdd} f^{\hgg \ \hbb }_{\ \hee}  \, . 
 \label{fck}
 \eea
These relations are necessary to ensure gauge invariance in any gauge theory (with or without massive vectors). Since they are components of the structure constants of a larger group, and since they control the gauge invariance of massive vector interactions, the Clebsch-Gordan coefficients $f^{\hgg \ \hbb }_{\ \hee}$ need to obey two further identities: 
\bea
f^{\haa \ \hgg}_{\ \hee} f^{\hee \ \hbb}_{\ \hdd}  -
f^{\haa \ \hbb}_{\ \hee} f^{\hee \ \hgg }_{\ \hdd}  &=& 
f^{\haa \ \hee}_{\ \hdd} \, f^{\hgg \ \hbb}_{\ \hee} \,, \nn\\ 
\Big(f^{\hbb \ \hee}_{\ \hgg} f_{\hee \ \hdd}^{\ \haa}
+ f^{\haa \ \hee}_{\ \hdd} f_{\hee \ \hgg}^{\ \hbb} 
 + f^{\ha \ \hbb}_{\ \hgg} f^{\ha \ \haa}_{\ \hdd}\Big)- (\haa\leftrightarrow\hbb)
&=& f^{\haa \ \hbb}_{\ \hee} f_{\hdd \ \hgg}^{\ \hee}  \, . 
\label{extraf} 
\eea
It is important to note that, for a given assignment of external masses,  at most three terms of the above seven-term identity can be non-zero.
Hence, the seven-term identity can be thought of as a compact notation for a set of
distinct three-term identities. These three-term identities will be the ones imposed on the numerator factors in a duality-satisfying amplitude presentation.  

We should also note that, depending on the field content, the relations in \eqn{extraf}
could be relaxed, in the sense of replacing $f^{\hgg \ \hbb }_{\ \hee}$
by another (more general) solution to \eqn{fck}.
This is the case when the  fields transforming in the matter representations are scalars or/and fermions.  However, if massive vectors transform in matter representations of the unbroken gauge group, then these extra relations are required by the consistency of the theory (as the massive vectors can arise only through a Higgs mechanism).

Color/kinematics duality for YM theories with gauge symmetry spontaneously broken by an adjoint Higgs field is 
implemented by requiring that the kinematic numerators of scattering amplitudes in these theories obeys identities that mirror the color identities 
in \eqn{fck} and \eqref{extraf}. Except for the Jacobi identity, these kinematic identities are pictorially shown in \fig{BrokenJacobiFigure}. 
Note that since these identities always break up into three-term identities, they can in practice be mapped to the usual 
three-term numerator identities considered in the framework of color/kinematics duality. Indeed, as it is well known (e.g. see refs.~\cite{Alday:2009zm,Boels:2010mj}), amplitudes in SYM theory on the Coulomb branch can be reinterpreted as amplitudes in a $(D+1)$ dimensional unbroken SYM theory (see \app{SYMdredSect} for a Lagrangian derivation of this). For a SYM theory on the Coulomb branch the kinematic identities in \fig{BrokenJacobiFigure} are simply obtained through a decomposition of the usual $(D+1)$ dimensional kinematic Jacobi identity into states with zero (massless states) and positive/negative (massive states) momentum in the $(D+1)$ direction.\footnote{Note that the type of kinematic algebra introduced here is more general; it need not be inherited from $(D+1)$ dimensions. For example, it applies to the explicitly broken YM + $\phi^3$ theory considered in~\sec{YMBreakScalar}.}

An important consideration for color/kinematics duality to give well-behaved double copies, is that that we construct amplitude presentations valid for arbitrary gauge groups and arbitrary breaking patterns. This is to prevent the color factors from obeying accidental algebraic relations beyond those of eqs.~\eqref{fck} and \eqref{extraf}, as it might happen for particular choices of gauge groups and gauge-symmetry breakings. 

It is useful to note that in the basis in which the mass matrix is diagonal, $m_\haa^\hbb = \delta_\haa^\hbb \, m_\haa$ (no sum), there is a direct correspondence between the masses of the complex fields and the non-vanishing structure constants involving the broken generators,
\bea
f^{\haa \ \hgg}_{\ \hbb} \neq 0 ~~~~ &\Leftrightarrow&  ~~~~ m_\haa+m_\hgg=m_\hbb\,, \nn \\
f^{\ha \ \hgg}_{\ \hbb} \neq 0 ~~~~ &\Leftrightarrow&  ~~~~ m_\hgg=m_\hbb\,.
\eea
Such relations arise from the proportionality relation between the mass and the charge with respect to the  preferred $U(1)$ generator in \eqref{massM}. 
As a trivial consequence of \eqref{massM}, mass and charge obey the same three-term identities
\bea
\hskip1cm 
q_\haa+q_\hgg=q_\hbb~~~~ &\Leftrightarrow&  ~~~~ m_\haa+m_\hgg=m_\hbb \,, \nn \\
\hskip1.9cm 
q_\hgg=q_\hbb ~~~~ &\Leftrightarrow&  ~~~~ m_\hgg=m_\hbb\,,
\label{chargeMassConservation}
\eea
which can be seen as charge/mass conservation for the trilinear interactions. 
The double-copy construction that we will spell out in \sec{2copySec} requires the masses in the (super)gravity theory 
to be equal to the ones in the two gauge-theory factors and relies on the charge/mass conservation at each vertex. Interestingly, the close relationship between masses and charges is similar to the double copy framework discovered in refs.~\cite{Monteiro:2014cda,Luna:2015paa}, where charges of gauge-theory classical solutions were interchanged with masses of classical gravitational solutions.

In sections \ref{secamp} and \ref{sec:loops}, we present tree- and loop-level amplitudes in spontaneously-broken SYM that 
exhibit color/kinematics duality.

 
\subsubsection{${SU}(N)$ Examples \label{examples}}

For the purpose of illustration, in this subsection we include two simple examples of spontaneous symmetry breaking. The simplest breaking pattern is  
\be
{SU}(N_1+N_2)\rightarrow {SU}(N_1)\times {SU}(N_2)\times {U}(1) \ .
\ee 
This pattern can be obtained by giving  a VEV
\be 
\langle \phi^0 \rangle = \vym \left( \begin{array}{cc} {1 \over N_1} I_{N_1} & 0 \\ 0 & - {1 \over N_2} I_{N_2} \end{array}  \right) \ ,
\ee
where we have absorbed a normalization constant in the VEV. As discussed,  we denote the corresponding generator as $t^0$  
(with a proper normalization factor) 
and the generators of the unbroken subgroup that commutes with $t^0$ as $t^{\ha}$,
with $a=1,2,\ldots , N_1^2 +N_2^2 -2$. 
The remaining ``non-hermitian'' generators can be divided into two conjugate sets,
\be 
(t^{(kl)})_{\hi}^{\ \hj} =  \delta_{\hi}^k \delta^{\hj}_l~~~~~ {\rm and} ~~~~~ (t_{(kl)})_{\hi}^{\ \hj} =  \delta_{\hi}^l \delta^{\hj}_k\,,  
\ee
where we introduced the composite index $\alpha = (kl)$ with $k = 1,\ldots, N_1$ and $l = N_1 + 1,\ldots, N_1 + N_2$. 
With this choice, the mass matrix is diagonal and has a single eigenvalue:
\be 
m = g \vym \Big({1 \over N_1} + {1 \over N_2} \Big) \ .
\ee
The theory can be represented by a quiver diagram with two nodes and two lines with opposite orientations connecting them. 
In the supersymmetric case each node corresponds to a massless adjoint vector multiplet and each link corresponds to a 
massive bifundamental vector multiplet. 

The simplest example with several masses involves the breaking pattern
\be 
{SU}(N_1+N_2 + N_3) \rightarrow{SU}(N_1)\times{SU}(N_2)\times{SU}(N_3)\times{U}(1)^2 \ .
\ee
It can be realized by choosing a scalar VEV with three diagonal blocks
\be 
\langle 
\phi^0 \rangle =  
V \left( \begin{array}{ccc} {v_1 \over N_1} I_{N_1} & 0 & 0 \\ 0 & {v_2 \over N_2} I_{N_2} & 0 \\
0 & 0 & - {v_1+v_2 \over N_3} I_{N_3} \end{array}  \right) 
\ .
\ee
In this case, the broken generators can be divided into six sets. The three upper-diagonal sets of generators are 
\bea &&
(t^{(kl)})_{\hi}^{\ \hj} =  \delta_{\hi}^k \delta^{\hj}_l\,, ~~~~~ (t^{(kr)})_{\hi}^{\ \hj} =  \delta_{\hi}^k \delta^{\hj}_r\,, ~~~~~ (t^{(lr)})_{\hi}^{\ \hj} =  \delta_{\hi}^l \delta^{\hj}_r\,,  
\eea
with $k=1,\ldots, N_1$; $l=1+N_1,\ldots, N_1+N_2$; $r=1+N_1+N_2,\ldots, N_1+N_2+N_3$, 
and the corresponding eigenvalues are
\be m_1 = g V  \Big({v_1 \over N_1} - {v_2 \over N_2} \Big) \ , \qquad m_2 = g V  \Big({v_1 \over N_1} + {v_1 + v_2 \over N_3} \Big) \ ,
\qquad  m_3 = g V  \Big({v_2 \over N_2} + {v_1 +v_2 \over N_3} \Big) \ , \ee
with $m_2 = m_1 + m_3$.
In this case, the quiver diagram has three nodes and six links pairwise connecting the nodes.


\subsection{Explicit breaking of the global group $G_k$ \label{YMBreakScalar}}

Returning to the YM-scalar theories, in ref.~\cite{Chiodaroli:2014xia} amplitudes in the generic Jordan family YMESG theories were constructed by double-copying the pure ${\cal N}=2$ SYM theory with a bosonic YM~+~$\phi^3$ theory (setting $N_f=0$), where the latter is described by the Lagrangian:
\bea
{\cal L}_{\rm YM+\phi^3} &=&-\frac{1}{4}F_{\mu\nu}^{\ha}F^{\mu\nu \ha}+\frac{1}{2}(D_\mu\phi^{A})^{\ha} (D^\mu\phi^{A})^{\ha} + 
\frac{1}{3!}\lambda g F^{ABC} f^{\ha \hb \hc} \phi^{A \ha}\phi^{B \hb}\phi^{C \hc} \nn \\
&& \null - \frac{g^2}{4} f^{\ha \hb \he} f^{\he \hc \hd}\phi^{A \ha}\phi^{B \hb}\phi^{A \hc}\phi^{B \hd}\,.
\label{YMscalarNf0}
\eea
The global $G_k$ symmetry acting on the $A,B,C$ indices becomes, through the double copy, a local (gauge) symmetry in the resulting supergravity theory. 
Since our goal is to describe the latter theory with broken gauge symmetry, it is natural to discuss the breaking of the $G_k$ 
symmetry before the double copy is taken. 

To reduce the global symmetry, $G_{k} \rightarrow G_k^{\rm red.}$, while preserving the $G_k$ symmetry at high energies,\footnote{This is necessary as, 
on the one hand, the unbroken and the spontaneously-broken phases of a supergravity theory (or any theory) are, from the perspective of the integrand,  the 
same at high energy  and on the other the high-energy limit of the supergravity integrand is given by the high-energy limit of the integrands of the two gauge theories.}
we shall follow a pattern  similar to the adjoint Higgs mechanism discussed in the previous section 
and break the symmetry by adding to the Lagrangian terms 
with dimension smaller than four (in $D=4$) -- { i.e.} quadratic and cubic terms.
To this end, we single out one generator, $T^0$, define  $G_k^{\rm red.}$ to be spanned by the generators of $G_k$ that commute with $T^0$, 
and decompose the adjoint representation of $G_k$ in representations of $G_k^{\rm red.}$,
\bea
\phi^{A \ha} = (\phi^{a \ha},\overline{\varphi}^{\alpha \ha},\varphi_{\ \alpha}^\ha)\, .
\label{decomposition}
\eea
The first field 
transforms in the adjoint representation of $G_k^{\rm red.}$ 
and the latter two transform in conjugate complex representations of  $G_k^{\rm red.}$.
Note that these latter fields carry an adjoint index of the $G_c$ gauge group and an index of a complex representation of $ G_k^{\rm red.}$ 
and are thus different from the fields $\varphi^{i \hi}$ which appeared in \sec{YMscalar1}. 

With this decomposition, the symmetry-breaking terms we introduce are 
\be
\delta_1 {\cal L} = - (m^2)_{\alpha}^{\ \beta} \, \overline{\varphi}^{ \alpha \ha }  \varphi_{\beta }^{\ \ha}  \ ,
\qquad
\delta_2 {\cal L} \propto F^{0\ \alpha}_{\ \beta}  f^{\hb \ha \hc} \phi^{0 \ha} \overline{\varphi}^{ \alpha \hb} \varphi_{ \beta}^{\ \hc} \ .
\label{sbterms}
\ee
We take  the mass matrix to be
 \be
m_{\alpha}^{\ \beta} = {i \over 2} \rho \lambda F^{0\ \alpha}_{\ \beta} 
\ ,
\label{gaugem}
\ee
where $\rho$ is a free real parameter, $F^{0\ \beta}_{\ \alpha}=- i {\rm Tr}([T^{0},(T^{\alpha})^\dagger ]\, T^{\beta})$, and $T^{\alpha}$ are (non-hermitian) generators of $G_k$ that do not commute with the $T^{0}$.\footnote{This pattern is akin to that in which a symmetry is spontaneously broken; the main difference here is that 
none of the fields in \eqn{YMscalarNf0}, including the one corresponding to $T^0$, have a vacuum expectation value.}
For the normalization of the cubic term, it is convenient to introduce the diagonal matrix 
\be
\Delta^{ab} = \delta^{ab} + (\sqrt{1+\rho^2} -1) \delta^{a0}\delta^{0b} \ . \label{defS}
\ee
With this notation, the Lagrangian with broken $G_k$ symmetry
that we will use  is
\bea
{\cal L}_{\rm YM+\cancel{\phi^3}} &=&-\frac{1}{4}F_{\mu\nu}^{\ha}F^{\mu\nu \ha}+\frac{1}{2}(D_\mu\phi^{a})^{\ha} (D^\mu\phi^{a})^{\ha} +  (\overline{D_\mu \varphi^{ \alpha}})^\ha  (D^\mu \varphi_{\alpha})^\ha 
- (m^2)_{\alpha}^{\ \beta} \, \overline{\varphi}^{ \alpha \ha }  \varphi_{\beta }^{\ \ha}  \nn \\
&& \null + 
\frac{1}{3!}g \lambda  F^{abc} f^{\ha \hb \hc} \phi^{a \ha}\phi^{b \hb}\phi^{c \hc}
+ g \lambda  \Delta^{ab}F^{a \ \beta}_{\ \alpha}f^{\hb \ha \hc} \phi^{b \ha} \overline{\varphi}^{ \alpha \hb} \varphi_{ \beta}^{\ \hc} 
\nn \\
&& \null +\frac{1}{2}g\lambda  F^{\alpha \ \gamma}_{\ \beta} f^{\ha \hb \hc}  \varphi_{\alpha}^{\ \ha} \overline{\varphi}^{ \beta \hb} \varphi_{ \gamma}^{\ \hc} +\frac{1}{2}g\lambda  F_{\alpha \ \gamma}^{\ \beta} f^{\ha \hb \hc}  \overline{\varphi}^{\alpha \ha} \varphi_{ \beta}^{\ \hb} \overline{\varphi}^{ \gamma \hc}  \nn \\  
&&\null - \frac{g^2}{4} f^{\ha \hb \he} f^{\he \hc \hd}\big( 
										\phi^{a \ha}\phi^{a \hc}+ 2\overline{\varphi}^{\alpha \ha}\varphi^{\ \hc}_\alpha\big)
                  							     \big(
							       			\phi^{b \hb}\phi^{b \hd}+ 2\overline{\varphi}^{\beta \hb}\varphi^{\ \hd}_\beta\big)\nn \\
&& \null + \frac{g^2}{2} f^{\ha \hb \he} f^{\he \hc \hd} \overline{\varphi}^{\alpha \ha}\varphi^{\ \hb}_\alpha  \overline{\varphi}^{\beta \hc}\varphi^{\ \hd}_\beta\,,
\label{YMscalarGlobalBroken}
\eea
where the structure constants $F^{a\ \beta}_{\ \alpha}$ and $F^{\alpha \ \gamma}_{\ \beta}$ are defined in the usual way,
\be
F^{a\ \beta}_{\ \alpha}= - i {\rm Tr}([T^{a},(T^{\alpha})^\dagger ]\, T^{\beta})
\qquad
F^{\alpha \ \gamma}_{\ \beta}=(F_{\gamma \ \alpha}^{\ \beta})^\dagger=- i {\rm Tr}([T^{\alpha},(T^{\beta})^\dagger ]\, T^{\gamma})
\ ,
\ee 
with $T^{a}$ being hermitian generators of $G_k^{\rm red}$.

We will motivate the symmetry-breaking terms through calculations in \sec{secamp} and \app{YMphi3dredSect}.  In \sec{secamp} we calculate amplitudes and show 
that color/kinematics duality requires that these terms be present.  Moreover, in \app{YMphi3dredSect} this Lagrangian is derived as the dimensional 
compactification/reduction and truncation of the unbroken $(D+1)$ theory. This does not imply that the amplitudes of the explicitly broken theory are 
equivalent to $(D+1)$-dimensional amplitudes; indeed, they are not, as there are no massive vectors in the Lagrangian \eqref{YMscalarGlobalBroken}. 
See \app{YMphi3dredSect} for more details.

Returning to the color/kinematics duality, we expect that it should be possible find amplitude presentations such that for each three-term Jacobi identity of $G_c$ there exists a three-term numerator identity. The latter requires $G_k$ relations which are decompositions of the Jacobi identity (decomposed following \eqn{decomposition}). Thus we have the following correspondences:
\bea
c_i-c_j=c_k \hskip2cm &\Leftrightarrow& \hskip2cm  n_i-n_j=n_k\,, \nn \\ \nn \\ 
\tf^{\hd \ha \hc} \tf^{\hc \hb \he} - \tf^{\hd \hb \hc} \tf^{\hc \ha \he}  = \tf^{\ha \hb \hc} \tf^{\hd \hc \he}  
 &\Leftrightarrow &   \left\{ \begin{array}{l}
 F^{d a c} F^{c b e} - F^{d b c} F^{c a e}  = F^{a b c} F^{d c e}
 \\
 F^{a \ \beta}_{\ \gamma } F^{b \ \gamma }_{\ \alpha}-  F^{b \ \beta}_{\ \gamma }F^{a \ \gamma }_{\ \alpha}  = F^{a b c} F^{c \ \beta }_{\ \alpha} \\
F^{a \ \gamma }_{\ \epsilon} F^{\epsilon \ \beta}_{\ \delta } - F^{a \ \beta }_{\ \epsilon} F^{\epsilon \ \gamma}_{\ \delta } =F^{a \ \epsilon}_{\ \delta} F^{\gamma \ \beta}_{\ \epsilon } \\
F^{\alpha \ \gamma }_{\ \epsilon} F^{\epsilon \ \beta}_{\ \delta } - F^{\alpha \ \beta }_{\ \epsilon} F^{\epsilon \ \gamma}_{\ \delta } =F^{\alpha \ \epsilon}_{\ \delta} F^{\gamma \ \beta}_{\ \epsilon } \\
4 F^{\epsilon \ [\alpha  }_{\ [\delta} F_{\gamma] \ \epsilon}^{\ \beta] }+ 2  F^{a \ \alpha }_{\ [\delta} F^{a \ \beta}_{\ \gamma] } 
= F^{\alpha \ \beta }_{\ \epsilon} F_{\delta \ \gamma}^{\ \epsilon } 
\end{array} \right\}\,.  ~~~
\label{coloralgebra2}
\eea
The last identity for the $F$'s has seven terms, but given fixed assignments of the free indices at most three terms contribute (the integer factors in the seven-term relation compensates for the antisymmetrization over indices).\footnote{The fact that the seven-term identity is not affected by the symmetry-breaking terms when insisting on color/kinematics duality is a rather non-trivial fact, as we shall see in \sec{secamp}.}

The Lagrangian in eq.~\eqref{YMscalarGlobalBroken} can be generalized to fields $\varphi_\alpha{}^{\hat a}$ that transform in a complex representation of the gauge group 
while preserving the numerator relations and, simultaneously, replacing the color Jacobi relations by the five identities in eqs.~\eqref{fck} and \eqref{extraf}. This can be done in 
several different ways, all leading to the same formal expression for the Lagrangian but each emphasizing different properties. 
For example, labeling by Greek hatted indices a complex potentially reducible representation of $G_c$, one may simply 
replace $\varphi_\alpha{}^{\hat a}\rightarrow \varphi_\alpha{}^{\hat \alpha}$ and  ${\bar \varphi}^\alpha{}^{\hat a}\rightarrow {\bar \varphi}^\alpha{}_{\hat \alpha}$ 
while making the corresponding replacements of indices on the $f^{{\hat a}{\hat b}{\hat c}}$ structure constants and requiring that the resulting coefficients 
obey the relations in eqs.~\eqref{fck} and \eqref{extraf}. This construction, which is quite general, does not require any correlation between 
the representation of the gauge and global symmetry groups.
One may also decompose the adjoint color indices into adjoint and complex representations of a subgroup (the latter being denoted by hatted Greek letters), 
assign to complex gauge and global indices, ${\hat \alpha}$  and $\alpha$, the $U(1)$ charge corresponding to the diagonal of the preferred generators 
$f^0{}_{\hat\beta}{}^{\hat\alpha}$ and  $F^0{}_{\beta}{}^{\alpha}$ and project onto the fields with vanishing charge. 
This restricts the gauge group to the chosen subgroup and introduces a correlation between the irreducible representations under this and global group, i.e. the fields 
carry only certain combinations of the irreducible components of the representations denoted by the ${\hat \alpha}$ and $\alpha$ indices. By construction, the remaining 
components of the $f^{\ha \hb \hc}$ structure constants obey the relations  in eqs.~\eqref{fck} and \eqref{extraf}.
Due to the closer similarity between the gauge and global symmetry representations carried by fields one may interpret the resulting Lagrangian\footnote{Alternatively, 
we could have constructed this Lagrangian directly as a YM-scalar Lagrangian which 
(1) has the same gauge group as the unbroken gauge group of a spontaneously-broken theory of the form~\eqref{brokenGeneral}, 
(2) contains additional scalar fields with the same masses as the fields in eq~\eqref{brokenGeneral} and conjugate gauge-group representations,
(3) has cubic couplings analogous to the ones of \eqref{YMscalarNf0}, and 
(4) has the cubic and quartic couplings selected by requiring relations between kinematic numerators that mirror relations between color factors.}
as a more refined example of color/kinematics duality:
\bea
{\cal L}'_{\rm YM+\cancel{\phi^3}} &=&-\frac{1}{4}F_{\mu\nu}^{\ha}F^{\mu\nu \ha}+\frac{1}{2}(D_\mu\phi^{a})^{\ha} 
(D^\mu\phi^{a})^{\ha} +  (\overline{D_\mu \varphi^{ \alpha}})_\haa  (D^\mu \varphi_{\alpha})^\haa 
- (m^2)_{\alpha}^{\ \beta} \, \overline{\varphi}^{ \alpha}_\haa  \varphi_{\beta }^{\ \haa}  \nn \\
&& \null + 
\frac{1}{3!}g \lambda  F^{abc} f^{\ha \hb \hc} \phi^{a \ha}\phi^{b \hb}\phi^{c \hc}
+ g \lambda  \Delta^{ab}F^{a \ \beta}_{\ \alpha} f^{\ha \ \hbb}_{\ \hgg} \phi^{b \ha} \overline{\varphi}^{ \alpha }_{\ \hbb} \varphi_{ \beta}^{\ \hgg} 
\nn \\
&& \null +\frac{1}{2}g\lambda  F^{\alpha \ \gamma}_{\ \beta} f_{\haa \  \hgg}^{\ \hbb}  \varphi_{\alpha}^{\ \haa} \overline{\varphi}^{ \beta}_{\hbb} \varphi_{ \gamma}^{\ \hgg} 
+\frac{1}{2}g\lambda  F_{\alpha \ \gamma}^{\ \beta} f^{\haa \ \hgg}_{\ \hbb}  \overline{\varphi}^{\alpha}_{\ \haa} \varphi_{ \beta}^{\ \hbb} \overline{\varphi}^{ \gamma}_{\ \hgg} 
\nn \\  
&&\null - \frac{g^2}{4} f^{\ha \hb \he} f^{\he \hc \hd} \phi^{a \ha}\phi^{a \hc} \phi^{b \hb}\phi^{b \hd} 
- g^2 f^{\ha \ \hgg}_{\ \haa} f^{\hb \ \hbb}_{\ \hgg} \phi^{a \ha}\phi^{a \hb} \overline{\varphi}^\alpha_{\ \hbb} \varphi_\alpha^{\ \haa}  \nn \\
&& \null - {g^2} f_{\ \hee}^{\haa \ \hbb } f^{\ \hee}_{\hgg \ \hdd} \overline{\varphi}^{\alpha}_{\ \haa} \varphi^{\ \hgg}_\alpha \overline{\varphi}^{\beta}_{\ \hbb} \varphi^{\ \hdd}_\beta + \frac{g^2}{2} f^{\he \ \haa}_{\ \hbb} f^{\he \ \hgg}_{\ \hdd} \overline{\varphi}^{\alpha}_{\ \haa}\varphi^{\ \hbb}_\alpha  \overline{\varphi}^{\beta}_{\ \hgg}
\varphi^{\ \hdd}_\beta  \, ,
\label{YMscalarGlobalBrokentrunc}
\eea
Indeed, explicit calculations summarized in \sec{secamp} confirm that the tree-level scattering amplitudes following from this Lagrangian obey color/kinematics 
duality. Note that the kinematic numerators of the amplitudes coming from the theory \eqref{YMscalarGlobalBrokentrunc} can be chosen to be the same as 
those of the theory \eqref{YMscalarGlobalBroken}, since the change of $G_c$ representations only affects the color factors of the amplitudes.\footnote{This is particularly 
clear in the construction of \eqref{YMscalarGlobalBrokentrunc} through projection.}

It is important to note that the introduction of the Lagrangian \eqref{YMscalarGlobalBrokentrunc} together with a specific correlation between the irreducible components 
of the complex gauge and global indices can be motivated from the double-copy construction between a spontaneously broken SYM and the current theory that we will 
define in \sec{2copySec}.  To guarantee the required properties of the fields $\varphi$, that they are double-copied with ${\cal N}=2$ SYM fields with the same mass, 
it is necessary to impose the condition
\be 
2 V f^{0 \ \haa}_{\ \hbb} \varphi^{\ \hbb}_\alpha = \lambda \rho F^{0 \ \beta}_{\ \alpha} \varphi^{\ \haa}_\beta  \ .
\label{mass_relation}
\ee
The general solution to this equation is that the fields $\varphi^{\ \haa}_\alpha$ have a block structure in the space of irreducible components of the $\alpha$ and $\haa$ indices
and all fields inside each block have equal mass. Identifying the gauge groups of the YM+$\cancel{\phi^3}$ and ${\cal N}=2$ SYM theories, eq.~\eqref{mass_relation} 
together with eqs.~\eqref{massM} and \eqref{gaugem}, implies that fields in the same representation of the gauge group have equal masses. 


\subsubsection{${SU}(N)$ Example \label{YMBreakScalarEx}}

An interesting  example involves the generation of several different (flavored) massive scalars. The global symmetry group is broken as
\be 
{SU}(N_k+N_f)\rightarrow {SU}(N_k)\times {U}(1)^{N_f}\, .
\ee
Different flavors carry different charges under the $U(1)$ factors. 
Such a breaking can be obtained by choosing $T^0$ to be 
\be 
T^0 =  \left( \begin{array}{cccc} 
 v_1 & \cdots & 0 & \! \!  0 \\
\vdots  & \ddots & \vdots & \! \!  \vdots  \\
 0 & \cdots & v_{N_f} & \! \!  0 \\ 
 0 & \cdots & 0 & \! \! v_0 I_{N_k}
\end{array}  \right) \ ;
\ee
tracelessness requires that $v_0=-\sum_{n=1}^{N_f} v_n/N_k$ and the $v_i$ are normalized as Tr$(T^0 T^0)=1$.

The symmetry generators of the original S$U(N_k+N_f)$ symmetry group can be divided into six sets (the generators in the second through fifth sets are broken):
\be
\begin{array}{llc} 
{SU}(N_k)~{\rm adjoint} & \hskip1cm T^a\,, &  \hskip-7mm (a=1,\ldots, N_k^2-1) \\
{SU}(N_k)~\text{fund.\,\&\,flavored} & \hskip1cm (T^{(kn)})_{i}^{\ j}=\delta_{i}^{k} \delta_{n}^{j} \,, & \hskip-7mm  \\
{SU}(N_k)~\overline{\rm fund}.\text{\,\&\,flavored}  & \hskip1cm (T_{(kn)})_{i}^{\ j}=\delta_{i}^{n} \delta_{k}^{j} \,, & \hskip-7mm\\
{U}(1)^{N_f}~~\text{bi-flavored} & \hskip1cm (T^{(nm)})_{i}^{\ j}=\delta_{i}^n \delta_{m}^{j} \,,& \hskip-7mm (n<m)\\
{U}(1)^{N_f}~~\text{bi-flavored} & \hskip1cm (T_{(nm)})_{i}^{\ j}=\delta_{i}^m \delta_{n}^{j} \,, & \hskip-7mm (n<m)\\
{U}(1)^{N_f}~~\text{un-flavored} & \hskip1cm (T^{(nn)})_{i}^{\ j}=\delta_{i}^n \delta_{n}^{j}-\frac{1}{N_k} I_{N_k}\,, & \hskip-7mm \text{(no sum)}
\end{array} \label{gen}
\ee
where  $k=N_f+1,\ldots, N_f+N_k$ are fundamental indices, and $n,m=1,\ldots, N_f$ are flavor indices. 
The eigenvalues of the corresponding mass matrix are
\be 
m_{(kn)} = \rho \lambda (v_0-v_n) \,, \qquad 
m_{(nm)}= \rho \lambda (v_n-v_m )\,,
\ee
where we use the convention that conjugate representations have masses of opposite signs just like the charges  (all physical quantities depend only on the squared masses).

Lagrangians discussed in earlier sections may be obtained from the  Lagrangian (\ref{YMscalarGlobalBroken}) and the generators (\ref{gen}) by restricting to a 
subset of its fields.
While this truncation is not always technically consistent (in the sense that the equations of motion of the fields that are truncated away contain sources 
depending only on the 
remaining fields) we may nevertheless define such a restricted theory. Its tree-level S matrix however cannot be obtained from that of the parent by simply 
restricting the external states to those of the daughter theory; rather, it is necessary to also eliminate all the (Feynman) graphs with truncated fields appearing 
on the internal lines.

By truncating away the $U(1)$-flavored modes corresponding to the generators $T^{(nm)}$, $T_{(nm)}$, $T^{(nn)}$ and $T^{0}$ we 
can recover a theory that is very similar to the one in \eqn{YMscalar}. Under this truncation the only surviving structure constants are
\be
F^{abc}~~~~~\text{and}~~~~~F^{a \ (jn)}_{\ (im)} = (T^a)^{\ j}_i \, \delta_{m}^{n}  - (T^a)^{\ n}_m \, \delta_{i}^{j}
\rightarrow (T^a)^{\ j}_i \, \delta_{m}^{n} \,,
\ee
where in the second structure constant the term, $- (T^a)^{\ n}_m \, \delta_{i}^{j}$, has been dropped due to the truncation.
The only difference between this theory and the one described by the Lagrangian in \eqn{YMscalar} is that while there the complex scalars 
in \eqn{YMscalar} transform in the fundamental representation of the gauge group, here the complex scalars are in the adjoint. The two Lagrangians may 
nevertheless be mapped into each other by identifying 
the gauge group generators in the adjoint representation and replacing them with the ones in the fundamental representation,\footnote{Such a replacement 
can also be done at the level of (Feynman) graphs.} { e.g.} $\overline{\varphi}^{\alpha \ha} \tf^{\ha \hc \hb} \varphi_{\beta}^{\hb}
\rightarrow \overline{\varphi}^{\alpha} t^{\hc} \varphi_{\beta}$. This is straightforward except for the quartic terms 
for which the color/kinematics-satisfying result is obtained as: %
\bea
&& \frac{g^2}{4} f^{\ha \hb \he} f^{\he \hc \hd}\Big[\big( \phi^{a \ha}\phi^{a \hc} +2 \overline{\varphi}^{\alpha \ha}\varphi^{\ \hc}_\alpha\big)\big( \phi^{b \hb}\phi^{b \hd} +2 \overline{\varphi}^{\beta \hb}\varphi^{\ \hd}_\beta\big) - 2 \overline{\varphi}^{\alpha \ha}\varphi^{\ \hb}_\alpha  \overline{\varphi}^{\beta \hc}\varphi^{\ \hd}_\beta\Big]  \\
&&~~~~~~= \frac{g^2}{4} f^{\ha \hb \he} f^{\he \hc \hd}\Big[\phi^{a \ha}\phi^{a \hc}\phi^{b \hb}\phi^{b \hd} 
- 4\overline{\varphi}^{\alpha \ha}\phi^{a \hb}\phi^{a \hc}\varphi^{\ \hd}_\alpha -4 \overline{\varphi}^{\alpha \ha}\varphi^{\ \hb}_\beta\overline{\varphi}^{\beta \hc}\varphi^{\ \hd}_\alpha +2\overline{\varphi}^{\alpha \ha}\varphi^{\ \hb}_\alpha\overline{\varphi}^{\beta \hc}\varphi^{\ \hd}_\beta \Big] \nn \\
&&  ~~~~~~\rightarrow \frac{g^2}{4} f^{\ha \hb \he} f^{\he \hc \hd}\phi^{a \ha}\phi^{a \hc}\phi^{b \hb}\phi^{b \hd} 
- g^2 \overline{\varphi}^{\alpha}\phi^{a}\phi^{a }\varphi_\alpha -g^2 \overline{\varphi}^{\alpha}t^{\he} \varphi_\beta \overline{\varphi}^{\beta} t^{\he}\varphi_\alpha +\frac{g^2}{2}\overline{\varphi}^{\alpha}t^{\he} \varphi_\alpha\overline{\varphi}^{\beta}t^{\he} \varphi_\beta \nn \,.
\eea
On the second line, a Jacobi relation was used to reorganize the $(\overline{\varphi} \varphi)^2$ terms. 
Using $\alpha=(im)$, $\beta=(jn)$, the manipulations above lead exactly to the Lagrangian in \eqn{YMscalar} with massive fundamental scalars and 
symmetry ${SU}(N_c)\times {SU}(N_k) \times {U}(1)^{N_f}$.

\subsection{The double copy for spontaneously-broken theories \label{2copySec}}
\newcommand{\ra}{{\rm a}}

Here we spell out a double-copy construction which combines the ingredients introduced in the previous sections 
to produce amplitudes in Yang-Mills-Einstein supergravities, some of which have spontaneously-broken gauge symmetry.
The case of unbroken Yang-Mills-Einstein supergravities is a review of ref.~\cite{Chiodaroli:2014xia}.

Let us assume that expressions for gauge-theory scattering amplitudes are available such that the 
kinematic numerators $\tilde{n}_i$ satisfy the same general Lie-algebraic relations as the corresponding color factors $c_i$. By general Lie-algebraic relations we mean relations that are not specific to a given gauge group or symmetry-breaking pattern, but are more generally valid, such as the Jacobi identity and commutation relation in \fig{BasicJacobiFigure}, and the kinematic relations for theories with broken symmetry in \fig{BrokenJacobiFigure}. 
The double-copy construction states that, regardless of the spacetime dimension, a valid (super)gravity amplitude is obtained by replacing color factors with numerators in a gauge-theory amplitude, and by replacing the gauge coupling with its gravitational counterpart:
\be
 c_i \rightarrow \tilde{n}_i\, ~~~~~{\rm and}~~~~~g \rightarrow  \frac{\kappa}{2} \,.
 \label{colorreplace}
 \ee 
This statement can be taken as a conjecture to which we will give non-trivial 
supporting evidence in the case of spontaneously-broken Yang-Mills-Einstein supergravities.
 
We note that if two different gauge theories are considered, and the numerators of the first theory are replacing 
the color factors of the second theory,\footnote{As indicated by their common graph label, if $\tilde{n}_i$ and $c_i$ belong 
to different theories they still need to dress the same cubic-diagram specified by the poles $1/D_i$ in their respective amplitudes,
thus ensuring that the mass spectra of the two theories are aligned.} then it is convenient to take 
the two gauge groups, and thus the color factors, to be identical. 
Since the double copy does not depend on the details of the color factors, there is no loss of generality. 

A familiar property of the double copy, which also holds for spontaneously-broken theories, is that it is sufficient 
for one set of numerators $\tilde{n}_i$ to be manifestly duality-satisfying, while the other needs not to obey the duality manifestly. 
This is because, once color factors are replaced by kinematic factors with the same algebraic properties, the second kinematic numerators can in principle~\cite{Chiodaroli:2014xia} be brought to a duality-satisfying form through generalized gauge transformations~\cite{Bern:2008qj,Bern:2010ue}.

Using \eqn{colorreplace} gravity scattering amplitude will take the same general form already given in \eqn{BCJformGravity}; however, the details will differ depending on the gauge theory and whether it is unbroken or (spontaneously) broken. To understand the precise outcome of this prescription, it is essential to identify the proper tensor products of the asymptotic states that appear in the various theories introduced.

Before we discuss the explicit theories introduced in previous sections let us look at the asymptotic states from a uniform formal perspective. 
In particular, we have introduced gauge theories where fields transform in massless adjoint representations and massive complex (conjugate) representations. Let us assemble the fields into sets, or multiplets, 
corresponding to these three types:
\be
\Big({\cal V},V(m^2),\overline{V}(m^2)\Big)\,,
\ee
where ${\cal V}$ is the set of massless fields and $m^2$ labels the massive ones. 
By this notation it is understood that there are distinct massive multiplets $V(m^2)$ and $\overline{V}(m^2)$ for each allowed mass $m$ in the spectrum. The gauge-group indices of the fields have been suppressed, and since they are asymptotic fields 
for all practical purposes we may think of them as having been stripped of their color dependence.

The asymptotic fields produced by the double copy \eqref{colorreplace} are then given by the 
gauge-invariant subset of tensor products of gauge-theory fields of the left $(L)$ and right $(R)$ theories. 
We obtain the supergravity states 
\be
\Big({\cal V}_L \otimes {\cal V}_R,V_L(m^2) \otimes V_R(m^2) ,\overline{V}_L(m^2) \otimes  \overline{V}_R(m^2)  \Big)\Big|_{\text{gauge invariant}}
\, ,
\label{VdoubleCopy}
\ee
where for each allowed $m^2$ there is a distinct pair of tensor products that contribute to the supergravity spectrum. 
It is important to note that the tensor-product structure of \eqn{VdoubleCopy} is not an independent prescription but rather follows from \eqn{colorreplace}. This is because the gauge-theory asymptotic states already have a double-copy structure, between the kinematic and color wave 
functions (e.g. $A^{\mu \ha} \sim \varepsilon^\mu c^{\ha}$ and $W^{\mu}_\haa \sim \varepsilon^\mu c_{\haa}$). After the replacement (\ref{colorreplace}) the gravitational theory inherits such a structure.
 
Connecting to previous work, one can associate to each field a charge  that is uniform within the multiplets ${\cal V},V(m^2),$ $\overline{V}(m^2)$ but otherwise distinct, as was explicitly 
done in theories constructed through orbifold projections in ref.~\cite{Carrasco:2012ca, Chiodaroli:2013upa}. 
For example, in our case this charge may  be 
given in terms of the $t^0$ generator.
From this point of view it is convenient to take the fields of the left and right theory to have opposite charges.
The consistency of the construction through orbifold projection then requires  the set of supergravity states to be given by the set of 
zero-charge bilinears constructed from the states of the two gauge theories, as in \eqn{VdoubleCopy}. 

Let us now be concrete and describe the asymptotic states that enter the double copy 
(\ref{VdoubleCopy}) for each of the theories of interest. 

\subsubsection{GR + YM = YM $\otimes$ (YM~+~$\phi^3$)}

\begin{table}[t]
\centering
\begin{tabular}{|l||c|c|}
\hline 
Gravity coupled to YM & Left gauge theory  & Right gauge theory 
\\
\hline \\[-13pt]
\hline 
${\cal N}=4$ YMESG theory & ${\cal N}=4$ SYM  & YM~+~$\phi^3$ 
\\
${\cal N}=2$ YMESG theory (gen.Jordan) & ${\cal N}=2$ SYM  & YM~+~$\phi^3$ 
\\
${\cal N}=1$ YMESG theory  & ${\cal N}=1$ SYM  & YM~+~$\phi^3$ 
\\
${\cal N}=0$ YME + dilaton + $B^{\mu\nu}$ &  YM  & YM~+~$\phi^3$ 
\\
${\cal N}=0$ ${\rm YM}_{\rm DR}$-E + dilaton + $B^{\mu\nu}$ &  ${\rm YM}_{\rm DR}$  & YM~+~$\phi^3$ 
\\
\hline
\end{tabular}
\small \caption[a]{\small The double-copy constructions that appeared in ref.~\cite{Chiodaroli:2014xia}. 
These give amplitudes in YME gravity theories for various amounts of supersymmetry, corresponding to different choices of the left gauge theory. 
The right theory labeled by YM~+~$\phi^3$ corresponds to the YM + scalar theory with $N_f\rightarrow 0$. 
The ${\cal N}=1$ YMESG theory is a particular truncation of a generic Jordan family ${\cal N}=2$ YMESG theory in which the scalar and one fermion is dropped from every 
nonabelian vector multiplet together with the vector field and one of the gravitini in the graviton multiplet.
The last row corresponds to dimensional reductions 
of a higher-dimensional left gauge theory; this row has the same bosonic content as the previous cases, given that the original theory lived in $D=10,6,4,4$ 
dimensions, respectively.}
\label{TheoryConstructions1}
\end{table}

The case of Yang-Mills-Einstein supergravities was first treated in~\cite{Chiodaroli:2014xia}, here we give a summary of that construction. The massive multiplets $V(m^2)$ are absent in the unbroken case. Following the discussion above, the massless multiplets of the pure-adjoint unbroken left gauge theories,
\be
\begin{array}{lll}
{{\cal N}=4}~{\rm SYM}: ~ &  {\cal V}_L =A^\mu \oplus \lambda^{1,2,3,4} \oplus \phi^{0,1,2,3,4,5}\,,   \\
{{\cal N}=2}~{\rm SYM}: &  {\cal V}_L =A^\mu \oplus \lambda^{1,2} \oplus \phi^{0,1}\,,\\
{{\cal N}=1}~{\rm SYM}: & {\cal V}_L =A^\mu \oplus \lambda\,,  \\
{\rm pure}~{\rm YM}: & {\cal V}_L =A^\mu\,,  \\
{\rm YM}_{\rm DR}: & {\cal V}_L =A^\mu \oplus  \phi^{a'}\,,
\end{array}
\ee
are to be double copied (\ref{VdoubleCopy}) with the right theory 
\be
{\rm YM}+ \phi^3: ~~~~~~~~{\cal V}_R =A^\mu \oplus  \phi^a\,. \hskip2.4cm \phantom{a} \ .
\ee
We recall that ${\rm YM}_{\rm DR}$ stands for the dimensional reduction of some higher-dimensional pure YM theory.
As explained in ref.~\cite{Chiodaroli:2014xia} the double copy of these left and right theories gives rise to amplitudes in
(super)gravity coupled to pure Yang-Mills theory. The supersymmetric ${{\cal N}=4,2}$ theories can be uplifted to $D=10,6$ dimensions, 
respectively, without spoiling the construction. Similarly the bosonic theories can be considered in any dimension. 

The tensor product between ${\cal V}_L$ corresponding to YM$_\text{DR}$ (for some higher dimension) and ${\cal V}_R$ corresponding to ${\rm YM}+ \phi^3$ 
is part of all of these gravitational theories; it is given by
\be
{\cal V}_L \otimes {\cal V}_R \rightarrow \big( h^{\mu \nu},\phi, B^{\mu \nu},A^{\mu a},A^{\mu a'},\phi^{aa'} \big) \,,
\label{spectrum1}
\ee
where $a$ is an $G_k$ index and $a'$ is either a $R$-symmetry index or an additional global index. 
The construction is summarized in \tab{TheoryConstructions1}.

\subsubsection{GR + \cancel{YM} = \cancel{YM} $\otimes$ (YM + $\cancel{\phi^3}$) \label{YMYMphi3}}

\begin{table}[t]
\centering
\begin{tabular}{|l||c|c|}
\hline 
Gravity coupled to \cancel{YM} & Left gauge theory  & Right gauge theory  
\\
\hline \\[-13pt]
\hline 
${\cal N}=4$ \cancel{YM}ESG & ${\cal N}=4$ S\cancel{YM}  & YM + $\cancel{\phi^3}$ 
\\
${\cal N}=2$ \cancel{YM}ESG (gen.Jordan) & ${\cal N}=2$ S\cancel{YM}  & YM + $\cancel{\phi^3}$ 
\\
${\cal N}=0$ $\cancel{\rm YM}_{\rm DR}$-E + dilaton + $B^{\mu\nu}$ &  $\cancel{\rm YM}_{\rm DR}$  & YM + $\cancel{\phi^3}$ 
\\
\hline
\end{tabular}
\caption[a]{\small New double-copy constructions corresponding to spontaneously-broken YME gravity theories for different 
amounts of supersymmetry. The dimensionally-reduced YM$_{\rm DR}$ theory must have at least one scalar to provide the 
VEV responsible for spontaneous symmetry breaking. See the caption of \tab{TheoryConstructions1} for further details.}
\label{TheoryConstructions2}
\end{table}

The multiplets of the pure-adjoint spontaneously-broken left gauge theories are as follows:
\be
\begin{array}{lll}
{{\cal N}=4}~{\rm S\cancel{\rm YM}}: ~ & {\cal V}_L =A^\mu \oplus \lambda^{1,2,3,4} \oplus \phi^{0,1,2,3,4,5}\,, & ~ V_L(m^2) = W^\mu \oplus \Lambda^{1,2,3,4} \oplus \varphi^{1,2,3,4,5}\,,  \\
{{\cal N}=2}~{\rm S\cancel{\rm YM}}: &{\cal V}_L =A^\mu \oplus \lambda^{1,2} \oplus \phi^{0,1}\,, & ~ V_L(m^2) = W^\mu \oplus \Lambda^{1,2} \oplus \varphi^{1}\,,\\
\cancel{\rm YM}_{\rm DR}: &{\cal V}_L =A^\mu \oplus  \phi^{0,a}\,, & ~ V_L(m^2) = W^\mu  \oplus \varphi^{a'} \,,\\
\end{array}
\ee
where for brevity have suppressed the mass dependence of the component fields in $V_L(m^2)$. Similarly,
for brevity, we do not display the set of conjugate fields $\overline{V}_L(m^2)$ since it gives no additional information. 

The above fields are to be double copied with the right theory asymptotic fields
\be
{\rm YM} + \cancel{\phi^3}: \hskip1cm {\cal V}_R =A^\mu \oplus  \phi^a\,, ~~~ V_R(m^2) = \varphi_\alpha\,. \hskip1cm \phantom{a}
\ee
This construction gives rise to amplitudes in (super)gravity coupled to pure spontaneously-broken Yang-Mills theory. 
The supersymmetric ${\cal N}=4,2$ theories can be uplifted to $D=9,5$ dimensions without spoiling the construction, 
and similarly the bosonic theories can be considered in any dimension. The various supergravity theories constructed in this section are collected in \tab{TheoryConstructions2}.

The spectra of the above supergravity theories share their bosonic part of the spectrum with the double-copy between the spontaneously-broken dimensionally-reduced YM theory and the YM theory coupled to $\phi^3$ scalar theory with broken global symmetry. The result, ($\cancel{\rm YM}_{\rm DR})\otimes({\rm YM} + \cancel{\phi^3}$), is shown in table~\ref{doublecopyspectrum1}.

\begin{table*}
\centering
\begin{tabular}{|c||c|c|c|}
\hline
 & ${\cal V}_R$  & \phantom{\Big|} ${V}_R(m^2)$ \phantom{\Big|}&  $\overline{V}_R(m^2)$   \\
\hline\\[-13pt]
\hline 
${\cal V}_L$  & $h^{\mu \nu},\phi, B^{\mu \nu},A^{\mu a},A^{\mu a'},\phi^{aa'}$ & $\emptyset$  &  $\phantom{\Big|}  \emptyset \phantom{\Big|} $  \\
\hline
${V}_L(m^2)$  & $\phantom{\Big|}  \emptyset \phantom{\Big|} $  &  $W^{\mu}_\alpha,\varphi^{a'}_\alpha$ &$ \emptyset$    \\
\hline
$\overline{V}_L(m^2)$  & $\phantom{\Big|}  \emptyset \phantom{\Big|} $ & $\emptyset$ &  $\overline{W}^{\mu \alpha},\overline{\varphi}^{a'\alpha}$  \\
\hline
\end{tabular}
\small \caption[a]{\small The spectrum of the double-copy of $\cancel{\rm YM}_{\rm DR}$ and ${\rm YM} + \cancel{\phi^3}$. The bosonic spectra of the ${\cal N}=4,2$ YMESG theories are similar. Here $a,\alpha$ are $G_k$ indices and $a'$ is an $R$-symmetry index. }
\label{doublecopyspectrum1}
\end{table*}


\section{Spontaneously-broken Yang-Mills-Einstein \\ supergravity theories \label{sec:YME}}

\renewcommand{\theequation}{3.\arabic{equation}}
\setcounter{equation}{0}

The double-copy construction described in the previous section should give amplitudes of large classes of Yang-Mills-Einstein theories, with or without supersymmetry, and with or without spontaneously-broken gauge symmetry.
Given a procedure to compute all the tree-level scattering amplitudes of a field theory, 
it is in principle possible to reconstruct its Lagrangian order by order in the 
number of fields. However, since gravitational Lagrangians of the type discussed here involve quantities depending non-polynomially on the scalar fields, this procedure can be impractical.

However, there exist a very special class of $\cN=2$ supergravity theories in four and five dimensions for  which  the full non-polynomial Lagrangian can be reconstructed  from  the three-point interactions.
Such theories provide the simplest examples of our construction and are reviewed in this section.

\subsection{Higgs mechanism in five-dimensional $\cN=2 $  YMESG theories}

$\cN=2$ Maxwell-Einstein supergravity theories describe the coupling of an arbitrary number $\nvec5$ of vector multiplets 
to $\cN=2$ supergravity. An $\cN=2$ vector multiplet in five dimensions consists of a vector field $A_\mu$, 
a symplectic Majorana spinor $\lambda^i$  and a real scalar $\phi$.
The bosonic part of the $\cN=2$ MESG theory in five dimensions can be written in the form \cite{Gunaydin:1983bi}
\begin{eqnarray}
e^{-1}\mathcal{L}&=&-\frac{1}{2}R- \frac{1}{4}
{\stackrel{\circ}{a}}_{IJ} F_{\mu\nu}^{I}F^{\mu\nu J}- \frac{1}{2}
g_{xy}(\partial_{\mu}\phi^{x})(\partial^{\mu}
\phi^{y}) +\frac{e^{-1}}{6\sqrt{6}}C_{IJK}
\varepsilon^{\mu\nu\rho\sigma\lambda}F_{\mu\nu}^{I}
F_{\rho\sigma}^{J}A_{\lambda}^{K} \ , \quad
\end{eqnarray}
where $A_\mu^I \, (I= 0,1,...\nvec5)$ denote the vector fields of the theory including 
the bare graviphoton $A^0_\mu$, and $F^I_{\mu\nu}$ are the corresponding abelian field strengths. 
The scalar fields are labeled as $\phi^x$ ($x,y,..=1,..,\nvec5)$, and $g_{xy}$ is the metric of the scalar manifold. 
The Lagrangian is completely determined by the  constant symmetric tensor $C_{IJK}$.  Using this tensor one defines a cubic form
\begin{equation}
\mathcal{V}(\xi)\equiv C_{IJK}\xi^{I} \xi^{J} \xi^{K}\,,
\end{equation}
in the ambient space coordinates $\xi^{I}$. The  $(\nvec5+1)$-dimensional ambient space spanned by the $\xi^I$ has the metric
\begin{equation}\label{aij}
a_{IJ}(\xi)\equiv -\frac{1}{3}\frac{\partial}{\partial \xi^{I}}
\frac{\partial}{\partial \xi^{J}} \ln \mathcal{V}(\xi)\,,
\end{equation}
while the $\nvec5$-dimensional scalar  manifold $\cM_5$ is the co-dimension one  hypersurface  given by the condition \cite{Gunaydin:1983bi}:
\begin{equation}
{\cal V} (h)=C_{IJK}h^{I}h^{J}h^{K}=1 \qquad \text{with} \qquad h^I = \sqrt{2 \over 3} \xi^I
\end{equation}
and is parameterized by the coordinates $\varphi^x$. 
The  metric $g_{xy}$  is the induced metric on the hypersurface $\cM_5$, whereas the ``metric" ${\stackrel{\circ}{a}}_{IJ}(\phi)$ 
in the kinetic-energy term of the vector fields is given by the
restriction of $a_{IJ}$ to the hypersurface $\cM_5$,
\be
g_{xy}(\phi) = \frac{3}{2}\left. \frac{\partial \xi^I}{\partial \phi^x}
\frac{\partial \xi^J}{\partial \phi^y} a_{IJ} \right|_{ {\cal V} = 1} \;, \qquad
{\stackrel{\circ}{a}}_{IJ}(\phi)=a_{IJ}|_{{\cal V}=1}\;.
\ee
The ambient space indices are lowered and raised with the metric ${\stackrel{\circ}{a}}_{IJ}(\phi)$ and its inverse.

Defining
\be \label{hIx}
h^I_x \equiv  -
\sqrt{\frac{3}{2}} \frac{\partial h^I}{\partial \phi^x}  \;, \ee
one finds the following identities:
\bea
&& h^I h_I =1 \ ,
\\
&& h^I_x h_I =h_{Ix} h^I =0 \ ,
\\
&& {\stackrel{\circ}{a}}_{IJ} = h_I h_J + h_I^x h_J^y g_{xy} \ . \label{idmetric}
\eea
Next, we consider a group ${\cal G}$ of symmetry  transformations acting on  the ambient space as
\be \delta_{\alpha} \xi^I = (M_r)^I_{\ J} \xi^J \alpha^r \ ,   \ee
where $M_r$  satisfy the  commutation relations
\be [ M_r, M_s ] = f_{rs}^{\ \ \ t} M_t \ . \ee
If ${\cal G}$ is a symmetry of the Lagrangian of the five-dimensional MESG theory, then its $C$-tensor is invariant under it, 
and it satisfies the relation
\be
(M_r)_{(I}^{~L} C_{JK)L} =0 \ .
\ee
The vector fields of the theory transform linearly under the action of ${\cal G}$,
\bea
\delta_{\alpha} A^I_\mu & = & (M_r)^I_{\ J} A^J_\mu \alpha^r \ ,
\eea
and ${\cal G}$ acts as isometries of the scalar manifold $\mathcal{M}_5$
\bea 
\delta_\alpha \varphi^x &=& K^x_r \alpha^r  \ ,
\eea
where  $K^x_r$ is a Killing vector  of $\mathcal{M}_5$ given by
\bea  
K^x_r &=& - \sqrt{ 3 \over 2} (M_r)^J_{\ I} h_J h^{I x} \ .
\eea
The   $h^I(\varphi^x)$ transform linearly under ${\cal G}$  just like the vector fields,
\bea
\delta_{\alpha} h^I (\varphi^x)& = & (M_r)^I_{\ J} h^J(\varphi^x)  \alpha^r \ .
\eea
Spin-$1/2$ fields undergo rotations under the maximal compact subgroup of the global symmetry group ${\cal G}$,
\be
\delta_\alpha \lambda^a_{\hat \imath}  =  L^{ab}_r \lambda^b_{\hat \imath} \alpha^r
\ , \qquad \text{with} \qquad  L^{ab}_r =  (M_r)^J_{\ I} h^{[a|}_J h^{I|b]} - \Omega^{ab}_x K^x_r \ , 
\ee
where $\Omega^{ab}_x $ is the spin connection of $\mathcal{M}_5$ and $a,b,..=1,2,..\nvec5$ denote the flat tangent space  indices.
The remaining fields (gravitini and graviton) are inert under the action of ${\cal G}$.

We should note that using the identity (\ref{idmetric})
one can write the kinetic term of the vector fields
as
\begin{equation}
e^{-1} \hat{\cal L}_{\mscr{vec}} =
-\frac{1}{4} \stackrel{\circ}{a}_{IJ} \mathcal{F}_{\mu\nu}^{I}
\mathcal{F}^{\mu\nu J}=
-\frac{1}{4} \mathcal{F}_{\mu\nu}^{0} \mathcal{F}^{\mu\nu 0} -\frac{1}{4}  g_{xy}
\mathcal{F}_{\mu\nu}^{x} \mathcal{F}^{\mu\nu y} \ ,
\end{equation}
where
\begin{eqnarray}
A_{\mu}^{0}&\equiv &h_I  A_{\mu}^{I}, \qquad A_{\mu}^{x} \ \equiv  \ h_I^x A_{\mu}^{I}  \ .
\end{eqnarray}
Supersymmetry rotates $A_\mu^0 $ into the gravitini and $A_{\mu}^{x} $ into gaugini.
Therefore in a given background the physical graviphoton and the physical gaugini are given by the linear combinations
$\langle h_I \rangle A_{\mu}^{I}$
and $\langle h_I^x \rangle A_{\mu}^{I}$, respectively.

Yang-Mills-Einstein supergravity theories are obtained by gauging a subgroup $K$ of full global symmetry group ${\cal G}$ of the corresponding MESG theories
\cite{Gunaydin:1984ak,Gunaydin:1984nt,Gunaydin:1999zx}.
A subset of the  vector fields, denoted as   $A^r_\mu$ must then transform in  the adjoint representation
of  $K$. We consider only gaugings of compact groups $K$ such that  the other non-gauge vector fields are spectator fields, i.e. they are  inert under $K$.
In this case the non-zero entries of the matrices $M_r$ are simply
\be
(M_r)^s_{\ t}=  f^{rst}  \ .
\label{group_generators}
\ee
Throughout the paper it will be convenient to formally introduce group structure constants in which the indices
can assume values outside the range corresponding to the adjoint vectors $A_\mu^r$, i.e. $f^{IJK}$.
Such structure constants will always vanish if one or more of the indices correspond to a spectator vector field.

The bosonic sector of the $\cN=2$ YMESG theory in five dimensions has the Lagrangian
\bea
 \label{gaugedL}
e^{-1} {\cal L} &=& -{R \over 2} -{1\over 4} \text{\textit{\aa}}_{IJ} \mathcal{F}^I_{\mu \nu} \mathcal{F}^{J \mu \nu }
- {1 \over 2} g_{xy} \mathcal{D}_\mu \varphi^x \mathcal{D}^\mu \varphi^y +  \no
{e^{-1} \over 6\sqrt{6}} C_{IJK} \epsilon^{\mu\nu\rho\sigma\lambda} \left\{ \vphantom{1\over 2}
 F^I_{\mu \nu} F^J_{\rho \sigma} A^K_{\lambda} \right. \\
&& \qquad \left. \null + {3 \over 2} g_s f^{K}{}_{ J' K'} F^I_{\mu \nu} A^J_\rho A^{J'}_\sigma A^{K'}_\lambda +
{3\over 5} g_s^2 A^I_\mu f^{J}{}_{ I' J'} A^{I'}_\nu A^{J'}_\rho f^{K}{}_{K' L'} A^{K'}_\sigma A^{L'}_{\lambda} \right\} \,,~~
\eea
where
\bea  \mathcal{D}_\mu \varphi^x = \partial_{\mu} \varphi^x + g_s A^r_\mu K^x_r \ , \\
 \mathcal{F}^I_{\mu\nu}= 2 \partial_{[\mu} A^I_{\nu]} + g_s f^I_{\ JK} A^J_\mu A^K_\nu \ .
\eea
To preserve supersymmetry, gauging also requires the introduction of a Yukawa-like term
\begin{equation}
\mathcal{L}'= -\frac{i}{2}g_s{\bar{\lambda}}^{i a}\lambda_{i}^{b}K_{r[a} h^{r}_{b]}\,,
\label{yukawa}
\end{equation}
into the Lagrangian. 
However, in five dimensions, $\cN =2$ YMESG theories  without tensor fields  do not have any scalar potential terms, and  
therefore all their vacua are Minkowskian.

One can break the non-abelian gauge symmetry to a subgroup by giving a VEV to some of the scalars while preserving full 
$\cN=2$ supersymmetry. In this paper we  study the double-copy construction of the amplitudes of spontaneously-broken YMESG theories obtained by gauging the compact isometries of the $\cN=2$ MESG theories belonging to the generic Jordan family. Their cubic forms are of the form
\be
N(\xi) = C_{IJK} \xi^I \xi^J \xi^K = \sqrt{2} \xi^0 \left( (\xi^1)^2 - (\xi^2)^2 - \cdots - (\xi^{\nvec5})^2 \right)\,,
\ee
corresponding to the $C$-tensor
\bea
C_{011} =\frac{\sqrt{3}}{2} , \qquad C_{0rs} = -\frac{\sqrt{3}}{2}  \delta_{rs} , \qquad r,s=2,\cdots , \nvec5\,, \label{C-tensor}
\eea
and the base point\footnote{The base point is the point where the scalar metric as well as the ``metric" of the kinetic energy term of the vector fields become the Kronecker delta symbol.} 
\be
c^I = (\frac{1}{\sqrt{2}},1,0,\cdots, 0) \ .
\ee
The global symmetry group of the MESG theories belonging to the generic Jordan family is $SO(1,1)\times SO(\nvec5,1)$. 
Since one can embed the adjoint representation of any simple group into the fundamental representation of an orthogonal group, $SO(\nvec5)$,  
one can obtain a YMESG theory with an arbitrary simple gauge group by gauging of the generic Jordan family of MESG theories. 
We should note however that in five dimensions these YMESG theories will have at least one spectator vector field  in addition to the graviphoton.

Starting from a YMESG theory belonging  to the generic  Jordan family with gauge group $K$ 
we will spontaneously break the gauge symmetry 
by giving a VEV to the scalar partner of a gauge field in the adjoint of $K$ following  \cite{Mohaupt:2001be, Louis:2003gj},
where the breaking of $SU(2)$ gauge group down to its $U(1)$ subgroup was studied. 
This can be achieved by expanding the Lagrangian around the VEV shifted base point
\be 
c_{V_s}^I = \Big( {1 \over \sqrt{2}} , 1  , \vs , 0, 0 \Big)\,,
\label{base5d} 
\ee
corresponding to giving a VEV to $h^2$. 
The resulting theory describes YMESG theory coupled to some massive BPS vector multiplets.  The vector fields acquire their
masses, via the Higgs mechanism, from the term that is quadratic in the vector fields in the covariantized kinetic-energy term for the scalar fields,
\bea
- {1 \over 2} g_{xy} \mathcal{D}_\mu \varphi^x \mathcal{D}^\mu \varphi^y = - \frac{1}{2}
g_{xy} \partial_{\mu}\phi^{x} \partial^{\mu} \phi^y - g_s g_{xy} A^{\mu r} K_r^x \partial_\mu \phi^{y}- \frac{g_s^2}{2} g_{xy} K^x_r K^y_s A^r_\mu A^{\mu s} \ ,
\eea
and the gaugini acquire their  masses through the Yukawa-like term (\ref{yukawa}).
To preserve $\cN=2$ Poincar\'{e} supersymmetry
the masses of the gauge fields and gaugini must be equal. At first glance the mass terms appear different.  
However, as was pointed out in \cite{Mohaupt:2001be},  the mass term for the gauge fields
can be written in the form
\be
\frac{g_s^2}{2} g_{xy} K^x_r K^y_s A^r_\mu A^{\mu s} =\frac{1}{2} A^x_\mu A^{\mu y}  g_s W_{xz}  g_s W_{wy} g^{wz} \ ,
\qquad
  W_{xy} = h^r_{[x} K_{r y]}  \ .
\ee
Comparing this with the mass term for the gaugini,
\be
 \frac{i}{2} (\bar{\lambda}^{i x} \lambda^y_i ) (g_s W_{xy}) \ ,
\ee
one sees that they have the same mass as required by supersymmetry.
Therefore under this Higgs phenomenon, the gauge field corresponding to each broken generator  ``eats" one scalar field,
and we end up with a massive BPS vector supermultiplet consisting of  a massive vector and two massive spinor fields.

\subsection{Higgs mechanism in four-dimensional $\cN=2$ YMESG theories}

Dimensional reduction of the five-dimensional
$\cN=2$ YMESG theory of the previous subsection leads  to a four-dimensional YMESG theory with an additional abelian spectator vector multiplet. Hence
the spectrum of  the resulting four-dimensional $\cN=2$ Yang-Mills-Einstein supergravity
with gauge group $K$ includes one graviton multiplet and $\nvec5 + 1$ vector multiplets.
Each four-dimensional vector multiplet consists of a vector $A^I_{\mu}$, two spin-$1/2$ fields $\lambda^{ I \hat i}$ and a complex scalar $z^I$.
In addition to $\text{dim}(K)$  vector multiplets in  the adjoint representation of $K$, we have
$(\nvec5 - \text{dim}(K)+ 1)$  spectator vector multiplets 
that do not partake in the gauging. As in the previous subsection the vectors furnishing the adjoint representation will be denoted as $A^r_\mu$.

The bosonic part of the four-dimensional $\cN=2$ Yang-Mills-Einstein Lagrangian can be written in the form \cite{deWit:1983rz,deWit:1984pk,deWit:1984px,Craps:1997gp,Cremmer:1984hj,Gunaydin:2005bf}\footnote{ For further references on the subject we refer to the excellent book by Freedman and Van Proeyen~\cite{Freedman:2012zz}.}
\be e^{-1} {\cal L}= -{1\over 2} R - g_{I \bar J} {\cal D}_\mu z^I {\cal D}^\mu \bar z^J + {1 \over 4}  \Im \, \cN_{AB} {\cal F}^A_{\mu \nu } {\cal F}^{B \mu \nu}
- {e^{-1} \over 8}
\epsilon^{\mu\nu\rho\sigma} \Re \, \cN_{AB} {\cal F}^A_{\mu \nu} {\cal F}^B_{\rho \sigma} + g_s^2 {\cal P}_4 \ , \label{L4d}  \ee
where the gauge covariant derivatives and the four-dimensional potential term ${\cal P}_4$ are given by 
\bea
\mathcal{D}_{\mu} z^{I} & \equiv   & \partial_{\mu} z^{I} + g_s A_{\mu}^{J}f_{JK}^{I} z^{K} \ , \\
\mathcal{F}_{\mu\nu}^{I}  &  \equiv   &    2\partial_{[\mu}A_{\nu]}^{I} + g_s f_{JK}^{I}A_{\mu}^{J}A_{\nu}^{K} \ , \\
{\cal P}_4  & \equiv  &  - {1 \over 2}  e^{\cal K} g_{IJ}f^{IKL}f^{JMN} z^K \bar z^L z^M \bar z^N \ . \label{P4}
\eea 

In the symplectic formulation, the target-space metric $g_{I \bar J}$ and the period matrix ${\cN}_{AB}$ are obtained from
an holomorphic prepotential $F$,
which depends on $\nvec5 + 2$ complex variables.  For YMESG theories  obtained by dimensional reduction,
the prepotential is expressed in terms of the five-dimensional $C$-tensor as
\be
F(Z^A) = - {2 \over 3 \sqrt{3}} C_{IJK} {Z^I Z^J Z^K \over Z^{-1}} \ ,
\ee
where $Z^{-1} \equiv  Z^{A=-1}$. 
The construction goes as follows.
The prepotential is associated to a (holomorphic) symplectic vector
\be v(z) = \left( \begin{array}{c} Z^A(z) \\ {\displaystyle\partial F  \over \displaystyle\partial Z^A}(z)\end{array} \right) \ , \ee
where the $Z^A(z)$ are $\tilde n+2$ arbitrary holomorphic functions of $\tilde n+1$ complex variables
$z^I$, which need to satisfy a non-degeneracy condition.
The specific choice for such holomorphic functions is related
to the choice of the physical scalars and will be discussed shortly.
The symplectic vector $v(z)$ defines
a K\"{a}hler potential ${\cal K}(z,\bar z)$,
\be e^{-\cal K} = -i \langle v, \bar v \rangle = - i \Big( Z^A {\partial \bar F \over \partial \bar Z^A} - \bar Z^A {\partial  F \over \partial Z^A}
\Big) \ .   \ee
One then introduces a second (non-holomorphic) symplectic vector,
\be V(z,\bar z) = \left( \begin{array}{c} X^A \\  F_A   \end{array} \right) = e^{{\cal K }\over 2} v(z) \ ,  \ee
and its target-space covariant derivatives,
\bea D_{\bar I}\bar X^A &=& \partial_{\bar I}\bar X^I + {1\over 2} (\partial_{\bar I} {\cal K})\bar X^A \ , \no \\
D_{\bar I} \bar F_A &=& \partial_{\bar I} \bar F_A + {1\over 2} (\partial_{\bar I} {\cal K}) \bar F_A \ . \eea
The scalar metric and the period matrix
are expressed in terms of the quantities above as
\bea g_{I \bar J} &=& \partial_I \partial_{\bar J} {\cal K} \ , \\
{\cal N}_{AB} &=& \big( F_A  \ D_{\bar I} \bar F_A \big) \big( X^B  \ D_{\bar I}\bar X^B \big)^{-1} \ .   
\eea

As in the previous subsection, we will focus on the generic Jordan family of Yang-Mills-Einstein supergravities
whose  $C$-tensor was given in \eqn{C-tensor} and only consider compact gaugings of the isometry group $SO(1,1) \times SO(\nvec5-1,1)$.
It is important to note that, thanks to their five-dimensional origin, the Lagrangians of the theories we are considering are uniquely specified by
the choice of $C$-tensor and by the compact gauge group $K$ that is a subgroup of the global symmetry of the five-dimensional theory.
This fact allows us to identify a theory simply by its three-point interactions as both the $C$-tensor and the gauge group appear explicitly
in the expressions for the three-point amplitudes.

The choice of five-dimensional base-point (\ref{base5d}) is equivalent 
to specifying the set of non-degenerate functions entering the symplectic vector $Z^A(z)$ as follows
\be Z^A(z) = \Big( 1 ,  \ {i \over 2} + z^0,  \  {i \over \sqrt{2} } + z^1  ,  \  {i \over \sqrt{2}}  \vs   + z^2 , \   z^3, \  \ldots , \  z^\nvec5   \Big)  \ , \ee
with real $\vs$. We have chosen to label $z^0$ and $z^1$ the  scalars belonging to the two universal spectator vector multiplets  and $z^2, z^3$,
$ \ldots$,
$z^{\text{dim}(K)+ 1}$ the scalars
transforming in the  adjoint of compact gauge group $K$. 

It should be noted that for theories in the generic Jordan family all base points can be brought
into this form with a $SO(\text{dim}(K))$ transformation.\footnote{
This $SO(\text{dim}(K))$ transformation will in general not belong to $K$ 
and can be thought of as a redefinition of the Lie algebra generators.}

For $\vs=0$ we obtain a Yang-Mills-Einstein supergravity with unbroken gauge group.
In contrast, a non-zero $\vs$ breaks the gauge symmetry group $K$ down to an unbroken subgroup 
$\Kun$. In general $\Kun$
will have at least a $U(1)$ factor since the choice of base point corresponds to an adjoint scalar acquiring an expectation value, i.e. a non-zero value of
$\vs$ takes us on the Coulomb branch of the theory, similarly to the gauge theory case discussed in \sec{Higgs}.
To write explicitly the Lagrangian of the spontaneously-broken theory we split the indices running over the vectors of the theory
$A,B=-1,1,\ldots, \nvec5$ as
\be 
A = \big( a , \alpha, \bar \alpha \big) \ ,   
\ee
so that the index $a$ runs over the gluons of the unbroken gauge-group $\Kun$ as well as the spectator vectors,
while $\alpha$ and $\bar \alpha$ run over two conjugate representations of the unbroken gauge group.
Consequently, the vector fields are written as\footnote{An alternative notation is to introduce projectors acting in the space spanned by the $A,B$ indices and to define spectators,
unbroken gauge fields and massive vectors accordingly as,
\bea
\tilde A^A_\mu = ({\cal P}_{0})^A_{\ B} A^B_\mu \ , \qquad
W^A_\mu = ({\cal P}_W)^A_{\ B} A^B_\mu \ , \qquad
\overline{W}^A_\mu = ({\cal P}_{\overline{W}})^A_{\ B} A^B_\mu \ . 
\nonumber
\eea
This approach is closer to the paper \cite{Chiodaroli:2013upa}.}
\be A^A_\mu = \Big( A^a_\mu ,  \ W_{\alpha \mu} , \  \overline{W}^{ \alpha}_{\mu}  \Big) \ . \ee
In general, the unbroken gauge group will not necessarily be semisimple and the indices $\alpha, \bar \alpha$ may give
a reducible representation.
Similarly, the scalars $z^I = x^I + i y^I$ are split as\footnote{ Indices $x,y,..$ in $\varphi_{x\alpha}$ etc. are not to be confused with the labels of $D=5$ scalar fields. }
\be x^I = \Big( x^i , \ \varphi_{x \alpha} , \ \ \overline{\varphi}_x^{\alpha}  \Big) \ , \qquad
y^I = \Big( y^i , \ \varphi_{y \alpha} , \ \ \overline{\varphi}_y^{\alpha}  \Big) \ . \label{scalarsplit} \ee
Under this split, the only non-zero entries of the structure constants $f^{ABC}$ are
\be
f^{ABC} \rightarrow (f^{abc},f^{a \bar \alpha \beta},f^{\alpha \bar \beta \gamma}, f^{\bar \alpha \beta \bar \gamma} )\,,
\label{fsplit}
\ee
and yield the structure constants of the unbroken gauge group, the representation matrices 
for the massive fields and tensors with three representation indices which will give multi-flavor couplings
involving three massive fields.\footnote{Such couplings can be non-zero only when the mass of one of the fields equals the sum of the other two.} It should be noted that, as in the case of the gauge theories of the previous section,
these objects all obey Jacobi-like relations.

Among the vector multiplets providing the adjoint representation of the unbroken gauge group $\Kun$,
there is always a preferred multiplet. The abelian vector of this multiplet, denoted  with $A^2_\mu$, gives
the $U(1)$ factor which is always part of the unbroken gauge group $\Kun$.
One of the scalar fields of the preferred multiplet, $\hh$, can be thought of as the Higgs field.
Note that this field is the imaginary part of a four-dimensional complex scalar, $z^2 = \cchi + i \hh$, because the
gauge symmetry breaking has a five-dimensional origin.

The next step is to rewrite the covariant
derivatives appearing in the Lagrangian before symmetry breaking as
\be
{\cal D}_\mu y^{I}
     = \left( \!\! \begin{array}{c}   D_\mu y^{i} + g_s {\overline{W}}_\mu f^i \varphi_y - g_s {\overline{ \varphi}}_y f^i W_\mu  \\[5pt]
    D_\mu{\varphi}_{y \alpha} - i (\tilde m {W}_{\mu})_\alpha + { g_s} {y}^{i} (f^i W_\mu)_\alpha +
     {g_s}  \overline{W}_\mu f_\alpha  \varphi_y  -  {g_s }  \overline{\varphi}_y  f_\alpha W_\mu  +
     {g_s } \varphi_{y}  f_\alpha W_\mu 
      \\[5pt]
    D_\mu{\overline{\varphi}}_{y}^{\alpha} + i { (\overline{W}_\mu \tilde m)}^{\alpha} - { g_s } {y}^{i} ( {\overline{W}}_\mu f^i)_\alpha
    + {g_s}  \overline{W}_\mu  f^\alpha \varphi_y - {g_s}  \overline{\varphi}_y f^\alpha W_\mu + {g_s} {\overline{\varphi}_y}  f^\alpha \overline{W}_\mu
     \end{array} \!\! \right) \ , \label{DyI}
\ee
where $D_\mu$ is the covariant derivative for the unbroken gauge group. We have introduced the Hermitian matrix $\tilde m$, which is proportional 
to the mass matrix $m$ for the massive fields, and is defined as follows,  
\be \tilde m^{\ \alpha}_{\beta} = i g_s {\vs } (f^{2  \ \alpha}_{\ \beta} ) =  \sqrt{1 - \vs^2} \ m^{\ \alpha}_{\beta} \ . \label{sugram} \ee
 Without any loss of generality we will take $m$ to be block-diagonal.
The derivative ${\cal D}_\mu x^I$ of the real part of $z^I$  has an analogous expression with the terms proportional to $m$ missing.
The covariant field strengths are rewritten as
\be
{\cal F}^A_{\mu \nu}
 = \left( \begin{array}{c}   {\cal F}^a_{\mu \nu} + 2 g_s {\overline{W}}_{[\mu} f^a W_{\nu]}  \\[5pt]
    2 D_{[\mu}{ W}_{\nu]\alpha} + 2  g_s  \overline{W}_{[\mu} f_\alpha {W}_{\nu]}  -  {g_s } \overline{W}_{\mu} f_\alpha W_\nu \\[5pt]
    2 D_{[\mu} \overline{ W}_{\nu]}^{\alpha} +  2 g_s  \overline{W}_{[\mu}  f^\alpha W_{\nu]}  -  {g_s } \overline{W}_{\mu} f^\alpha \overline{W}_\nu
     \end{array} \right) \ . \label{FAmnsugra}
\ee
According to the above index decomposition, both period matrix and scalar metric are split into blocks as
\bea {\cN}_{AB} = \left( \begin{array}{ccc}  \cN_{ab} &    \cN_{a}^{\ \beta} & \cN_{a \beta} \\
                           \cN^\alpha_{\ b} & 0 & \cN^\alpha_{\ \beta} \\
                            \cN_{\alpha b} &   \cN_{\alpha}^{\ \beta} & 0 \\ \end{array} \right) \ , \qquad
 {g}_{I \bar J} = \left( \begin{array}{ccc}  g_{ij} &   g_{i}^{\ \beta} & g_{i \beta}  \\
                          g^\alpha_{\ j} & 0 & g^\alpha_{\ \beta} \\
                          g_{\alpha j} &  g_{\alpha}^{\ \beta} & 0
                            \end{array} \right) \label{blocks} \ . \eea
The full four-dimensional bosonic Lagrangian after dimensional reduction
can be obtained by plugging (\ref{DyI}), (\ref{FAmnsugra}) and (\ref{blocks}) into (\ref{L4d}), and plugging (\ref{scalarsplit}) and (\ref{fsplit})
into (\ref{P4}).
In analogy with our previous paper \cite{Chiodaroli:2014xia}, we then take the following steps after dimensional reduction to four dimensions:
\begin{enumerate}
 \item We dualize the graviphoton field $F^{-1}_{\mu \nu}$. Since this field is a spectator (as long as we are not considering $R$-symmetry gaugings),
 this dualization does not interfere with the gauging procedure.
 \item We employ a linear field  redefinition to canonically normalize the bosonic Lagrangian at the base point  and to
  render the supersymmetry transformations diagonal in the sense that the indices $A,B$ of the fields are not mixed by supersymmetry.
 Such a redefinition involves only spectator fields together with the
 preferred abelian vector field $A^2_\mu$ and takes the following form,
 \bea 
  A^{-1}_\mu &=&  - {\sqrt{1-\vs^2}\over 4} \big( {A^{-1}_\mu}' + {A^0_\mu}' + \sqrt{2} {A^1_\mu}'  \big) \no \ , \\
  A^{0}_\mu & = &  {1 \over 2 \sqrt{1-  \vs^2}} \big( {A^{-1}_\mu}' + {A^0_\mu}' - \sqrt{2} {A^1_\mu}'  \big) \no \ , \\
  A^{1}_\mu & = &  {1 \over \sqrt{2- 2 \vs^2}} \big(  {A^{-1}_\mu}' - {A^0_\mu}' + \sqrt{2} \vs {A^2_\mu}'  \big) \no\ , \\
  A^{2}_\mu & = &  {1 \over \sqrt{2- 2 \vs^2}} \big( \vs {A^{-1}_\mu}' - \vs {A^0_\mu}' + \sqrt{2} {A^2_\mu}'  \big) \no\ , \\
  x^{1} & = & {x^1}' + \vs {\cchi}' \ , \no \\
  \cchi & = & \vs {x^1}' +  {\cchi}' \ , \no \\
  y^{1} & = & {y^1}' + \vs {\hh}' \ ,\no \\
  \hh & = & \vs {y^1}' + {\hh}' \ , \no \eea \bea 
  \varphi_{x\alpha} & = & \sqrt{1 + \vs^2} \varphi'_{x\alpha} \ , \qquad  \varphi_{y\alpha}  =  \sqrt{1 + \vs^2} \varphi'_{y\alpha} \ ,  \no \\
  \overline{\varphi}^{\alpha}_{x} & = & \sqrt{1 + \vs^2} {\overline{\varphi}'}^\alpha_{x} \ , 
  \qquad \overline{\varphi}_{y}^\alpha  =  \sqrt{1 + \vs^2} {\overline{\varphi}'}_{y}^\alpha \ .\label{redefinition}
  \eea
\item We pick the standard $R_\xi$ gauge and introduce the gauge-fixing term
\be {\cal L}_{gf}=-{1 \over \xi} \overline{G}_\alpha  G^\alpha \ , \qquad G_\alpha = D^\mu W'_{\alpha \mu} + i \xi (m \varphi'_y)_\alpha  \label{id} \ . \ee
If we choose the unitarity gauge, $\xi \rightarrow \infty$, the scalar field $\varphi'_{y \alpha}$
acquires an infinite mass and can be integrated out.
\end{enumerate}
The final expansions for the scalar metric and period matrix which will be used in the Feynman-rule computation can be found in \app{appex}.
For notational simplicity we do not put a prime on the fields which appear in the final Lagrangian.

\section{Tree-level scattering amplitudes \label{secamp}}
\renewcommand{\theequation}{4.\arabic{equation}}
\setcounter{equation}{0}

\def\gsugra{{g_s}}

\subsection{Gauge theory amplitudes}

In this subsection we evaluate  three- and  four-point amplitudes 
in the gauge theories discussed in \sec{sec.ck2copy}. Three-point amplitudes 
will be the building-blocks used to construct 
three-points supergravity amplitudes using the double-copy prescription and, in the $\cN=2$ case, will lead to the identification 
of the complete supergravity Lagrangian.
Four-point amplitudes will enable us to study the constraints imposed by color/kinematics duality. Amplitudes in this section
will be written using a metric with mostly-minus signature.

\subsubsection{Three points\label{threeptg}}

The completely-massless three-point amplitudes that follow from the YM-scalar theory described in  \sec{YMBreakScalar}
are, up to field redefinitions, the same as the ones already considered in \cite{Chiodaroli:2014xia}. 
We therefore focus on amplitudes with massive fields. 

In the single-flavor case the only non-vanishing amplitudes have two massive and one massless external states; they are:\footnote{
We use the following conversion between structure constants of different normalizations:
\be 
\tilde f^{\ha \hb \hc} = \sqrt{2} i f^{\ha \hb \hc} \ , \qquad \tilde F^{abc} = \sqrt{2} i F^{abc} \ . 
\ee}
\bea 
{\cal A}_3\big( 1 \phi^{a \ha}, 2 \varphi_{\alpha}^\haa, 3 \overline{\varphi}^{\beta}_\hbb \big) &=& 
-{i \over 2} g \lambda \tilde F^{b \ \alpha}_{\ \beta} \Delta^{ab} \tilde f^{\ha \hbb}_{\ \ \haa} \label{amp3ptN0b} \ , \\
{\cal A}_3\big( 1 A^{\ha}_\mu, 2 \varphi_{\alpha}^{ \haa}, 3 \overline{\varphi}^{\beta}_{\hbb} \big) &=& \sqrt{2} i  g (k_2 \cdot \epsilon_1) 
\delta_\beta^\alpha \tilde f^{\ha \hbb}_{ \ \ \haa} \ .
   \eea
In the multi-flavor case, we also have a non-zero amplitude with three massive fields; it is:
\bea 
{\cal A}_3\big( 1 \varphi_{\alpha}^{ \haa}, 2 \varphi_{\beta}^{ \hbb}, 3 \overline{\varphi}^{\gamma}_{\hgg} \big) 
&=& - {i \over 2} \lambda g \tilde F^{\alpha \ \beta}_{\ \gamma} \tilde f_{\haa \ \hbb}^{\ \hgg} \ . 
\eea

Inspecting the Lagrangian of the spontaneously-broken YM-scalar theory described in \sec{Higgs}, it is easy to 
see that the three-point amplitudes are:
\bea {\cal A}_3\big( 1 \phi^{\hat a \si}, 2 \varphi_{ \haa }, 3 \overline{\varphi}^{ \hbb} \big) &=& 
 - \sqrt{2} i  g m \delta^{\si 0} \tilde f^{\ha \ \haa}_{\ \hbb} \  , \\
{\cal A}_3\big( 1 A^{\ha}, 2 \varphi_{ \haa }, 3 \overline{\varphi}^{ \hbb} \big) &=& 
\sqrt{2} i  g (k_2 \cdot \Pvec_1)  \tilde f^{\ha \ \haa}_{\ \hbb} \ , \\
{\cal A}_3\big( 1 \phi^{\hat a \si}, 2 W_{ \haa}, 3 \overline{W}^{ \hbb} \big) &=& 
\sqrt{2} i  g m \delta^{\si 0} (\Pvec_2 \cdot \Pvec_3)  \tilde f^{\ha \ \haa}_{\ \hbb} \ , \\
{\cal A}_3\big( 1 A^{\ha}, 2 W_{ \haa }, 3 \overline{W}^{ \hbb} \big) &=& 
-\sqrt{2} i  g \Big( (k_2 \cdot \Pvec_1) (\Pvec_2 \cdot \Pvec_3) + 
(k_1 \cdot \Pvec_3) (\Pvec_1 \cdot \Pvec_2) - (k_1 \cdot \Pvec_2) (\Pvec_1 \cdot \Pvec_3) \Big) \tilde f^{\ha \ \haa}_{\ \hbb} \ . \no \\
\eea
As for the YM-scalar theory, when more than one flavor is present, there are two additional non-zero amplitudes,
\bea  
{\cal A}_3 \big( 1 W_{\haa}, 2 W_{ \hbb }, 3 \overline{W}^{ \hgg} \big) &=&
- \sqrt{2}  i g \Big( (k_2 \cdot \Pvec_1) (\Pvec_2 \cdot \Pvec_3) + 
(k_1 \cdot \Pvec_3) (\Pvec_1 \cdot \Pvec_2) - (k_1 \cdot \Pvec_2) (\Pvec_1 \cdot \Pvec_3) \Big)  \tilde f^{\haa \ \hbb}_{\ \hgg}  , \no  \\ 
{\cal A}_3 \big( 1 W_{\haa}, 2 \varphi_{ \hbb }, 3 \overline{\varphi}^{ \hgg} \big) &=&
 \sqrt{2}  i g  (k_2 \cdot \Pvec_1 )  \tilde f^{\haa \ \hbb}_{\ \hgg} \ . 
\eea
We will see that, in the $\cN=2$ case, these building blocks will 
be sufficient to identify the supergravity obtained from the double-copy prescription.

\subsubsection{Four points}

Using the four-point amplitudes we can study the constraints imposed by color/kinematics duality on the theories constructed in \sec{sec.ck2copy}. 
We start from the YM-scalar theory with explicitly-broken global symmetry and Lagrangian given by (\ref{YMscalarGlobalBrokentrunc})
and compute first the amplitude between 
two massive and two massless scalars. To have a non-zero amplitude, the two masses must be equal:
\bea && {\cal A}_4 \big( 1 \phi^{a \ha}, 2 \phi^{b \hb}, 3 \varphi_{\alpha}^{\haa}, 4 \overline{\varphi}^{\beta}_{\hbb} \big) = \no \\
&& \quad  -{i\over 2} g^2 \left\{ 
\tilde f^{\ha}{}_ {\hgg}{}^{\hbb} \tilde f^{\hb}{}_{\haa}{}^{\hgg} 
\Big( {{\lambda^2 \over 2}\tilde F^{c \ \gamma}_{\ \beta} \tilde F^{d \ \alpha}_{\ \gamma} \Delta^{ac}\Delta^{bd}
\over (k_1+k_4)^2 - m^2} + \delta^\alpha_\beta \delta^{ab}  \Big) + 
\tilde f^{\hb  \hbb}_{\ \ \hgg} \tilde f^{\ha \hgg}_{\ \ \haa} 
\Big( {{\lambda^2 \over 2} \tilde F^{d \ \gamma}_{\ \beta} \tilde F^{c \ \alpha}_{\ \gamma} \Delta^{ac}\Delta^{bd} \over (k_1+k_3)^2 - m^2} + \delta^\alpha_\beta \delta^{ab}  \Big) \right. \no \\
&& \quad  \left. +  
{\tilde f^{\ha \hb \hc} \tilde f^{\hc \hbb}_{\ \ \haa} \over (k_1+k_2)^2} \big( {\lambda^2 \over 2}\tilde F^{abc}  \Delta^{cd} \tilde F^{d \ \alpha}_{ \ \beta} 
+ 2 (k_1 \cdot k_3 - k_1 \cdot k_4)   
\delta^\alpha_\beta \delta^{ab}  \Big) \right\} \ .\eea
The numerator factors are naturally organized by the power of $\lambda$. 
The ${\cal O}(\lambda^0)$ parts of the numerator factors are the same as in the massless theory and obey the kinematic Jacobi relations. 
Imposing color/kinematics duality at ${\cal O}(\lambda^2)$ and taking $\Delta^{ab}$ to be invertible leads to the requirement
\be 
 F^{a \ \gamma}_{\ \beta}  F^{b \ \alpha}_{\ \gamma} -  F^{b \ \gamma}_{\ \beta}  F^{a \ \alpha}_{\ \gamma} + F^{abc}  F^{c \ \alpha}_{\ \beta} = 0 \ , 
\label{KJac}
\ee 
i.e. the tensors $F^{abc}, F^{a \ \alpha}_{\ \beta}$ can be seen as the structure constants and representation matrices of 
the unbroken global symmetry group, respectively.
Similarly, imposing color/kinematics duality on the amplitude with one massless and three massive scalars leads to the identity,
\be F^{a \ \gamma }_{\ \epsilon} F^{\epsilon \ \beta}_{\ \delta } - F^{a \ \beta }_{\ \epsilon} F^{\epsilon \ \gamma}_{\ \delta } =F^{a \ \epsilon}_{\ \delta} 
F^{\gamma \ \beta}_{\ \epsilon } \ .
\label{KJac2} 
\ee

\begin{figure}[t]
      \centering
      \includegraphics[scale=0.9]{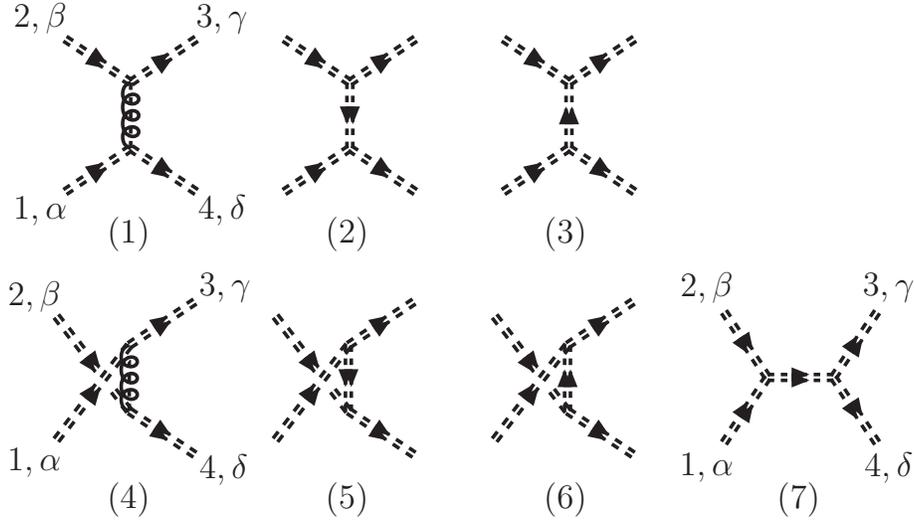}
      \caption{ Seven separate contributions to tree amplitudes with four massive scalars. Dashed lines with arrows denote complex (massive) scalars. In diagram (1) and (4) the exchanged particle is a sum of a massless scalar and a gluon.   \label{fig4mass} }
\end{figure}

We next turn 
to amplitudes with four massive fields. 
The terms with four such fields in \eqn{YMscalarGlobalBroken} may appear mysterious; 
let us assume therefore a generic dependence on such fields (constrained by symmetries) 
and see that the coefficients are fixed by color/kinematics duality as stated in that equation. 
Thus, we assume that the Lagrangian contains the contact terms\footnote{Additionally, we could consider a contact term of the form 
$\overline{\varphi}^\alpha f^{\hee} \varphi_\alpha \overline{\varphi}^\beta f_{\hee} \varphi_\beta$, 
but it is possible to show that it gives vanishing contribution to all amplitudes entering the double-copy construction in the next section.}
\be 
{g^2 \over 2} \big( b_2 \overline{\varphi}^\alpha f^{\ha} \varphi_\alpha \overline{\varphi}^\beta f^{\ha} \varphi_\beta + 
b_1 \overline{\varphi}^\alpha f^{\ha} \varphi_\beta \overline{\varphi}^\beta f^{\ha} \varphi_\alpha 
+ b_3 f_{\ \haa \hbb}^\hee \varphi_{\alpha}^{\ \haa} \varphi_{\beta}^{\ \hbb} f^{\ \hgg \hdd}_\hee \overline{\varphi}^{\alpha}_{\ \hgg} \overline{\varphi}^{\beta}_{\ \hdd} + 
b_4 \overline{\varphi}^\alpha f^{\hee} \varphi_\beta \overline{\varphi}^\beta f_{\hee} \varphi_\alpha\big)  \ . \\
\label{contactterms}
\ee
The scattering amplitude of four massive scalars can be cast in the form
\be {\cal A}_4 \big( 1 \varphi_{\alpha}^{\haa}, 2 \varphi_{\beta}^{\hbb}, 3 \overline{\varphi}^{\gamma}_{\hgg}, 4 \overline{\varphi}^{\delta}_{\hdd} \big) = 
-i g^2 \left( {n_1 c_1 \over D_1 } + {n_2 c_2 \over D_2 } + {n_3 c_3 \over D_3 } +{n_4 c_4 \over D_4 } +{n_5 c_5 \over D_5 } +{n_6 c_6 \over D_6 } +
{n_7 c_7 \over D_7 } \right) \ , \label{4massform_gauge} 
\ee
where the terms contributing to the amplitude are shown in figure \ref{fig4mass}; they correspond to decomposing each of the s, t and u channels 
following the representation of the intermediate state. When more than one mass is present, graphs with different internal mass are regarded as distinct.
The color factors are given by
\bea 
c_1 = f^{\ha  \hdd}_{\ \ \haa} \tilde f^{\ha  \hgg }_{\ \ \hbb} \ , & \quad c_2 =  \tilde f^{\  \hdd}_{\hee \ \haa} \tilde f^{\hee  \hgg}_{\ \ \hbb} \ ,  & 
\quad c_3 = \tilde f^{\hee  \hdd}_{\ \ \haa} \tilde f^{\  \hgg}_{\hee \ \hbb} \ , \qquad  c_4 = \tilde f^{\ha  \hgg}_{\ \ \haa} \tilde f^{\ha  \hdd}_{\ \ \hbb} \ , \no \\
c_5 = \tilde f^{\hee  \hgg}_{\ \ \haa} \tilde f^{\  \hdd}_{\hee \ \hbb} \ ,  &   
\quad c_6 = \tilde f^{\  \hgg}_{\hee \ \haa} \tilde f^{\hee  \hdd}_{\ \ \hbb}  \ , &  \quad c_7 = f^{\hgg  \hdd}_{\  \ \hee} \tilde f^{\hee}_{\ \haa \hbb} \ , 
\eea
while the  (massless and massive) inverse propagators are
\bea 
D_1 = (k_1+k_4)^2 \, , && D_2=D_3=(k_1+k_4)^2 - (m_1-m_4)^2  \,,   \no \\ 
D_4 = (k_1+k_3)^2 \,  , && D_5 =D_6 = (k_1+k_3)^2 - (m_1-m_3)^2   \,, \quad D_7 = (k_1+k_2)^2 - (m_1+m_2)^2 \,. \no \\ \label{4massprops} 
\eea

The numerator factors have the following expressions,
\bea n_1 &=&   (k_1 \cdot k_3 - k_1 \cdot k_2) 
\delta^\alpha_\delta \delta^\beta_\gamma - {i\over 4} 
\lambda^2 \tilde F^{b \ \alpha}_{\ \delta} \tilde F^{c \ \beta}_{\ \gamma} \Delta^{ba} \Delta^{ac} - {i \over 2} (k_1+k_4)^2 \big(  b_1 \delta^\alpha_\gamma \delta^\beta_\delta +
 b_2 \delta^\alpha_\delta \delta^\beta_\gamma \big) \ , \no \\
n_2 &=&  {1 \over 4} \lambda^2 \tilde F^{\epsilon \ \alpha}_{\ \delta} \tilde F^{\ \ \beta}_{\epsilon \gamma} 
- {i \over 2} b_4 \big( (k_1+k_4)^2 - (m_1-m_4)^2 \big) \delta^\alpha_\gamma \delta^\beta_\delta\no \ , \\ 
 n_3 &=& {1 \over 4} \lambda^2 \tilde F^{\ \ \alpha}_{\epsilon \delta} \tilde F^{\epsilon \ \beta}_{\ \gamma} 
- {i \over 2} b_4 \big( (k_1+k_4)^2 - (m_1-m_4)^2 \big) \delta^\alpha_\gamma \delta^\beta_\delta \no \ ,  \\
n_4 &=&  (k_1 \cdot k_4 - k_1 \cdot k_2) 
\delta^\alpha_\gamma \delta^\beta_\delta - {i \over 4} 
\lambda^2 \tilde F^{b \ \alpha}_{\ \gamma} \tilde F^{c \ \beta}_{\ \delta} \Delta^{ba} \Delta^{ac} - {i \over 2} (k_1+k_3)^2 \big(b_2 \delta^\alpha_\gamma \delta^\beta_\delta + 
 b_1 \delta^\alpha_\delta \delta^\beta_\gamma \big) \ , \no \\
n_5 &=&  {1 \over 4} \lambda^2 \tilde F^{\ \ \alpha}_{\epsilon \gamma} \tilde F^{\epsilon \ \beta}_{\ \delta} 
- {i\over 2} b_4 \big((k_1+k_3)^2 - (m_1-m_3)^2 \big) \delta^\alpha_\delta \delta^\beta_\gamma \ ,\no \\
n_6 &=&  {1 \over 4} \lambda^2 \tilde F^{\epsilon \ \alpha}_{\ \gamma} \tilde F^{\ \ \beta}_{\epsilon \delta} 
- {i \over 2} b_4 \big((k_1+k_3)^2 - (m_1-m_3)^2 \big) \delta^\alpha_\delta \delta^\beta_\gamma \ , \no \\
n_7 &=&  {1 \over 4} \lambda^2  \tilde F^{ \alpha \beta }{}_{ \epsilon} \tilde F^{\epsilon}_{\ \gamma \delta} 
- {i \over 2} b_3  \big ((k_1+k_2)^2 - (m_1+m_2)^2 \big) 
(\delta^\alpha_\gamma \delta^\beta_\delta - \delta^\alpha_\delta \delta^\beta_\gamma ) \ . \label{nums}
\eea
Note that this amplitude vanishes unless the masses of the external scalars are related as\footnote{This relation holds because of our
choice of masses and gauge-theory representations for the theory with explicitly-broken global symmetry.}
\be 
m_1 + m_2 = m_3 + m_4  \ .  
\ee
We start by looking at the kinematic counterpart of the color seven-term relation in \eqn{extraf}. As explained before, this color identity 
is to be thought of as a set of three-term identities. Consequently, different three-term numerator 
identities need to be imposed for the various possible choices of 
masses for the external particles.\footnote{In principle, imposing several three-term relations 
on the numerator factors is different from imposing a single seven-term relation. 
The former choice is natural in our approach as the various graphs entering the amplitude presentation 
have a definite value of the mass for each internal or external line. Hence, graphs with different external masses are distinct 
and must be treated separately. Taking into account the possible values of the external mass, 
one finds that the color seven-term relation always reduces to three-term relations, and the  
corresponding three-term relations need to be imposed on the numerator factors.} 
We start by taking all masses to be equal. 
In this case, the color factors corresponding to massive $t-$ and $u-$channel exchanges vanish, and the seven terms relation collapses to
\be c_1 - c_4 + c_7 = 0 \ . \label{relcoleqm}\ee
We proceed to impose the corresponding numerator relation
\be n_1 - n_4 + n_7 =0 \ . \label{relnumeqm} \ee
At the ${\cal O}(\lambda^0)$ order  we obtain the condition
\be
\Big\{  2 \big( 1 -b_2 - b_1  \big) (k_1 \cdot k_3) + \big( -1 -b_2 - b_3  \big) (k_1 \cdot k_2)  \Big\} \delta^{\alpha}_\delta
\delta^\beta_\gamma - (3 \leftrightarrow 4) = 0 \ ,
\label{rel1}
 \ee
which can be solved by 
\be b_3 = -1  -b_2 \ , \qquad b_1 =  1 - b_2 \ . \label{solcontact} 
\ee
The constraint at ${\cal O}(\lambda^2)$ is
\bea
&& \lambda^2 \Big(2 \tilde F^{b \ \alpha}_{\ [\delta} \tilde F^{c \ \beta}_{\ \gamma]} \Delta^{ba} \Delta^{ac} 
+ \tilde F^{\ \alpha \beta}_\epsilon \tilde F^{\epsilon}_{\ \gamma \delta}
  \Big)
=
 8 m^2 \delta^{\alpha}_\gamma
\delta^\beta_\delta - (3 \leftrightarrow 4)   \ .
\label{rel2}
\eea
The mass terms appear in this relation because the masses are chosen to be proportional to $\lambda$. 
This ${\cal O}(\lambda^2)$ identity forces us  to pick one of the kinematic group generators, $F^{0 \ \beta}_{\ \alpha}$, 
to be proportional to the mass matrix; indeed, this has been our choice in section \ref{YMBreakScalar}.
In particular, the proportionality relation between $F^{0 \ \beta}_{\ \alpha}$ and the mass, together with the relations (\ref{KJac2}), 
implies
\bea
F^{\alpha \ \gamma}_{\ \beta} \neq 0 ~~~~ &\Leftrightarrow&  ~~~~ m_\alpha+m_\gamma=m_\beta \,,  \no \\
F^{a \ \gamma}_{\ \beta} \neq 0 ~~~~ &\Leftrightarrow&  ~~~~ m_\gamma =m_\beta\,.  \label{Fvanish} \eea
The right-hand side of (\ref{rel2}) is cancelled by  
the terms depending on $\rho$ in the matrix $\Delta$ on the left side 
and the remainder is just a part of the the last identity for the kinematic algebra in (\ref{coloralgebra2}),
\be 2 \tilde F^{b \ \alpha}_{\ [\delta} \tilde F^{c \ \beta}_{\ \gamma]} \Delta^{ba} \Delta^{ac}  %
+ \tilde F^{\ \alpha \beta}_\epsilon \tilde F^{\epsilon}_{\ \gamma \delta}
 = 0 \ . \label{sevenpiece1}\ee 
Next, we consider the cases in which masses are pairwise equal, $m_1 = m_3$ and $m_2 = m_4$ with $m_1 \neq m_2$. In this case the seven-term identity reduces to\footnote{There
are two distinct cases, according to whether $m_1-m_4 = + m_{\text{int}}$ or $m_1-m_4 = - m_{\text{int}}$ 
for some possible mass of the $t$-channel particle.} 
\be c_2 - c_4 + c_7 = 0 \  , \qquad c_3 - c_4 + c_7 = 0 \ .  \ee
Imposing the corresponding numerator identities and repeating the procedure above lead to the extra condition on the contact-term parameters
\be b_1 = b_4 \ , \ee
together with the relations 
\bea && - \tilde F^{b \ \alpha}_{\ \gamma} \tilde F^{c \ \beta}_{\ \delta} \Delta^{ba} \Delta^{ac}  + \tilde F^{\epsilon \ \alpha}_{\ \delta} 
\tilde F^{\ \ \beta \ }_{\epsilon \gamma}  
+ \tilde F^{\ \alpha \beta}_\epsilon \tilde F^{\epsilon}_{\ \gamma \delta}  = 0 \ , \no \\
&&
 - \tilde F^{b \ \alpha}_{\ \gamma} \tilde F^{c \ \beta}_{\ \delta} \Delta^{ba} \Delta^{ac}  + \tilde F^{\epsilon \ \beta}_{\ \gamma} 
\tilde F^{\ \ \alpha \ }_{\epsilon \delta}  
+ \tilde F^{\ \alpha \beta}_\epsilon \tilde F^{\epsilon}_{\ \gamma \delta}  = 0 \ .
\label{sevenpiece2}  \eea
In particular, all numerator relations can be satisfied if we choose $b_1  = b_4  = 0 $, $ b_2 =1$ and $b_3 = -2 $ as in the Lagrangian (\ref{YMscalarGlobalBrokentrunc}).

Finally, there are several three-term color relations corresponding to the case in which all external masses are different,
\be c_2 - c_5 + c_7 = 0 \ , \qquad  c_3 - c_5 + c_7 = 0 \ , \qquad  c_2 - c_6 + c_7 = 0 \ , \qquad  c_3 - c_6 + c_7 = 0 \ .  \ee
The corresponding numerator relations, combined with (\ref{sevenpiece1}), (\ref{sevenpiece2}) and (\ref{Fvanish}), 
are equivalent to a seven-term relation for the global (kinematic) group structure constants,
\bea &&
2 \tilde F^{b \ \alpha}_{\ [\delta} \tilde F^{c \ \beta}_{\ \gamma]} \Delta^{ba} \Delta^{ac} 
+ 4 \tilde F^{\epsilon \ [\alpha}_{\ [\delta} \tilde F^{\ \ \beta ]}_{\epsilon \gamma]}  
+ \tilde F^{\ \alpha \beta}_\epsilon \tilde F^{\epsilon}_{\ \gamma \delta} = 0 \label{KJac3} \ . \eea

There exist one more amplitude with four massive scalars,
\be {\cal A}_4 \big( 1 \varphi_{\alpha}^{\haa}, 2 \varphi_{\beta}^{\hbb}, 3 {\varphi}_{\gamma}^{\hgg}, 4 \overline{\varphi}^{\delta}_{\hdd} \big) \ . \ee
Its numerator factors obey an additional relation provided that 
\be F^{\alpha \ \gamma }_{\ \epsilon} F^{\epsilon \ \beta}_{\ \delta } - F^{\alpha \ \beta }_{\ \epsilon} F^{\epsilon \ \gamma}_{\ \delta } =F^{\alpha \ \epsilon}_{\ \delta} F^{\gamma \ \beta}_{\ \epsilon } \\
\label{rel3} \ . \ee
Equations (\ref{KJac}), (\ref{KJac2}), (\ref{KJac3}) and (\ref{rel3}) are equivalent to requiring that the 
$F$-tensors can be combined to give the structure constants of a larger global symmetry group, 
which is broken by the masses of some of the fields (and by the gauge-group representations). Indeed, this has been our approach in deriving the Lagrangian (\ref{YMscalarGlobalBroken}).

Next, we analyze scalar amplitudes in the spontaneously-broken YM-scalar theory reviewed in \sec{Higgs}. 
We specialize to the case in which the theory has only two real adjoint scalars (i.e. it can be seen as the bosonic part of 
the spontaneously-broken pure $\cN=2$ SYM theory). 
There are two non-zero amplitudes with two massive and two massless scalars,
\bea 
&& \!\!\!\!\! {\cal A}_4 \big( 1 \phi^{0 \ha}, 2 \phi^{0 \hb}, 3 \varphi_{\haa}, 4 \overline{\varphi}^{\hbb} \big) = \no \\
&&   -{ i } g^2 \left\{ \tilde f^{\ha \ \hgg}_{\ \hbb} \tilde f^{\hb \ \haa}_{\ \hgg} 
 { k_1 \cdot k_4 + 2 m^2   \over (k_1+k_4)^2 - m^2}  + 
\tilde f^{\hb \ \hgg}_{\ \hbb} \tilde f^{\ha \ \haa}_{\ \hgg} {k_1 \cdot k_3 + 2 m^2 \over (k_1+k_3)^2 - m^2} 
+  \tilde f^{\ha \hb \hc} \tilde f^{\hc \ \haa}_{\ \hbb} { k_1 \cdot k_3 - k_1 \cdot k_4 \over (k_1+k_2)^2}  \right\} \,, \no \\[10pt]
&& \!\!\!\!\! {\cal A}_4 \big( 1 \phi^{1 \ha}, 2 \phi^{1 \hb}, 3 \varphi_{\haa}, 4 \overline{\varphi}^{\hbb} \big) = \no \\
&&   -{ i } g^2 \left\{ \tilde f^{\ha \ \hgg}_{\ \hbb} \tilde f^{\hb \ \haa}_{\ \hgg} 
 { k_1 \cdot k_4 + 2 k_1 \cdot k_2     \over (k_1+k_4)^2 - m^2}  + 
\tilde f^{\hb \ \hgg}_{\ \hbb} \tilde f^{\ha \ \haa}_{\ \hgg} {k_1 \cdot k_3 + 2 k_1 \cdot k_2  \over (k_1+k_3)^2 - m^2} 
+  \tilde f^{\ha \hb \hc} \tilde f^{\hc \ \haa}_{\ \hbb} { k_1 \cdot k_3 - k_1 \cdot k_4 \over (k_1+k_2)^2}  \right\} \,, \no \\  \eea
where $\phi^{0}$ is the fluctuation of the field responsible for symmetry breaking (i.e.  the fluctuations of the field which acquires a VEV), $\phi^1$ is the other real scalar in the theory and the masses of 
the two massive scalars need to be equal to have a non-zero amplitude.
These amplitudes manifestly display color/kinematics duality, as the numerator factors obey the same relations as the corresponding color factors. 

Finally, we consider scalar amplitudes with four massive fields, 
\be A \big( 1 \varphi_{ \haa}, 2 \varphi_{\hbb}, 3 \overline{\varphi }{}^{\hgg}, 4 \overline{\varphi }{}^{\hdd} \big) \ . \ee
The amplitude can be organized in the form (\ref{4massform_gauge})
with inverse propagators (\ref{4massprops}), color factors
\bea 
c_1 = f^{\ha \ \haa}_{\ \hdd} \tilde f^{\ha \ \hbb }_{\ \hgg} \ , & \quad c_2 = \tilde f^{\hee \ \haa}_{\  \hdd} \tilde f^{\ \ \hbb}_{\hee  \hgg} \ , & 
\quad c_3 = \tilde f^{\ \ \haa}_{\hee  \hdd} \tilde f^{\hee \ \hbb}_{\  \hgg} \ , \qquad  c_4 = \tilde f^{\ha \ \haa}_{\  \hgg} \tilde f^{\ha \ \hbb}_{\  \hdd} \ , \no \\
c_5 =  \tilde f^{\ \ \haa}_{\hee  \hgg} \tilde f^{\hee \ \hbb}_{\  \hdd} \ ,  &   
\quad c_6 = \tilde f^{\hee \ \haa}_{\ \hgg} \tilde f^{\  \ \hbb}_{\hee  \hdd}  \ , &  \quad c_7 = f_{\hgg  \hdd}^{\  \ \hee} \tilde f_{\hee}^{\ \haa \hbb} \ , 
\label{color_zero}
\eea
and numerator factors
\bea 
\tilde n_1 &=& \tilde n_2 = \tilde n_3 =   - \big( k_1 \cdot k_2  -  m_1 m_2  - k_1 \cdot k_3  - m_1 m_3 \big) \ , \no \\
\tilde n_4 &=& \tilde n_5 = \tilde n_6 =  - \big(  2 k_1 \cdot k_2 - 2 m_1 m_2 + k_1 \cdot k_3 + m_1 m_3 \big) \ , \no \\
\tilde n_7 &=&  \big (k_1 \cdot k_2 + 2 k_1 \cdot k_3 -  m_1 m_2 + 2 m_1 m_3 \big) \ . \label{numstilde}
\eea
It is immediate to verify that these numerators obey the same three-term relations as the corresponding color factors.

\subsection{Supergravity amplitudes \label{secsugraamp}}

In this section we compare, in explicit examples, the result of the double-copy construction described in section \ref{2copySec} 
with three- and four-point amplitudes computed from the expected supergravity Lagrangian derived in \sec{sec:YME} and 
find the map between the Lagrangian and double-copy fields.

One of the gauge-theory factors entering the construction is  the spontaneously-broken $\cN=2$ SYM theory. The bosonic 
part of the Lagrangian is shown in section \ref{Higgs}. We list here the bosonic fields in four dimensions:
\be 
\big( A^\ha_\mu \ , \phi^{\ha a'} \ , W_{\haa \mu} \ , \varphi_{\haa} \ , \overline{W}_\mu^\haa \ ,  \overline{\varphi}^\haa \big) \ , \qquad a'=1,2  \ .  
\ee
The other gauge-theory factor is the YM-scalar theory discussed in section \ref{YMBreakScalar}. Its field content is
\be 
\big( A^\ha_\mu \ , \phi^{1\ha} \ , \phi^{a\ha}  \ , \varphi^{\haa}_\alpha \ ,   \overline{\varphi}^\alpha_\haa \big)  \ .   
\ee
We will verify that the double-copy of these theories yields the spontaneously-broken generic Jordan family YMESG theory with general gauge group. 

To identify the result of the double-copy construction as one of the supergravities discussed in \sec{sec:YME}, we  want the theory to have an uplift 
to five dimensions. To this end, we need to single out a particular adjoint scalar which does not enter the trilinear couplings in 
\eqn{YMscalarGlobalBroken} and hence can combine with the four-dimensional gluons to produce the gluons of the five-dimensional theory. 
We will denote this scalar as $\phi^{1 \ha}$. In contrast, the scalars corresponding to non-vanishing $F^{abc}$ will be denoted as $\phi^{a \ha}$, where the index $a$ 
runs over the multiplets transforming in the adjoint representation of the unbroken gauge group $\tilde K$ ($a=2,3,\ldots , \text{dim}(\tilde K) +1$)
and  can also include extra spectator fields, when present. 
 With a slight change of notation from \sec{YMBreakScalar}, 
 the global-group generator proportional to the masses will be denoted as 
$F^{2 \ \beta}_{\ \alpha}$. The corresponding scalar field will be called $\phi^2$, while $\phi^3,\phi^4, \ldots$ will 
be the other massless scalars partaking to the trilinear interactions controlled by  
the $F^{abc}$ tensors. This shift of indices is necessary to ``align" the  gauge-theory global 
indices with the supergravity gauge adjoint indices, as the supergravity always 
has at least two spectator multiplets.

\subsubsection{Three-point amplitudes and double-copy field map \label{themap}}

We begin by finding the three-point amplitudes of two massive scalars and a massless non-spectator scalar 
in a spontaneously-broken generic Jordan family YMESG theory. There are two such amplitudes,
\bea 
{\cal M}_3(1 \varphi_\alpha, 2 {\overline{\varphi}}^\beta, 3 {y^2}) &=& - i\Big({\kappa \over 2} \Big) \frac{ \sqrt{2} \gsugra  m}{\sqrt{1 - V_s^2}} \tilde F^{2 \ \alpha}_{\ \beta} \ , 
\label{a1}\\ 
{\cal M}_3(1 \varphi_\alpha, 2 {\overline{\varphi}}^\beta, 3 {y^a}) &=& - i \Big({\kappa \over 2} \Big) { \sqrt{2} \gsugra  m} \tilde F^{a \ \alpha}_{\ \beta}  \ .
\label{a2}
\eea
The first amplitude involves the scalar of the preferred vector multiplet which contains the fluctuations of the field acquiring a VEV, 
while the second involves the other massless scalars transforming in the adjoint representation of the unbroken gauge group. 
Note that we need both massive scalars to have the same mass in order for the amplitude to be non-zero.

It is natural to expect that these amplitudes are reproduced by the double copy 
\bea 
{\cal A}_3(1 \varphi, 2 {\overline{\varphi}}, 3 {\phi}^0)\Big|_{\cN=2} \otimes {\cal A}_3(1 \varphi, 2 {\overline{\varphi}}, 3 {\phi}^2)\Big|_{\cN=0} &=& - {i \over \sqrt{2}} \Big({\kappa \over 2} \Big) \lambda m \sqrt{1 + \rho^2} 
\tilde F^{a \ \alpha}_{\ \beta}  \ , \label{b1} \\
{\cal A}_3(1 \varphi, 2 {\overline{\varphi}}, 3 {\phi}^0)\Big|_{\cN=2} \otimes {\cal A}_3(1 \varphi, 2 {\overline{\varphi}}, 3 {\phi}^a)\Big|_{\cN=0} &=& - {i \over \sqrt{2}} \Big({\kappa \over 2} \Big) \lambda m 
\tilde F^{a \ \alpha}_{\ \beta}  \ ,
\label{b2}
\eea
where we have used eqs.~(\ref{amp3ptN0b}) and (\ref{defS}). This double-copy can be constructed because eq.~\eqref{mass_relation}
guarantees that the massive fields have equal masses. The massless scalar $\phi^0$ in the  $\cN=2$ theory  is the 
fluctuation of the field that acquires the VEV,  and the scalar $\phi^2$ of the  $\cN=0$ theory is the scalar corresponding to 
the $U(1)$ generator related to the mass.

The amplitudes \eqref{a1}, \eqref{a2} are equal to the amplitudes \eqref{b1}, \eqref{b2} provided that we identify 
\be
\Big({\kappa \over 2} \Big) \lambda = 2 g_s  \ , \qquad \rho = {V_s \over \sqrt{1 - V_s^2}} \ , \qquad f^{ABC}_{\text{sugra}} = F^{ABC} \label{parameter} \ .
\ee
This identification, together with the relation between the gauge-theory mass and preferred $U(1)$ generator (\ref{gaugem}), 
leads precisely to the expression for the mass in the spontaneously-broken supergravity (\ref{sugram}). The 
other supergravity amplitudes with two massive fields are:
\bea 
{\cal M}_3(1 \varphi_\alpha, 2 {\overline{\varphi}}^\beta, 3 {y^0})&=&  \sqrt{2} i \Big({\kappa \over 2} \Big) m^2  \delta^\alpha_\beta \no  \ ,\\
{\cal M}_3(1\varphi_\alpha, 2{\overline{\varphi}}^\beta, 3 {A^-}) &=&   \frac{i}{\sqrt{2}} \Big({\kappa \over 2} \Big) m \, \Pvec_3\cdot(k_1-k_2) \delta^\alpha_\beta \no
\ ,\\
{\cal M}_3(1\varphi_\alpha, 2{\overline{\varphi}}^\beta, 3{A^0})&=& -  \frac{i}{\sqrt{2}} \Big({\kappa \over 2} \Big)  m  \, \Pvec_3\cdot(k_1-k_2) \delta^\alpha_\beta \no
\ ,\\
{\cal M}_3(1\varphi_\alpha, 2{\overline{\varphi}}^\beta, 3{A^a})&=& {i \over \sqrt{2}} \Big({\kappa \over 2} \Big) \lambda \, \Pvec_3\cdot (k_1- k_2) \Delta^{ab} \tilde F^{b \ \alpha}_{\ \beta} 
\no \ ,\\
{\cal M}_3(1\varphi_\alpha, 2{\overline{W}}^\beta, 3 {x^a}) &=& - {i \over \sqrt{2}} \Big({\kappa \over 2} \Big)  {\lambda} \, \Pvec_2\cdot (k_1- k_3) \Delta^{a b} \tilde F^{b \ \alpha}_{\ \beta}
\no \ ,\\
{\cal M}_3(1{\overline{\varphi}}^\alpha, 2 { W}_\beta, 3 {x^a}) &=& - {i \over \sqrt{2}} 
\Big({\kappa \over 2} \Big) \lambda \, \Pvec_2\cdot (k_1- k_3) \Delta^{ab} \tilde F^{b \ \beta}_{\ \alpha}
\no\ , \\
{\cal M}_3(1{W}_\alpha, 2{\overline{W}}^\beta, 3{A^-})  &=&  -  \frac{i}{\sqrt{2}} \Big({\kappa \over 2} \Big) m  \, \Pvec_3\cdot (k_1- k_2) \, \Pvec_1\cdot\Pvec_2 \delta^\alpha_\beta \no
\ ,\\
{\cal M}_3(1{W}_\alpha, 2{\overline{W}}^\beta, 3{A^0})  &=&  \frac{i}{\sqrt{2}}  \Big({\kappa \over 2} \Big)  m \, \Pvec_3\cdot (k_1- k_2) \, \Pvec_1\cdot\Pvec_2  \delta^\alpha_\beta \no
\ ,\\
{\cal M}_3(1{W}_\alpha, 2{\overline{W}}^\beta, 3{A^a})    &=& -{i \over 2 \sqrt{2}} \Big({\kappa \over 2} \Big)  \lambda   \big(
              \Pvec_3\cdot (k_1- k_2) \Pvec_1\cdot\Pvec_2 + \text{cyclic} 
    \big) \Delta^{ab} \tilde F^{b \ \alpha}_{ \ \beta}  \no
\no \ , \\
{\cal M}_3(1{W_\alpha}, 2{\overline{W}}^\beta, 3{x^0}) &=& - {  \sqrt{2}} i\Big({\kappa \over 2} \Big) \,\LCivita(k_1, k_2, \Pvec_1,     \Pvec_2) \delta^\alpha_\beta \no
\ ,\\
{\cal M}_3(1{W_\alpha}, 2{\overline{W}}^\beta, 3{y^0}) &=& - {\sqrt{2}} i \Big({\kappa \over 2} \Big)  {( m^2 \Pvec_1\cdot \Pvec_2 +   
\Pvec_1\cdot p_2 \,  \Pvec_2\cdot p_1)} \delta^\alpha_\beta \no
\ , \\
{\cal M}_3(1{W_\alpha}, 2{\overline{W}}^\beta, 3{y^a}) &=&   {i \over \sqrt{2}} \Big({\kappa \over 2} \Big)  m {\lambda} \, \Pvec_2\cdot\Pvec_3 \, \Delta^{ab} \tilde F^{b \ \alpha}_{\ \beta} 
\no \ ,\\
{\cal M}_3(1 \varphi_\alpha, 2 {\overline{W}}^\beta, 3 {A^-})&=&   
 \sqrt{2} i \Big({\kappa \over 2} \Big) \, \LCivita(k_2, k_3, \Pvec_2,    \Pvec_3) \delta^\alpha_\beta \no
\ ,\\
{\cal M}_3(1 \varphi_\alpha, 2 {\overline{W}}^\beta, 3 {A^0})&=&   
 \sqrt{2} i \Big({\kappa \over 2} \Big) \, \LCivita(k_2, k_3, \Pvec_2,    \Pvec_3)   \delta^\alpha_\beta \no
\ ,\\
%
{\cal M}_3(1{\overline{\varphi}}^\beta, 2{ W}_\alpha, 3 {A^0}) &=& 
 \sqrt{2} i \Big({\kappa \over 2} \Big)  \, \LCivita(k_2, k_3, \Pvec_2,    \Pvec_3) \delta^\alpha_\beta  \ ,
\eea
where the Levi-Civita tensor is normalized as $\LCivita^{0123}=-1$.
These amplitudes can be reproduced by the double-copy prescription with the following field identification:
\bea \!\!\! & \!\!\!\!\!\!\!\!\! \begin{array}{rclcrcl}
 \frac{1}{\sqrt{2}} (A^{-1}_\pm -  A^0_\pm) &=&  \phi^0 \big|_{\cN=2} \otimes A_\pm \big|_{\cN=0}\,,  & \quad & 
 A^a_\pm  &=& A_\pm \big|_{\cN=2} \otimes \phi^a \big|_{\cN=0}\,,  \\
\pm i  \frac{1}{\sqrt{2}}(A^{-1}_\pm +A^0_\pm ) &=& \phi^1 \big|_{\cN=2} \otimes A_\pm \big|_{\cN=0}\,, &&
\pm \alpha i A^1_\pm   &=&   A_\pm \big|_{\cN=2} \otimes  \phi^1 \big|_{\cN=0}\,,  \\
  \frac{1}{\sqrt{2}} (y^0+i x^0) &=&  A_+ \big|_{\cN=2} \otimes A_- \big|_{\cN=0}\,, &&
\frac{1}{\sqrt{2}} (y^0 - i x^0) &=&  A_- \big|_{\cN=2} \otimes A_+ \big|_{\cN=0}\,,  \\
\varphi_\alpha &=&  \varphi  \big|_{\cN=2} \otimes \varphi_\alpha \big|_{\cN=0}\,, && W_\alpha &=&  W \big|_{\cN=2} \otimes \varphi_\alpha \big|_{\cN=0}\,,  \\
 y^a &=& \phi^0 \big|_{\cN=2} \otimes \phi^a \big|_{\cN=0}\,, &&
x^a &=&
-\phi^1 \big|_{\cN=2} \otimes \phi^a \big|_{\cN=0}\,, \\
 y^1 &=& \phi^0 \big|_{\cN=2} \otimes  \phi^1 \big|_{\cN=0}\,, 
&& x^1 &=&
  -\phi^1 \big|_{\cN=2} \otimes \phi^1 \big|_{\cN=0}\,.  \end{array} \no \qquad \\  \label{fieldmap}
\eea
The field $\phi^1$ is a distinguished spectator scalar in the YM-scalar theory which does not enter the trilinear couplings and is necessary for the theory 
to have a five-dimensional uplift, and the index $a$ runs over the vector multiplets transforming in the adjoint representation 
of the unbroken gauge group plus extra spectator fields, when present. The free parameter $\alpha= \pm 1$ reflects the 
symmetry $\phi^1 \rightarrow - \phi^1$ of the $\cN=0$ gauge-theory factor.
Note that the spectator vectors $A^{-1,0,1}_\mu$ and spectator scalars $x^{0,1}, y^{0,1}$ are always present 
due to the choice of compact gauging and to the requirement of a five-dimensional uplift.

In the case in which the supergravity has more than one flavor of massive vectors, additional multi-flavor amplitudes become possible,
\bea {\cal M}_3 (1W_\alpha, 2 W_\beta, 3 \overline{W}^\gamma ) &=& { i \over 2 \sqrt{2}}  \Big({\kappa \over 2} \Big)   \lambda   \big(
              \Pvec_3\cdot (k_1- k_2) \Pvec_1\cdot\Pvec_2 + \text{cyclic} 
     \big) \tilde F^{\alpha  \beta}_{ \ \ \gamma} \ , \no \\
     {\cal M}_3 (1W_\alpha, 2 \varphi_\beta, 3 \overline{\varphi}^\gamma ) &=&- { i \over \sqrt{2}}  \Big({\kappa \over 2} \Big)  \lambda  
              \Pvec_1 \cdot  k_2   \tilde F^{\alpha  \beta}_{ \ \ \gamma} \ . \eea
They are reproduced by a double-copy construction with multi-flavor gauge theories.

\subsubsection{Four-point amplitudes}

To test the identification of parameters and fields constructed in \sec{themap} we construct selected 
four-point amplitudes with two and four massive fields and
compare them with the double-copy construction. We start with the supergravity amplitude with four massive scalars. Using the results from the previous sections, 
it can be expressed in the following form,
\be {\cal M}_4 \big( 1 \varphi_{\alpha}, 2 \varphi_{\beta}, 3 \overline{\varphi}^{\gamma}, 4 \overline{\varphi}^{\delta} \big) = 
-i \Big({\kappa \over 2}\Big)^{\! 2} \! \left( {n_1 \tilde n_1 \over D_1 } + {n_2 \tilde n_2 \over D_2 } + {n_3 \tilde n_3 \over D_3 } +{n_4 \tilde n_4 \over D_4 } +{n_5 \tilde n_5 \over D_5 } +{n_6 \tilde n_6 \over D_6 } +
{n_7 \tilde n_7 \over D_7 } \right) , \label{4massform} \ee
where the numerators are given by (\ref{nums}) and (\ref{numstilde}). 
It is instructive to verify that all poles in the above amplitude correspond to the exchange of a particle 
of the theory. 
Specifically, for a given assignment of external masses, aside from two massless channels, 
there are three massive channels with square masses $(m_1+m_2)^2, (m_1-m_3)^2$ and $(m_1-m_4)^2$. 
One can  see that, thanks to the relations (\ref{Fvanish}), the numerators $n_2,n_3,n_5,n_6,n_7$ in 
(\ref{nums}) are either zero or proportional to inverse propagators
when the mass of the intermediate channel is not one of the masses of the particles in the theory (i.e. one 
the eigenvalues of the matrix  $m_{\alpha}^{\ \beta}$).

We have verified that the expression (\ref{4massform}) reproduces the one from a Feynman-rule computation, once the field map (\ref{fieldmap}) is employed. 
The  expression for the general four massive scalar amplitude substantially simplifies
in the simplest case in which the supergravity has a $SU(2)$ gauge group which is spontaneously-broken 
to its $U(1)$ subgroup.
In this case only one flavor of massive fields is present, $\alpha,\beta \equiv 1$, and the structure constants become
\be 
F^{\alpha \ \gamma}_{\ \beta} \equiv  0 \ , \qquad F^{2 \ \beta}_{\ \alpha} \equiv  i \, . 
\ee
It is also convenient to absorb the $\rho$-dependent factor in the definition of $\lambda$,
\be 
\tilde \lambda = \sqrt{1 + \rho^2} \lambda \,.
\ee
The amplitude has a simple expression,
\be
{\cal M}_4(1\varphi,2\varphi,3{\overline{\varphi}},4{\overline{\varphi}}) = 
 {i \over 2}  \Big({\kappa \over 2} \Big)^2  (\tilde \lambda^2- 4 k_1\cdot k_2)
\left[1+2\frac{k_1\cdot k_2-m^2}{(k_2+k_3)^2}+2\frac{k_1\cdot k_2-m^2}{(k_2+k_4)^2}\right] \,.
\ee
In this particular case, all non-zero amplitudes with two massive fields have simple expressions, and we list here some of them: 
\bea
{\cal M}_4(1{x^0},2{\overline{\varphi}}, 3{x^0},4\varphi) \!\!\!&=\!\!\!&
 i \Big({\kappa \over 2} \Big)^2 \left[
 -2 m^2 + 4 \frac{k_1\cdot k_4 k_3\cdot k_4 }{(k_1 + k_3)^2}
 \right]
\ ,\no \\
{\cal M}_4(1{y^0},2{\overline{\varphi}}, 3{y^0},4\varphi) \!\!\!&=\!\!\!&
i \Big({\kappa \over 2} \Big)^2 \left[
-2m^2 
-\frac{m^4}{k_2\cdot k_3}
-\frac{m^4}{k_3\cdot k_4}
+4 \frac{ k_2\cdot k_3  k_3\cdot k_4}{(k_2+k_4)^2}
 \right]
\ , \no \\
{\cal M}_4(1{x^1},2{\overline{\varphi}}, 3{x^1},4\varphi) \!\!\!&=\!\!\!&
i \Big({\kappa \over 2} \Big)^2 \left[-2m^2 +4 \frac{ k_2\cdot k_3  k_3\cdot k_4}{(k_2+k_4)^2} \right]
\ , \no  \\
{\cal M}_4(1{y^0},2{\overline{\varphi}}, 3{y^2},4\varphi) \!\!\!&=\!\!\!& {i \over \sqrt{2}} \Big({\kappa \over 2} \Big)^2 \tilde \lambda \,  m
\left[
2 
+\frac{m^2}{k_2\cdot k_3}
+\frac{m^2}{k_3\cdot k_4}
\right] \no \ , \\
{\cal M}_4(1{y^1},2{\overline{\varphi}}, 3{y^1},4\varphi) \!\!\!&=\!\!\!& i \Big({\kappa \over 2} \Big)^2 \left[-(k_2+ k_4)^2 + 
4 \frac{k_2\cdot k_3 k_3\cdot k_4}{(k_2+k_4)^2}\right] \ ,
\no \\
{\cal M}_4(1{y^2},2{\overline{\varphi}}, 3{y^2},4\varphi) \!\!\!&=\!\!\!& 
i \Big({\kappa \over 2} \Big)^2 \left[
-{\tilde \lambda^2 \over 2} -2 m^2  
-\frac{m^2 \tilde \lambda^2}{2k_2\cdot k_3}
-\frac{ m^2 \tilde \lambda^2}{2 k_3\cdot k_4}
+ 4 \frac{k_2\cdot k_3 k_3\cdot k_4}{(k_2+k_4)^2}
\right] \ ,
\no \\
{\cal M}_4(1{x^2},2{\overline{\varphi}}, 3{x^2},4 \varphi)\!\!\!&=\!\!\!& i \Big({\kappa \over 2} \Big)^2 \left(
  {\tilde \lambda^2 \over 2}- (k_2 +  k_4)^2 
+ {\tilde \lambda^2 \over 2 } \frac{k_3\cdot k_4}{k_2\cdot k_3} 
+ {\tilde \lambda^2  \over 2} \frac{ k_2\cdot k_3}{k_3\cdot k_4} 
+4 \frac{k_2\cdot k_3 k_3\cdot k_4}{(k_2+k_4)^2}\right)  .   \no \\
\eea
There also exist amplitudes which vanish due to non-trivial cancellations. Among them there are
\bea
{\cal M}_4(1 {x^1},2 {\overline{\varphi}}, 3 {x^2},4 \varphi) &=& 0 \ ,\no
\\
{\cal M}_4(1 {y^0},2 {\overline{\varphi}}, 3 {y^1},4 \varphi) &=& 0 \ , \no 
\\
{\cal M}_4(1 {y^1},2 {\overline{\varphi}}, 3 {y^2},4 \varphi) &=& 0 \ .
\eea
We have explicitly checked that the result of the double-copy calculation for the amplitudes listed above matches the given expressions.

As an interesting aside, we note that the double copy of spontaneously-broken YM theory with itself,
namely \cancel{YM}\,$\otimes$\,\cancel{YM}, is a valid construction in the current treatment. 
To understand what the result might be, let us take spontaneously-broken gauge theories with all masses equal and consider the scattering of
four massive scalars. The kinematic numerator factor $\tilde n_7$ in \eqn{numstilde} is nonvanishing when
$(k_1 + k_2)^2 = (2m)^2$ and thus the graph 7 in \fig{fig4mass} exhibits a pole for such momentum configuration.
This pole does not contribute to the gauge-theory amplitude due to the vanishing of the color factor $c_7$ in \eqn{numstilde}. 
However, through the double copy, this pole features in the corresponding supergravity amplitude and signals the existence of a state of mass $2m$
in the spectrum. Such state is not part of the naive spectrum -- the gauge invariant part of the tensor product of 
the two gauge-theory spectra; unitarity requires it to be included.
The argument can be repeated starting from higher-point gauge-theory amplitudes and leads to the extension 
of the naive spectrum by an infinite number of states with equally spaced masses, $m_n = n m$ with $n$ integer. 
These states also carry maximum spin two.
Following from the discussion in \app{SYMdredSect} the amplitudes generated by the double-copy construction S\cancel{YM}\,$\otimes$\,S\cancel{YM} should belong to $(D+1)$-dimensional Kaluza-Klein supergravity.

\section{Loop amplitudes \label{sec:loops}}
\renewcommand{\theequation}{5.\arabic{equation}}
\setcounter{equation}{0}

Here we work out one of the simpler one-loop amplitudes in explicitly broken YM~+~$\phi^3$ in a form that obeys color/kinematics duality. Then, using the corresponding amplitudes in spontaneously-broken SYM, the double copy gives the one-loop four-vector amplitude in spontaneously-broken YMESG.

\subsection{One-loop massless-scalar amplitude in broken YM~+~$\phi^3$ theory \label{YMphi3_1L}}

Consider the one-loop amplitude for four massless external scalars in the explicitly broken YM~+~$\phi^3$ theory. We write the complete amplitude in the cubic-diagram form \eqref{BCJformYM}, decomposed over the massless and massive internal modes,
\begin{equation}
 {\cal A}^{1\text{-loop}}_4 =  g^{4} \,
\sum_{{\cal S}_4} \sum_{i\in{\rm\{box,tri,bub\}}} \int  \frac{d^D \ell}{(2 \pi)^D}
 \frac{1}{S_i}
   \left(\frac{n_i c_i}{D_i}+\sum_\alpha \frac{n_{i,\alpha}\, c_i}{D_{i,\alpha}}+\sum_\alpha \frac{\overline{n}_{i,\alpha}\, c_i}{D_{i,\alpha}}\right)\,,
\hskip .7 cm 
\label{1LoopCK}
\end{equation}
where the first sum runs over the permutations ${\cal S}_4$ of all four external leg labels. The second sum runs over the three 
listed integral topologies (box, triangle, internal bubble diagrams), and the corresponding symmetry factors 
are $S_{\rm box}=8$, $S_{\rm tri}=4$ and $S_{\rm bub}=16$. The summation index $\alpha$ labels the massive modes,
with mass $\pm m_\alpha$, inside the loop diagrams. The numerator without an $\alpha$ index corresponds to massless modes in the loop.

In the canonical order of the external legs, $(1,2,3,4)$, the color factors are
\bea
c_{\rm box}&=& \tf^{\hat b {\hat a}_1 \hat c} \tf^{\hat c{\hat a}_2 \hat d} \tf^{\hat d{\hat a}_3\hat e} \tf^{\hat e{\hat a}_4\hat b}\,, \nn \\
c_{\rm tri}&=&  \tf^{{\hat a}_1 {\hat a}_2 \hat c} \tf^{\hat b \hat c \hat d} \tf^{\hat d{\hat a}_3\hat e} \tf^{\hat e{\hat a}_4\hat b}\,, \nn \\
c_{\rm bub}&=&  \tf^{{\hat a}_1 {\hat a}_2 \hat c} \tf^{\hat b \hat c \hat d} \tf^{\hat d \hat e \hat b} \tf^{\hat e {\hat a}_3 {\hat a}_4}\,.
\eea
The denominator factors in the canonical ordering are given by
\bea
&&D_{\rm box}=\ell_1^2\ell_2^2\ell_3^2\ell_4^2\,,~~D_{\rm tri}=s \ell_2^2\ell_3^2\ell_4^2\,,~~D_{\rm bub}=s^2 \ell_2^2\ell_4^2\,, \nn \\
&&D_{{\rm box},\alpha}=(\ell_1^2-m_\alpha^2)(\ell_2^2-m_\alpha^2)(\ell_3^2-m_\alpha^2)(\ell_4^2-m_\alpha^2)\,, \nn \\ &&D_{{\rm tri},\alpha}=s (\ell_2^2-m_\alpha^2)(\ell_3^2-m_\alpha^2)(\ell_4^2-m_\alpha^2)\,,\nn \\
&&D_{{\rm bub},\alpha}=s^2 (\ell_2^2-m_\alpha^2)(\ell_4^2-m_\alpha^2)\,,~~~
\label{Denominators}
\eea
where $\ell_i=\ell-(k_1+\ldots+k_i)$. 

\begin{figure}[t]
      \centering
      \includegraphics[scale=0.79]{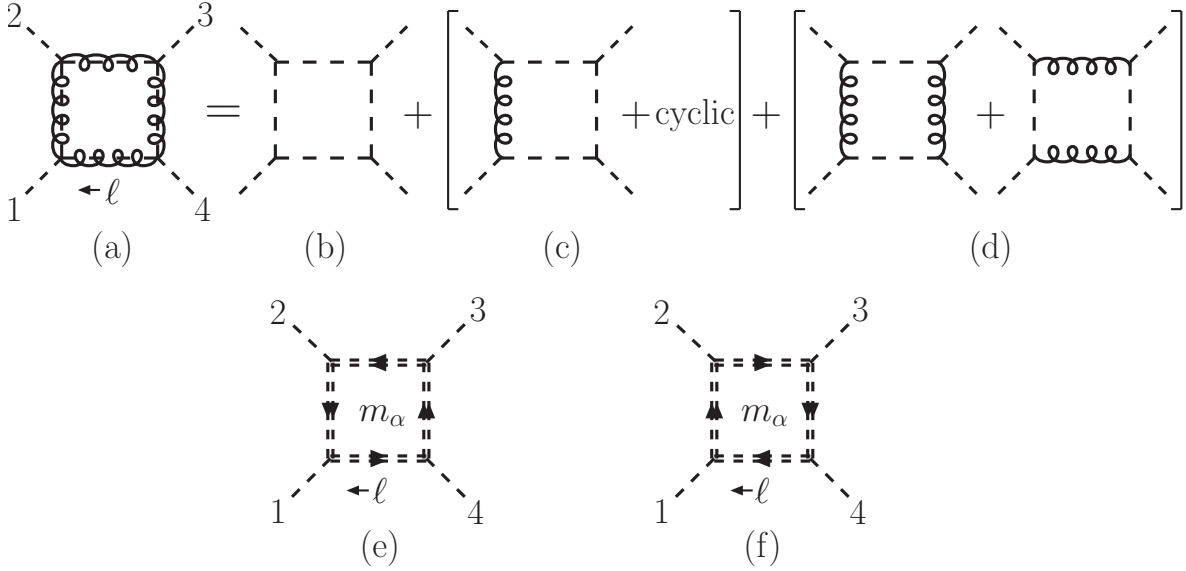}
      \caption[a]{\small The different types of box diagrams that contribute to the one-loop amplitude with four massless scalars in explicitly broken YM~+~$\phi^3$ theory. Diagram (a) denotes the sum of all box diagrams with massless internal states; the contribution (b) is of order $\lambda^4$, (c) is of order $\lambda^2$, and (d) is of order $\lambda^0$. Additionally, there are two conjugate diagrams (e), (f) with internal scalars of mass $m_\alpha$.  Dashed lines denote scalar fields, double lines of these corresponds to massive scalars, while curly lines denote vector fields. Note that quartic-scalar interactions are implicitly included in these diagrams, according to their power in the $\lambda$ coupling.
               \label{YMphi3diagramsNewFigure} }
\end{figure}

In ref.~\cite{Chiodaroli:2014xia} the massless contributions to this amplitude were worked out using the 
unitarity method~\cite{Bern:1994zx}; we quote the result again, in a slightly different form. We write the box 
numerator corresponding to massless fields, \fig{YMphi3diagramsNewFigure}(a), as
\be
n_{\rm box}=n^{(4)}_{\rm box}+n^{(2)}_{\rm box}+n^{(0)}_{\rm box}\,,
\label{fullNum}
\ee
where the superscript denotes the order in $\lambda$.
The ${\cal O}(\lambda^4)$ contribution of the box numerator, shown in \fig{YMphi3diagramsNewFigure}(b), is entirely expressed in terms of the structure constants of the global group,
\be
n^{(4)}_{\rm box}(1,2,3,4;\ell)= \frac{\lambda^4}{4} F^{b   a_1  c} F^{  c a_2  d} 
F^{  d a_3  e} F^{  e{ a}_4 b} \,. \label{numA}
\ee
The ${\cal O}(\lambda^2)$ numerator contributions, shown in \fig{YMphi3diagramsNewFigure}(c), is given by
\bea
n^{(2)}_{\rm box}(1,2,3,4;\ell)\!&=&\!
\frac{\lambda^2}{24} \Big\{ (N_\phi +D- 2) \big(F^{ a_1{ a}_4   b} F^{  b a_3 a_2}  (\ell_2^2 + \ell_4^2) + F^{  a_1{ a}_2   b} F^{  b a_3 a_4} (\ell_1^2 + \ell_3^2)\big) \nn \\ && \null +24 \big(s  F^{  a_1{ a}_4   b} F^{  b a_3 a_2}  +t F^{  a_1{ a}_2   b} F^{  b a_3 a_4} \big) +
   \delta^{a_3 a_4} \Tr_{12}  (6 \ell_3^2 - \ell_2^2 - \ell_4^2) \nn \\ &&
  \null  + \delta^{a_2 a_3} \Tr_{14} (6 \ell_2^2 - \ell_1^2 - \ell_3^2)  +
      \delta^{a_1 a_4} \Tr_{23} (6 \ell_4^2 - \ell_1^2 - \ell_3^2)  \label{numB}   \\ &&
      \null + \delta^{a_1 a_2} \Tr_{34} (6 \ell_1^2 - \ell_2^2 - \ell_4^2)  
      + (\ell_1^2 + \ell_2^2 + \ell_3^2 + \ell_4^2) (\delta^{a_2 a_4} \Tr_{13} + \delta^{a_1 a_3} \Tr_{24}) \Big\}\,,  \no 
 \eea
and the ${\cal O}(\lambda^0)$ numerator contributions, shown in \fig{YMphi3diagramsNewFigure}(d), is
\bea
n^{(0)}_{\rm box}(1,2,3,4;\ell)\!&=&\!\frac{1}{24} \Big\{  \delta^{ a_1 a_2}  \delta^{ a_3 a_4}  \big[24 t (t - 2 \ell_1^2 - 2 \ell_3^2)  +2 (N_\phi +D- 2) (3 \ell_1^2 \ell_3^2 - \ell_2^2 \ell_4^2) \nn \\ && 
\hskip 2.5cm \null  + (N_\phi +D+ 14) \big(t (\ell_1^2 + \ell_2^2 + \ell_3^2 + \ell_4^2) - u (\ell_1^2 + \ell_3^2)\big) \big]  \nn \\ &&
      \null ~ \,\,\,\,+ \delta^{ a_2 a_3} \delta^{ a_1 a_4}   \big[24 s (s - 2 \ell_2^2 - 2 \ell_4^2)+2 (N_\phi +D- 2)(3 \ell_2^2 \ell_4^2 - \ell_1^2 \ell_3^2)   \nn \\ &&
\hskip 2.5cm \null      + (N_\phi +D+ 14) \big(s (\ell_1^2 + \ell_2^2 + \ell_3^2 + \ell_4^2) - u (\ell_2^2 + \ell_4^2)\big)\big]  \nn \\ &&
     \null ~ \,\,\,\,+ \delta^{ a_1 a_3} \delta^{ a_2 a_4}  \big[2(N_\phi +D- 2)  (\ell_1^2 \ell_3^2 + \ell_2^2 \ell_4^2)  \nn \\ && 
  \hskip 2.5cm   \null - (N_\phi +D+ 14) \big(s (\ell_1^2 + \ell_3^2) + t (\ell_2^2 + \ell_4^2)\big)\big]\Big\} \, .
\label{numC}
\eea
As before $\ell_i=\ell-(k_1+\ldots+k_i)$ and in \eqn{numB} 
we use the shorthand notation $\Tr_{ij}=  F^{b  a_i c} F^{c{a}_j b}$. The parameter $N_\phi=\delta^{ab}\delta_{ab}$ is the number of massless scalars in the $D$-dimensional theory.

Finally, the numerators of the massive diagrams, \fig{YMphi3diagramsNewFigure}(e) and \fig{YMphi3diagramsNewFigure}(f), are conjugates of each other. We explicitly give the one corresponding to \fig{YMphi3diagramsNewFigure}(e), 
\begin{align}
n_{{\rm box},\alpha}(1,2,3,4;\ell)=\frac{\lambda^4}{4} & \widehat{F}^{a_1 \ \beta}_{ \ \ \alpha} \widehat{F}^{a_2 \ \gamma}_{ \ \ \beta} \widehat{F}^{a_3 \ \delta}_{ \ \ \gamma} \widehat{F}^{a_4 \ \alpha}_{ \ \ \delta}  \hskip 6.5cm (\text{no sum } \alpha)  \nn \\ 
\null +\frac{\lambda^2}{24}  & \Big\{ N_\alpha \big[F^{ a_1{ a}_4   b} F^{  b a_3 a_2}  (L_2 + L_4) + F^{  a_1{ a}_2   b} F^{  b a_3 a_4} (L_1 + L_3)\big] \nn \\ & \null 
+ \delta^{a_3 a_4} \, \widehat{\Tr}_{12;\alpha} \, (6 L_3 - L_2 - L_4) + \delta^{a_2 a_3} \, \widehat{\Tr}_{14;\alpha} \, (6 L_2 - L_1 - L_3) \nn \\ &
\null  + \delta^{a_1 a_4} \, \widehat{\Tr}_{23;\alpha}  \, (6 L_4 - L_1 - L_3)  + \delta^{a_1 a_2} \, \widehat{\Tr}_{34;\alpha}\, (6 L_1 - L_2 - L_4)  \nn  \\ &
\null + (L_1 + L_2 + L_3 + L_4) (\delta^{a_2 a_4}\,  \widehat{\Tr}_{13;\alpha} + \delta^{a_1 a_3} \, \widehat{\Tr}_{24;\alpha}) \Big\}  \nn \\ 
\null  + \frac{N_\alpha }{24} & \Big\{  \delta^{ a_1 a_2}  \delta^{ a_3 a_4}  \big[2 (3 L_1 L_3 - L_2 L_4) + t (L_1 + L_2 + L_3 + L_4) - u (L_1 + L_3) \big]  \nn \\ &
\null + \delta^{ a_2 a_3} \delta^{ a_1 a_4}   \big[ 2  (3 L_2 L_4 - L_1 L_3)   +  s (L_1 + L_2 + L_3 + L_4) - u (L_2 + L_4)\big]  \nn \\ &
\null + \delta^{ a_1 a_3} \delta^{ a_2 a_4}  \big[2   (L_1 L_3 + L_2 L_4)  - s (L_1 + L_3) + t (L_2 + L_4)\big]\Big\} \, ,
\label{massive_in_loop}
\end{align}
where $L_i=\ell_i^2-m_\alpha^2=(\ell-(k_1+\ldots+k_i))^2-m_\alpha^2$ are the inverse propagators of the box diagram, $\widehat{F}^{a \ \beta}_{ \ \alpha}\equiv  \Delta^{ab}F^{b \ \beta}_{ \  \alpha}$ and we use the shorthand notation $\widehat{\Tr}_{ij;\alpha}=  \widehat F^{a_i \  \beta}_{\ \alpha}  \widehat F^{{a}_j \  \alpha}_{\ \beta}$.

Since the mass depends on the index $\alpha$ we do not yet sum over this index; the numerator has to first be combined with the correct denominator factor (this is akin to not integrating over the loop momenta in numerators when they are not yet combined with their denominators). The parameter $N_\alpha$ is introduced to count the degeneracy of massive scalars corresponding to mass $m_\alpha$. By default $N_\alpha=1$ since we may formally consider all masses distinct, but it is useful to keep this parameter around should one choose to do the bookkeeping differently. As seen, every term in \eqn{massive_in_loop} depends on the index $\alpha$.

The conjugated box numerator, corresponding to \fig{YMphi3diagramsNewFigure}(f), can now be obtained by the identification
\be
 \overline{n}_{{\rm box},\alpha}(1,2,3,4;\ell)=  n_{{\rm box},\alpha}(4,3,2,1;-\ell)\,,
\ee
or, alternatively, by raising/lowering all the gauge-group indices in $n_{{\rm box},\alpha}$ corresponding to matter representations.

The box numerators have been constructed so to manifestly obey color/kinematics duality and satisfy all physical unitarity cuts (this includes all cuts that have no contributions from tadpoles or external bubbles, which are singular diagrams on shell).\footnote{
Unlike massless theories, in massive theories bubbles on external lines and tadpoles do not automatically vanish in dimensional regularization. For the current purpose, one may nonetheless ignore them by considering renormalized gauge-theory amplitudes in a particular scheme (and defining the double copy in this renormalization scheme). 

Graphs with a massive bubble on a massless external leg have the same color structure as the corresponding tree-level graphs (with the bubble removed) and as such they receive contributions from counterterms that renormalize the gauge-theory action. Since these bubbles integrate to constants, one may choose counterterms that completely cancel them (this is different from the $\overline{\rm MS}$ scheme). 

Cubic tadpole graphs typically vanish in a gauge theory because of the color factors, but the kinematic part of them is not automatically zero for massive tadpoles, and thus can potentially contribute to the gravity amplitude through the double copy. The existence of such a tadpole would indicate the instability of Minkowski vacuum,  and thus we remove them by choosing appropriate gravity counterterms (for the gauge-theory numerators this implies that we simply drop all tadpoles). 

The resulting gravity amplitude inherits the momentum dependence of the underlying gauge-theory amplitude while dropping certain renormalization-dependent constant factors, which from the gravity perspective are absorbed into the definition of the gravity action.} 
Note that in the $m_\alpha\rightarrow 0$ limit, there is a term-by-term map between the numerator in \eqn{massive_in_loop} and the terms in eqs. (\ref{numA})--(\ref{numC}), after the gluon loop contributions in the latter expressions have been excluded. This is consistent with the discussion in appendix \ref{YMphi3dredSect}.

The numerator factors for the remaining contributing diagrams, the triangles and internal bubbles, are given by the kinematic Lie algebra relations. For the massless numerators we have
\bea
n_{\rm tri}(1,2,3,4;\ell)&=&n_{\rm box}(1,2,3,4;\ell)-n_{\rm box}(2,1,3,4;\ell)\,, \nn \\
n_{\rm bub}(1,2,3,4;\ell)&=&n_{\rm tri}(1,2,3,4;\ell)-n_{\rm tri}(1,2,4,3;\ell) \ ,
\label{masslessTriBib}
\eea
and for the massive ones, we have
\bea
n_{{\rm tri},\alpha}(1,2,3,4;\ell)&=&n_{{\rm box},\alpha}(1,2,3,4;\ell)-n_{{\rm box},\alpha}(2,1,3,4;\ell)\,, \nn \\
n_{{\rm bub},\alpha}(1,2,3,4;\ell)&=&n_{{\rm tri},\alpha}(1,2,3,4;\ell)-n_{{\rm tri},\alpha}(1,2,4,3;\ell) \,,
\label{massiveTriBib}
\eea
and similarly for the conjugate ones, $\overline{n}_{i,\alpha}$.

\subsection{One-loop four-vector Yang-Mills-gravity amplitudes}

\begin{figure}[t]
      \centering
      \includegraphics[scale=0.79]{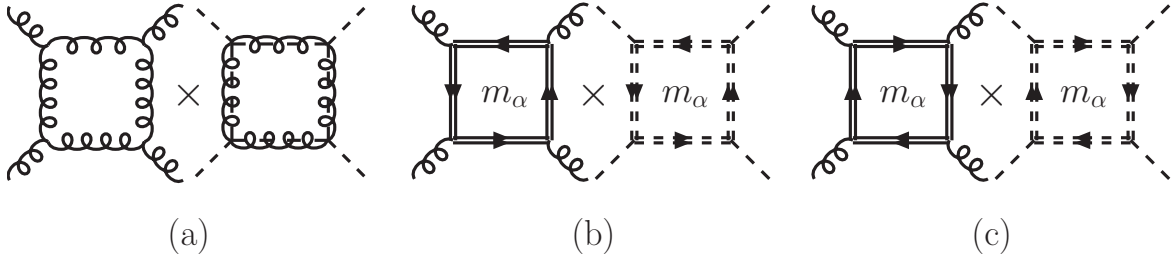}
      \caption[a]{\small The three types of box diagrams in the one-loop four-vector spontaneously-broken YMESG amplitude.  The contributions are, (a) the graviton and massless vector, (b) the massive $W_\alpha$ vectors, and (c) the massive $\overline{W}^\alpha$ vectors. These are given by double copies between spontaneously-broken SYM (left factors) and explicitly broken YM~+~$\phi^3$ theory (right factors).  The remaining triangle and bubble contributions are obtained from the boxes through the Jacobi and commutation relations of color/kinematics duality. See \fig{YMphi3diagramsNewFigure} for notation. \label{OneLoopGRFigureNew}}
\end{figure}

The double-copy procedure, inherent in color/kinematics duality, provides a straightforward way to construct loop amplitudes in spontaneously-broken YMESG theory. For example, \fig{OneLoopGRFigureNew} 
illustrates how to obtain the different types of contributions -- massless graviton and vector multiplets, massive $W_\alpha$ and $\overline{W}^\alpha$ multiplets -- as double copies between spontaneously-broken SYM numerators and the explicitly broken YM~+~$\phi^3$ numerators computed in the previous section.

The complete one-loop amplitude with four massless external vectors in spontaneously-broken YMESG theory, is given by the double-copy form \eqref{BCJformGravity},
\begin{equation}
 {\cal M}^{1\text{-loop}}_4 = \left(\frac{\kappa}{2}\right)^{4} 
\sum_{{\cal S}_4} \sum_{i\in{\rm\{box,tri,bub\}}} \int  \frac{d^D \ell}{(2 \pi)^D}
 \frac{1}{S_i}
   \left(\frac{n^{\rm SYM}_{i} \, n_{i}}{D_i}+\sum_\alpha \frac{n^{\rm SYM}_{i,\alpha} \, n_{i,\alpha}}{D_{i,\alpha}}+\sum_\alpha \frac{\overline{n}^{\rm SYM}_{i,\alpha} \,\overline{n}_{i,\alpha}}{D_{i,\alpha}}\right)\,,
\hskip .7 cm 
\label{1LoopDC}
\end{equation}
where the sums, symmetry factors, and denominator factors, are the same as in eqs. (\ref{1LoopCK}) and (\ref{Denominators}). As before, 
the $n_i$ are the numerators of explicitly broken YM~+~$\phi^3$ theory given in \sec{YMphi3_1L}, and the $n^{\rm SYM}_i$ are numerators 
of spontaneously-broken SYM theory and we identify the combination $\kappa \lambda/2$ with the supergravity gauge coupling $g$.

For the one-loop amplitudes, the spontaneously-broken $D$-dimensional SYM numerators are given by $(D+1)$-dimensional SYM numerators 
with the last component of the loop momentum interpreted as mass: $\ell \rightarrow (\ell, \pm m_\alpha)$. We may write the numerators as
\bea
n^{\rm SYM}_{i} &\equiv & n^{\rm SYM}_{i}(1,2,3,4;\ell;0)\,, \nn \\
n^{\rm SYM}_{i,\alpha} &\equiv  &n^{\rm SYM}_{i}(1,2,3,4;\ell;m_\alpha)\,, \\
\overline{n}^{\rm SYM}_{i,\alpha} &\equiv  &n^{\rm SYM}_{i}(1,2,3,4;\ell;-m_\alpha)\,, \nn
\eea
meaning that the numerators for massive and massless internal states are described by the same function, only the value of $m_\alpha$ differs between them.\footnote{Note that the numerators of spontaneously-broken SYM are thus straightforward to obtain from the massless theory in $(D+1)$ dimensions, in contrast to the numerators of explicitly broken YM~+~$\phi^3$ theory which in general have a more complicated relation to the massless numerators.} 

Specifying to the maximally supersymmetric case, the box numerator of spontaneously-broken ${\cal N}=4$ SYM is given by
\be
{n}^{{\cal N}=4 \,\, {\rm SYM}}_{\rm box}(1,2,3,4;\ell; m)= i s t A^{{\rm tree}}(1,2,3,4) = \frac{\spb{1}.{2}\spb{3}.{4}}{\spa{1}.{2}\spa{3}.{4}}\delta^{(8)}\Big(\sum \eta_i^\alpha |i\rangle\Big)\ ,
\ee
independent of the mass parameter $m$, whether zero or not, in agreement with ref.~\cite{Alday:2009zm}.
The corresponding triangle and bubble numerators vanish for this theory.
Plugging this into \eqref{1LoopDC} gives  the four-vector amplitude in 
spontaneously-broken ${\cal N}=4$ YMESG theory.

The ${\cal N}=2$ SYM one-loop numerator factors may be written as
the difference between ${\cal N}=4$ SYM and numerator factors for one adjoint ${\cal N}=2$ hypermultiplet running in the loop, 
\be
{n}^{{\cal N}=2 \,\, {\rm SYM}}_{i}(1,2,3,4;\ell; m)={n}^{{\cal N}=4 \,\, {\rm SYM}}_{i}(1,2,3,4;\ell; m)-2  {n}_{i}^{{\cal N}=2,{\rm mat.}}(1,2,3,4;\ell; m) \ .
\ee

Color/kinematics-satisfying one-loop numerator factors due to one adjoint hypermultiplet running in the loop may be found in refs.~\cite{Carrasco:2012ca,Nohle:2013bfa,Ochirov:2013xba,Johansson:2014zca}. A manifestly ${\cal N}=2$-supersymmetric box numerator valid for $D$-dimensional loop momenta was given in ref.~\cite{Johansson:2014zca}; 
introducing the mass-dependence we find
\bea
{n}_{\rm box}^{{\cal N}=2,{\rm mat.}}(1,2,3,4,\ell;m)
      \!\!&=& \!\!\, (\kappa_{12}+\kappa_{34}) \frac{(s-\ell_s)^2}{2s^2}
        + (\kappa_{23}+\kappa_{14}) \frac{\ell_t^2}{2t^2}
        + (\kappa_{13}+\kappa_{24}) \frac{st+(s+\ell_u)^2}{2u^2} \nn \\
    &&\null + (\mu^2+m^2) \Big( \frac{\kappa_{12}+\kappa_{34}}{s}
                     +\frac{\kappa_{23}+\kappa_{14}}{t}
                     +\frac{\kappa_{13}+\kappa_{24}}{u} \Big) \nn\\
      &&
      \null +2i \epsilon(k_1,k_2,k_3,\ell)\frac{\kappa_{13}-\kappa_{24}}{u^2} \ ,\label{Neq2matter}
\eea
where $\ell_s=2 \ell \cdot (k_1+k_2)$, $\ell_t=2 \ell \cdot (k_2+k_3)$ and $\ell_u=2 \ell \cdot (k_1+k_3)$. The numerator 
factors of other box integrals are obtained by relabeling. The parameter $\mu$ is the component of the loop momenta that 
is orthogonal to four-dimensional spacetime, and $\epsilon(k_1,k_2,k_3,\ell)=k_1^\mu k_2^\nu k_3^\rho \ell^\lambda \eps_{\mu\nu\rho\lambda}$ is the Levi-Civita invariant. The external multiplet dependence is captured by the variables~$\kappa_{ij}$, 
\be
   \kappa_{ij} = -\frac{ \spb{1}.{2} \spb{3}.{4} }{ \spa{1}.{2}\!\spa{3}.{4} }
   \delta^{(4)}\Big(\sum \eta_i^\alpha |i\rangle\Big) \spa{i}.{j}^{2} (\eta^3_i \eta^4_i)(  \eta^3_j \eta^4_j)   \ .
\label{kappa}
\ee
As before, the triangle and bubble numerators are given by the kinematic Jacobi relations,
\bea
{n}^{{\cal N}=2 \,\, {\rm SYM}}_{\rm tri}(1,2,3,4;\ell;m)\!&=&\!{n}^{{\cal N}=2 \,\, {\rm SYM}}_{\rm box}(1,2,3,4;\ell;m)
-{n}^{{\cal N}=2 \,\, {\rm SYM}}_{\rm box}(2,1,3,4;\ell;m)\,, \nn \\
{n}^{{\cal N}=2 \,\, {\rm SYM}}_{\rm bub}(1,2,3,4;\ell;m)\!&=&\!{n}^{{\cal N}=2 \,\, {\rm SYM}}_{\rm tri}(1,2,3,4;\ell;m)
-{n}^{{\cal N}=2 \,\, {\rm SYM}}_{\rm tri}(1,2,4,3;\ell;m) \, \label{tri_and_bub},~~~
\eea
which have no mass-dependence, since the mass term in \eqn{Neq2matter} is totally symmetric.

Plugging the ${\cal N}=2$ SYM numerators, together with the YM~+~$\phi^3$ numerators \eqref{fullNum}, into equation~\eqref{1LoopDC} gives  
a four-vector amplitude in the spontaneously-broken ${\cal N}=2$ YMESG theory.
The parameter $N_\phi$ in eq.~\eqref{fullNum} and \eqref{numC} is identified with the number of massless vector multiplets (i.e. the number of massless 
vector fields excluding those in the graviton multiplet). 
One may verify the construction by observing that the unitarity cuts of these amplitudes match the direct evaluation of cuts in terms of tree diagrams.

It is not difficult to integrate the resulting expression and find the divergence of the four-vector amplitude. As usual, the 
masses do not enter the UV divergence, which is the same as that of the unbroken theory; it is naturally organized in 
the powers of $\lambda$. 

\begin{itemize}

\item The ${\cal O}(\lambda^0)$ part of the amplitude is the same as in the MESG theory with the same field content. The four-vector amplitude diverges in four dimensions and, as a term in the effective action, the divergence is proportional to the square of the vector field stress tensor \cite{Fischler:1979yk, Fradkin:1983xs}.

\item The ${\cal O}(\lambda^2)$ part of the amplitude is finite in four (and five) dimensions; it is given by a combination of the four- and six-dimensional box integrals
with  tree-level color structures.

\item Since the ${\cal O}(\lambda^4)$ part of the YM~+~$\phi^3$ numerators is momentum-independent, the divergence at this order is proportional to the divergence of the four-gluon amplitude in the ${\cal N}=2$ SYM theory. In the UV limit the masses drop out and the sum 
over the index $\alpha$ leads to a factor of the index of the adjoint representation, $T(A) \delta^{ab} = F^{acd} F^{bcd}$. 

\end{itemize}

Next consider the maximally-helicity-violating (MHV) amplitude in YM$_{\rm DR}$ theory, which is the generalization of the bosonic part of SYM theories. For four-dimensional external states, the one-loop numerator factors may again be written as the difference between SYM numerators and numerator factors for scalar matter running in the loop, 
\bea
{n}^{{\rm YM}_{\rm DR}}_{i}(1,2,3,4;\ell; m)&=&{n}^{{\cal N}=4 \,\, {\rm SYM}}_{i}(1,2,3,4;\ell; m)-4 {n}_{i}^{{\cal N}=2,{\rm mat.}}(1,2,3,4;\ell; m) \nn \\
 && \null +(2+N'_\phi)\, {n}_{\rm box}^{{\rm YM}_{\rm DR},{\rm mat.}}(1,2,3,4;\ell;m)\,.
\label{N0scalar}
\eea
where $N'_\phi$ is the number of real scalars in the loop (also counting the Goldstone boson). In the gravity theory this number gives the number of real vector fields in the graviton multiplet. 

A box numerator for a four-vector amplitude with a single scalar running in the loop in the YM$_{\rm DR}$ theory, valid for $D$-dimensional loop momenta and four-dimensional external states, was given in ref.~\cite{Johansson:2014zca},
\begin{align}
{n}_{\rm box}^{{\rm YM}_{\rm DR},{\rm mat.}}&(1,2,3,4,\ell;m)= \nn
\\         & - (\kappa_{12}+\kappa_{34})
          \Big( \frac{\ell_s^4}{4 s^4} - \frac{\ell_s^2 (2 L + 3 \ell_s)}{4 s^3}
              + \frac{2 L \ell_s + \ell_s^2 - 2 M^2}{2 s^2}
              - \frac{2 L - \ell_s + s}{4 s} \Big) \nn \\
      & - (\kappa_{23}+\kappa_{14})
          \Big( \frac{\ell_t^4}{4 t^4}
              - \frac{\ell_t^2 (2 L - \ell_s - \ell_u + t)}{4 t^3}
              - \frac{M^2}{t^2} \Big)  \nn \\
      & - (\kappa_{13}+\kappa_{24})
          \Big( \frac{\ell_u^3 (\ell_u + 3 s)}{4 u^4}
              - \frac{\ell_u( \ell_u (2 L - \ell_s) - \ell_s^2 + \ell_t^2
                            + 4 s (L + \ell_u + 2 M))}{4 u^3}  \nn \\
      & ~~~~~~~~~~~~~~~~~~~~~~~~~~~~~~~
              - \frac{\ell_s^2 - \ell_t^2 + 3 \ell_u^2 + 4 L t
                     + 8 M  (\ell_u - s + M)}{8 u^2}
              - \frac{\ell_s - s}{4 u} \Big)  \nn \\ 
      & - 2i \epsilon(k_1,k_2,k_3,\ell) (\kappa_{13}-\kappa_{24})
          \frac{\ell_u^2 - u \ell_u - 2 M u}{u^4} \,.
\label{N0scalarBox}
\end{align}
with
\be
L=\ell^2-m^2\,~~~~~~{\rm and}~~~~~~ M=\mu^2+m^2\,.
\ee
In the above non-supersymmetric expressions it is understood that only the vector components of $\kappa_{ij}$ should be kept; that is, in \eqn{N0scalar} and \eqn{N0scalarBox} we take 
\be
   \kappa_{ij} \rightarrow \frac{ \spb{1}.{2} \spb{3}.{4} }{ \spa{1}.{2}\!\spa{3}.{4} }\spa{i}.{j}^{4} (\eta^1_i \eta^2_i \eta^3_i \eta^4_i)( \eta^1_j \eta^2_j \eta^3_j \eta^4_j)   \,.
\label{kappa2}
\ee
As before, the box numerator was constructed to obey color/kinematics duality, thus the triangle and bubble numerators are given by the kinematic Jacobi relations,
\bea
{n}^{{\rm YM}_{\rm DR}}_{\rm tri}(1,2,3,4;\ell;m)\!&=&\!{n}^{{\rm YM}_{\rm DR}}_{\rm box}(1,2,3,4;\ell;m)
-{n}^{{\rm YM}_{\rm DR}}_{\rm box}(2,1,3,4;\ell;m)\,, \nn \\
{n}^{{\rm YM}_{\rm DR}}_{\rm bub}(1,2,3,4;\ell;m)\!&=&\!{n}^{{\rm YM}_{\rm DR}}_{\rm tri}(1,2,3,4;\ell;m)
-{n}^{{\rm YM}_{\rm DR}}_{\rm tri}(1,2,4,3;\ell;m) \,.
\eea

Plugging the YM$_{\rm DR}$ numerators together with the YM~+~$\phi^3$ numerators \eqref{fullNum}, \eqref{masslessTriBib}, \eqref{massiveTriBib} 
in \eqn{1LoopDC} gives the four-vector MHV amplitude in a spontaneously-broken YM$_{\rm DR}$-Einstein theory.

\section{$\cN=4$  supergravity theories \label{Neq4}}

In this section we discuss the application of our results to construction of 
the amplitudes of $\cN=4$ Maxwell-Einstein and Yang-Mills-Einstein supergravity theories. We begin with a review of the Lagrangians of these theories.\footnote{Note that the conventions for labeling various quantities used in this section are independent of the conventions used earlier for $\cN=2$ supergravity theories.}

\subsection{ $\cN=4$ Maxwell-Einstein and Yang-Mills-Einstein supergravity theories}
\renewcommand{\theequation}{6.\arabic{equation}}
\setcounter{equation}{0}

The $\cN=4$ Maxwell-Einstein supergravity theories describe the coupling of $\cN=4$ supergravity to $\cN=4$ vector multiplets. Their construction and  various gaugings  in five dimensions were studied in  \cite{AT85,RO86,Dall'Agata:2001vb,Schon:2006kz}. Our review will follow mainly \cite{Dall'Agata:2001vb}.

The pure $\cN=4$ supergravity  in five dimensions contains one graviton $e_\mu{}^m$, four gravitini $\psi_\mu^i$, six
vector fields $(A_\mu^{ij},a_\mu)$, four spin $1/2$ fermions $\chi^i$ and one
 real scalar field $a$:
\begin{equation}
\Big(\,\,e_\mu{}^m\,,~\psi_\mu^i\,,~A_\mu^{ij}\,,~a_\mu\,,
~\chi^i\,,~ a\,\Big) \,.
\label{n=4_sugra}
\end{equation}
Here, $\mu,\nu,\dots$  ($m,n,\dots)$ denote the five-dimensional curved (flat)  indices and
 the  $i,j=1,\ldots,4$ are the indices of the fundamental
representation of the $R$-symmetry group $USp(4)$. The vector field $a_\mu$ is a $USp(4)$ singlet and
the vector fields $A_\mu^{ij}$ transform in the ${\bf 5}$ of $USp(4)$, i.e.,
\begin{equation}
A_\mu^{ij} \ = \ -A_\mu^{ji}~,\qquad A_\mu^{ij}\,\Omega_{ij} \ = \ 0\,,
\end{equation}
where $\Omega_{ij}$ is the symplectic metric of $ USp(4)\cong SO(5)$.
On the other hand an $\cN=4$ vector
multiplet contains the fields
\begin{equation}
\Big(\,A_\mu\,,~\lambda^i\,,~\phi^{ij}\,\Big),
\end{equation}
where $A_\mu$ is a vector field, $\lambda^i$ denotes four spin $1/2$ fields,
and the $\phi^{ij}$ are scalar fields in the {\bf 5} of
$USp(4)$
\be
\phi^{ij}=-\phi^{ji} \qquad \phi^{ij} \Omega_{ij}=0 \ .
\ee

The total
field content of the $\cN=4$ MESG theory with $n$ vector multiplets can be labelled as follows
\begin{equation}
\Big(\,e_\mu{}^m\,,~\psi_\mu^i\,,~A_\mu^\tI\,,~a_\mu\,,~\chi^i\,,~\la^{ia}\,,~
\si\,,~\phi^x\,\Big) \ ,
\label{fullcontent}
\end{equation}
where the index $a=1,\ldots,n$ counts
the number of $\cN=4$ vector multiplets whereas the indices 
 $\tI, \tJ,...=1,\ldots,(5+n)$  collectively denote
the vector fields $A_\mu^{ij}$ of supergravity multiplet and the vector fields coming from  the vector multiplets. The indices
$x,y,..=1,\ldots,5n$ denotes the scalar fields in the $n$ vector
multiplets.
The $USp(4)$
indices are raised and lowered with the symplectic metric $\Omega_{ij}$ and its inverse $\Omega^{ij}$:
\begin{equation}
T^i \ = \ \Omega^{ij}\,T_j~,\quad T_i \ = \ T^j\,\Omega_{ji}\,,
\end{equation}
and the  $a,b$ indices are raised and lowered with $\delta^{ab}$.

The scalar manifold spanned by the $(5n+1)$ scalar fields is \cite{AT85}
\begin{equation}
    \label{cM}
\cM \ = \ \frac{SO(5,n)}{SO(5)\times SO(n)}\times SO(1,1)\,,
\end{equation}
where the $SO(1,1)$ factor corresponds to the $USp(4)$-singlet scalar field $\sigma$
of the supergravity multiplet.
The metric of the coset part $\cG / \cH=\frac{SO(5,n)}{SO(5)\times SO(n)}$ of the scalar manifold  $\cM$ parametrized by the $5n$ scalars 
is denoted as $g_{xy}$ and the corresponding $SO(5)\times SO(n)$ ``vielbeins'' as $ f^a_{yij}$
\begin{equation}
g_{xy} \ = \ \frac14\,f_x^{ija}\,f^a_{yij}\,,
\end{equation}

 An equivalent description uses  coset $\cG / \cH $ representatives
$\,L_\tI{}^A\,$
where $\tI$ denotes a $\cG=SO(5,n)\,$
index, and $A=(ij,a)$ is a $\cH=SO(5)\times SO(n)\,$ index.
Denoting the inverse of $\,L_\tI{}^A\,$ by $\,L_A{}^\tI\,$,
\[
\,L_\tI{}^A\, \,L_B{}^\tI\ = \delta_{B}^{A} \ , \] 
one can define  the
vielbeins on $\cG/\cH$ and the composite $\cH$-connections
as follows:
\begin{equation}
L^{-1}\partial_\mu L \ = \ Q^{ab}_\mu \,\mathfrak{T}_{ab} + Q^{ij}_\mu \,
\mathfrak{T}_{ij} + P^{aij}_\mu \, \mathfrak{T}_{aij},
\end{equation}
where $\,(\mathfrak{T}_{ab},\mathfrak{T}_{ij})\,$ are the
generators of the Lie algebra $\mathfrak{h}$
of $\cH$, and $\mathfrak{T}_{aij}$ denotes the  generators of the coset part of
the
Lie algebra $\mathfrak{g}$ of $\cG$. More explicitly the composite $SO(n)$ and $USp(4)$ connections are given by
\begin{equation}\label{Q}
Q^{ab}_\mu \ = \ L^{\tI a}\partial_\mu L_\tI{}^b=-Q^{ba}_\mu \qquad\mathrm{and}\qquad Q^{ij}_\mu \ = \
L^{\tI ik}
\partial_\mu L_{\tI k}{}^j=Q^{ji}_\mu \ .
\end{equation}
   Furthermore we have
\begin{equation}\label{P}
P^{aij}_\mu \ = \  L^{\tI a}
\partial_\mu L_{\tI}{}^{ij}=-\frac{1}{2}f_{x}^{aij}\,\partial_\mu \phi^x \,.
\end{equation}

The   Lagrangian of the
five-dimensional  $\cN=4$ MESG theory  is   reproduced in \app{N4Lagrangian} following  \cite{AT85,Dall'Agata:2001vb}.
Its bosonic part can be written as:
\bea
e^{-1}\,\cL_{Bosonic} &=& -\frac12 R -\frac14\Sigma^2\,a_{\tI\tJ}\,F_{\mu\nu}^\tI
F^{\mu\nu\tJ}-\frac14\Sigma^{-4}\,G_{\mu\nu}G^{\mu\nu} \\
&&-\frac12\,(\prt_\mu\si)^2-\frac12 g_{xy}\prt_\mu \phi^{x}\prt^\mu \phi^y
+\frac{\sqrt2}{8}e^{-1}\,C_{\tI\tJ}\,\epsilon^{\mu\nu\rho\si\la}\,
F_{\mu\nu}^\tI F_{\rho\si}^\tJ\,a_\la \,, \nnu
\eea
where  \begin{equation}
\Sigma \ = \ e^{\frac{1}{\sqrt 3}\si}.
\end{equation}
and the abelian field strengths of vector fields are defined as
\begin{equation}
F_{\mu\nu}^\tI\ = (\prt_\mu A_\nu^\tI-\prt_\nu A_\mu^\tI)~,\quad
G_{\mu\nu}\ = (\prt_\mu a_\nu-\prt_\nu a_\mu)~,
\end{equation}

The main constraints imposed by supersymmetry are \footnote{ The indices $\tI ,\tJ,\ldots$ are raised and lowered by $a_{\tI\tJ}$ and its inverse, e.g.  $
L_{\tI}^A\ = \ a_{\tI\tJ}\,L^{\tJ A} $.}
\begin{equation}
a_{\tI\tJ} \ = \ L_\tI^{ij}L_{\tJ ij}+L_\tI^aL_\tJ^a~,\quad C_{\tI\tJ}
\ = \
L_\tI^{ij}L_{\tJ ij}-L_\tI^aL_\tJ^a\,,
\end{equation}
where $C_{\tI\tJ}$ is the constant $SO(5,n)$ invariant metric.

Five-dimensional $\cN=4$ MESG theories can be truncated to $\cN=2$ MESG theories with or without $\cN=2$ hypermultiplets. To understand the structure of truncations we note that the pure $\cN=4$ supergravity theory can be truncated to $\cN=2$ supergravity coupled to a single vector multiplet by discarding two of the $\cN=2$ gravitino supermultiplets where each gravitino multiplet contains a gravitino, two vectors and one spin $1/2$ field. The remaining vector multiplet involves the $SO(n,5)$ singlet vector $a_\mu$. On the other hand an $\cN=4$ vector multiplet decomposes into an $\cN=2$ vector multiplet plus an $\cN=2$ hypermultiplet which has four scalars. One can discard either the $\cN=2$ hypermultiplet or the $\cN=2$ vector multiplet in truncation. If one throws away the $\cN=2$ hypers from all the $\cN=4$ multiplets the resulting theory is an $\cN=2$ MESG theory belonging to the generic Jordan family with the scalar manifold
\[ \mathcal{M}_{V_{(n+1)}} = \frac{SO(1,1)\times SO(n,1)}{SO(n)} \ , \]
which  is unique modulo the embedding of $\cN=2$ $R$-symmetry group $SU(2)$ inside $USp(4)$. On the other hand if one throws away $m$ of the $\cN=2$ vector multiplets  and keeps the corresponding hypermultiplets  the resulting theory is an $\cN=2$ MESG theory coupled to $m$ hypermultiplets with the moduli space:
\[ 
\mathcal{M}_{V_{(n-m+1)}} \times \mathcal{V}_{H_m} = \frac{SO(1,1)\times SO(n-m,1)}{SO(n-m)} \times \frac{SO(m,4)}{SO(m)\times SO(4)}\,.
\]

The $\cN=2$ MESG theory sector of all these truncations is of the generic Jordan type.  The $F\wedge F\wedge A$ term
\begin{equation}
\frac{\sqrt2}{8}e^{-1}\,C_{\tI\tJ}\,\epsilon^{\mu\nu\rho\si\la}\,
F_{\mu\nu}^\tI F_{\rho\si}^\tJ\,a_\la \,,
\end{equation}
of the $\cN=4$ MESG theory reduces to
\begin{equation}
\frac{\sqrt2}{8}e^{-1}\,C_{RS}\,\epsilon^{\mu\nu\rho\si\la}\,
F_{\mu\nu}^R F_{\rho\si}^S\,a_\la \,,
\end{equation}
where $R,S,..=1,2,..(n-m+1)$. If we denote the singlet vector $a_\mu$ as $A^0_\mu$ this implies that the $C$-tensor of the $\cN=2$ MESG theory is simply given by
\[ 
C_{0RS} = \frac{\sqrt{3}}{2} C_{RS}\,,
\] 
where $C_{RS}$ is proportional to the constant  $SO(n-m,1)$ invariant metric, namely
\bea
C_{011}& =& \frac{\sqrt{3}}{2}\,, \nn \\
C_{0rs}& = &- \frac{\sqrt{3}}{2} \delta_{rs} \,, \quad r,s,.. = 1,2,...(n-m)\,.
\eea

Four-dimensional $\cN=4$ MESG theories and their gaugings were first studied  in \cite{deRoo:1985jh,deRoo:1986yw}. Their most general gaugings both in four and five dimensions using the embedding tensor formalism was given more recently~\cite{Schon:2006kz}.
Under  dimensional reduction the   five-dimensional $\cN=4$ MESG theory with $n$ vector multiplets  leads to the four-dimensional MESG theory with $(n+1)$ vector multiplets and  the scalar manifold
\be
\mathcal{M}_4 = \frac{SO(6,n+1)}{SO(6)\times SO(n+1)} \times \frac{SU(1,1)}{U(1)}\,.
\ee
The $SU(1,1)$  symmetry acts via  electric and magnetic dualities and in the symplectic section that descends directly from five dimensions via dimensional reduction the Lagrangian is invariant under  the five-dimensional U-duality group. These $\cN=4$ MESG theories in four dimensions can be truncated to $\cN=2$ MESG theories belonging to the generic Jordan family with or  without hypermultiplets. Truncation to maximal $\cN=2$ MESG theory with $(n+1)$ vector multiplets without hypers is unique modulo the embedding of the $\cN=2$ $R$-symmetry group $U(2)$ inside $\cN=4$ $R$-symmetry group $SO(6)=SU(4)$. The resulting theory has the scalar manifold
\be
\frac{SO(n+1,2)}{SO(n+1)\times SO(2)} \times \frac{SU(1,1)}{U(1)} \,.
\ee
If one retains $m$,  $\cN=2$ hypermultiplets and $(n+1-m)$ vector multiplets in the truncation the resulting theory is a MESG theory coupled to $m$ hypermultiplets with the scalar manifold
\be
\frac{SO(n+1-m,2)}{SO(n+1-m)\times SO(2)} \times \frac{SU(1,1)}{U(1)} \times \frac{SO(m,4)}{SO(m)\times SO(4)} \,.
\ee

Most general gaugings of $\cN=4$ supergravity theories coupled to $\cN=4$ vector multiplets were studied  in \cite{Schon:2006kz} using the embedding tensor formalism. In this paper we will only focus on gaugings that lead to $\cN=4$ supergravity coupled to Yang-Mills gauge theories with a compact gauge group that allow Minkowski vacua only.
For this we will follow the work of \cite{Dall'Agata:2001vb}  on gaugings of five-dimensional $\cN=4$ MESG theories.
As was shown by the authors of  \cite{Dall'Agata:2001vb}  gauging with tensors requires an abelian gauge group whose gauge field is the singlet vector $a_\mu$. Furthermore   gauging a semisimple subgroup of the global symmetry group $SO(5,n)$ by itself does not require coupling to any tensors and allows Minkowski vacua only.
To gauge a semisimple subgroup $K_S$ of $SO(5,n)$ one identifies the subset of vector fields $A_\mu^I$  that transform in the adjoint representation of $K_S$ with the remaining vector fields being spectators. Since such gaugings do not have tensors we can formally use the same index $\tI,\tJ,..=1,2,...,n+5$ to collectively denote the $K_S$ gauge fields plus the  spectators with the understanding that the structure constants $f_{\tI \tJ}^{~~\tK}$  of the gauge group vanishes when any one of the indices corresponds to the spectator vector fields. In this paper we restrict ourselves  to gaugings of a compact subgroup $K$ of $SO(n)$ global symmetry which do not involve any tensor fields and will use this formal trick to simplify the formulas.

The  gauging of a subgroup $K$ requires that all derivatives acting on fields that transform non-trivially under $K$  be covariantized.
This is implemented by the following substitutions in the Lagrangian
$A_\mu^\tI$ in the
standard way:
\bea F^{\tI}_{\mu\nu}& \lra& \cF^{\tI}_{\mu\nu} \ = \ F^{\tI}_{\mu\nu}
+\gs\,A_\mu^{\tJ}
f_{\tJ \tK}^I A_\nu^{\tK}\,, \nnu \\
\partial_\mu L^{\tI}_A & \lra& \fD_\mu L^{\tI}_A \ = \ \prt_\mu L^{\tI}_A  +\gs  A_\mu^{\tJ} f_{\tJ \tK}^I L^{\tK}_A.
\eea
The composite $USp(4)$ and $SO(n)$ connections,
the vielbein $\,P_{\mu ij}^a\,$ as well as the derivatives $\mathcal{D}_{\mu}$ acting on fermions are also modified by
 the new $\gs$ dependent contributions as reviewed in \app{N4Lagrangian}  where we also reproduce the Lagrangian of the $\cN=4$ YMESG theory in five dimensions following  \cite{Dall'Agata:2001vb}.
The bosonic part of the Lagrangian of the YMESG theory has the form  \cite{Dall'Agata:2001vb} :
\bea
e^{-1}\,\cL_{\rm YMESG} &=& -\frac12 R -\frac14\Sigma^2\,a_{\tI\tJ}\,\cF_{\mu\nu}^\tI
\cF^{\mu\nu\tJ}-\frac14\Sigma^{-4}\,G_{\mu\nu}G^{\mu\nu} \\
&&\null-\frac12\,(\prt_\mu\si)^2-\frac12\,\cP_\mu^{aij} \cP^\mu_{aij}\nnu
+\frac{\sqrt2}{8}e^{-1}\,C_{\tI\tJ}\,\epsilon^{\mu\nu\rho\si\la}\,
\cF_{\mu\nu}^\tI \cF_{\rho\si}^\tJ\,a_\la  \nnu \\ &&
\null-\gs^2 \left(
-\frac92\,S_{ij}\Delta^{ij}+\frac12\,T_{ij}^a\,T^{aij} \right)\,, \nnu
\label{Lagrange2}
\eea
where
\bea
S_{ij} &=& -\frac29 \Sigma^{-1} L^{\tJ}_{(i|k|} f_{\tJ\tI}^\tK L_{\tK}^{kl}
L^{\tI}_{|l|j)},\\
T_{ij}^a &=& -\Sigma^{-1} L^{\tJ a}  L^{\tK}_{(i}{}^{k}f_{\tJ \tK}^{\tI} L_{\tI |k|j)},\\
\cP_{\mu ij}^a &=& P_{\mu ij}^a -\gs
A_\mu^\tJ L^{\tK}_{ij}f_{\tJ\tK}^\tI L_{\tI}^a \,.\label{compco3b}
\eea

The $\cN=4$ Yang-Mills-Einstein supergravity with  a compact gauge group $K$ that is a subgroup of $SO(n)$  can be truncated to $\cN=2$ Yang-Mills-Einstein supergravity with the same gauge group that belongs to the generic Jordan family  discussed in \sec{sec:YME}. 
This truncation is unique for a given compact gauge group $K$, modulo the equivalence class of
embeddings of $K$ in $SO(n)$ and $R$ symmetry group $SU(2)$ inside $USp(4)$, and assuming that the number of spectator vector multiplets is
the same in the truncated theory as the original ${\cal N} = 4$ theory.
Conversely one can extend a YMESG theory  belonging to the generic Jordan family to an $\cN=4$ YMESG theory with the same gauge group. These results hold true also for the corresponding YMESG theories in four dimensions so long as one works in the symplectic section that descends directly from five dimensions. The four-dimensional YMESG theories have one extra spectator vector multiplet coming from the supergravity multiplet in five dimensions.

The spontaneous symmetry breaking mechanism of $\cN=2$ YMESG theories induced by giving a VEV to some of the scalars in the vectors multiplets can be extended to the $\cN=4$ YMESG theories for compact gauge groups $K$ that are subgroups of $SO(n)$  both in five as well as in four dimensions. For example the  $\cN=4$ supersymmetric Yang-Mills theory with gauge group  $SU(2)$ spontaneously-broken down to $U(1)$ subgroup by giving a VEV to one of the scalars leads to a massless gauge multiplet and two massive BPS vector multiplets which can be written as complex fields carrying opposite $U(1)$ charges. In four dimensions these charged  BPS vector multiplets have  5 massive complex scalars and four massive fermions \cite{ Fayet:1978ig}.
A massive $\cN=4$ BPS  vector multiplet decomposes into a massive BPS $\cN=2$ vector multiplet plus a massive $\cN=2$ BPS hypermultiplet. Therefore a spontaneously-broken $\cN=4$ YMESG theory can be truncated to a spontaneously-broken $\cN=2$ YMESG theory by throwing away the massive hypermultiplets. The spontaneous symmetry breaking by giving a VEV to one of the scalars in a gauge vector multiplet breaks the $R$-symmetry from $SO(6)$ down to $SO(5)=USp(4)$ in four dimensions and from $USp(4)$ down to $SO(4)$ in five dimensions.

\subsection{More on double copies with $\cN = 4$ supersymmetry}
\renewcommand{\theequation}{7.\arabic{equation}}
\setcounter{equation}{0}

In the double-copy construction of the amplitudes of $\cN=2$ MESG theories one gauge-theory copy is $\cN=2$ supersymmetric and the other copy has no supersymmetry. 
If one replaces the $\cN=2$ gauge-theory factor with an $\cN=4$  supersymmetric theory one obtains the amplitudes of an $\cN=4$ MESG theory both in five as well as in four dimensions. The fields of four-dimensional $\cN=4$ MESG theory and YMESG theory in terms of those of $\cN=4$ SYM and of the pure YM theory coupled to scalars in a specific way can be obtained by restricting to the product $ {\cal V}_L \otimes  {\cal V}_R $ in \sec{YMYMphi3} which we give in  Table~\ref{Neq4MESG theory}.

\begin{table*}
\centering
\begin{tabular}{|c||c|c|c|}
\hline
 ${\cal V}_R$$\backslash$${\cal V}_L$& $A_\mu$  & $\lambda^i$ & $ \phantom{\Big|}  \phi^{[ij]}$  \\
\hline\\[-13pt]
\hline
$A_\nu$  &  $h_{\mu\nu}\oplus \sigma \oplus \phi$ &  $\phantom{\Big|}  \Psi_{\mu}^{i} \oplus \psi^i$&  $A_\nu^{[ij]}$ \\
\hline
$\phi^c$  & $A_\mu^{c}$   & $\psi^{i,c}$  & $\phantom{\Big|} \phi^{[ij], c}$     \\
\hline
\end{tabular}
\caption[a]{\small The spectrum of the $D=4$, ${\cal N}=4$ Maxwell-Einstein and Yang-Mills-Einstein supergravity theories from the double-copy construction: one ${\cal N}=4$ supergravity multiplet given by the
second, third and fourth entries of the second line and as many vector multiplets as the range of the index $c$ of the scalar fields in the non-supersymmetric gauge-theory factor given in the third line. }
\label{Neq4MESG theory}
\end{table*}

The double-copy construction yields the superamplitudes of $\cN=4$ MESG theory in terms of the $\cN=4$ SYM  superamplitudes and the amplitudes of the dimensional reduction of pure YM theory from $D=4+n_V$, where $n_V$ is the number of vector multiplets. In the same sense as from a Lagrangian point of view we can truncate these $\cN=4$ supergravity superamplitudes to a combination of $\cN=2$ superamplitudes corresponding to vector and hypermultiplets that describe the  amplitudes of $\cN=2$ MESG theory coupled to hypermultiplets corresponding to the quaternionic manifold $\frac{SO(m,4)}{SO(m)\times SO(4)}$. Special cases of such amplitudes that arise in ${\cal N}=2$ MESG theories which are orbifolds of ${\cal N}=8$ supergravity were discussed in~\cite{Carrasco:2012ca,Chiodaroli:2014xia}.

Similarly, the superamplitudes of $\cN=4$ YMESG theories can be obtained as double copies  by replacing the $\cN=2$ supersymmetric
gauge-theory factor by an $\cN=4$ supersymmetric gauge-theory factor while keeping the $\cN=0$ gauge copy as in
\sec{YMscalar_unbroken}, with only the $\phi$ fields  but not the $\varphi$ fields. (Keeping both the fields $\phi$ and 
$\varphi$ leads to a theory that does not obey color/kinematics duality.)

Similarly to the unbroken symmetry case, scattering amplitudes with manifest ${\cal N}=4$ supersymmetry and spontaneously-broken gauge symmetry can be constructed rather straightforwardly by replacing the spontaneously-broken ${\cal N}=4$ SYM theory in place of the spontaneously-broken ${\cal N}=2$ SYM theory in \sec{2copySec}. These 
amplitudes are expected to describe spontaneously-broken YMESG theories  that preserve  all $\cN=4$ supersymmetries. 
Some of them, such as the anomalous amplitudes discussed in \cite{Carrasco:2013ypa} for MESG theories, are particularly easy to find from the expressions found in sec.~\ref{sec:loops}. 
Apart from the rational contribution present in the MESG and unbroken YMESG theories, they will also acquire nontrivial dependence on the non-zero masses.
However, spontaneous partial supersymmetry breaking is possible  in $\cN=4$ YMESG theories \cite{deRoo:1986yw,Wagemans:1987zy,Horst:2012ub}. In particular the  work of  \cite{Horst:2012ub} studies in depth the breaking of  $\cN=4$ supersymmetry down to $\cN=2$ supersymmetry in $\cN=4$ supergravity theories.  Detailed study of the double-copy construction of the amplitudes of spontaneously  broken $\cN=4$ YMESG theories with or without partial supersymmetry breaking is beyond the scope of this paper. It will be studied in a separate work where  we will discuss explicitly these amplitudes and their comparison with a direct Feynman graph-based calculations.

\section{Conclusions and outlook}
\renewcommand{\theequation}{7.\arabic{equation}}
\setcounter{equation}{0}

In this paper we have extended color/kinematics duality and the double-copy construction to gauge and gravity theories that are 
spontaneously broken by an adjoint scalar VEV. 

As demonstrated in earlier work~\cite{Chiodaroli:2014xia}, abelian and non-abelian gauge theories that couple to (super)gravity 
provide a rich class of theories for which both spectra and interactions appear to exhibit a double-copy structure.  The tree-level $S$ matrices and
the loop-level integrands of these YMESG theories can be constructed in terms of the tree-level $S$ matrices and
the loop-level integrands of particular matter-coupled YM theories. Color/kinematics duality is the main agent behind the consistency 
of this construction.

In the presence of a non-abelian gauge symmetry it is particularly natural to consider spontaneous symmetry breaking.
We observe that the gravity double-copy structure is present at the level of the spectrum of spontaneously-broken YMESG theories. 
The YMESG spectra can be expressed as the tensor product of the spectrum of two types of gauge theories: a spontaneously-broken YM theory and a 
YM theory coupled to massive scalars charged under a global symmetry. In the latter YM~+~$\phi^3$ theory, 
the scalar fields have acquired mass as a consequence of an explicit breaking of the global symmetry.

The double-copy construction is shown to work for the interacting fields given that the two gauge-theory factors obey color/kinematics duality of a form specific to broken gauge theories. As discussed in \sec{Higgs}, in addition to the Jacobi relation and commutation relation, there are new types of color-factor relations in a gauge theory spontaneously broken by an adjoint scalar VEV. Color/kinematics duality then requires that corresponding kinematic identities 
are satisfied by the kinematic numerators of the diagrammatic expansion of an amplitude. 
With the appropriate definition of the numerator factors,  the spontaneously-broken YM theory inherits 
color/kinematics duality from the corresponding unbroken $(D+1)$-dimensional theory. 
For the explicitly-broken YM~+~$\phi^3$ theory, color/kinematics duality acts as a highly non-trivial constraint on 
the terms in the Lagrangian that are introduced to break the global symmetry. These 
terms exhibit certain similarities with terms appearing in spontaneously-broken gauge theories, 
but the details differ significantly. While we do not discuss it in the current work, it should be interesting to understand these
terms as originating from some limiting case (perhaps a double-scaling limit) of a spontaneously-broken gauge theory.

Using the above gauge-theory ingredients, and building on our earlier work~\cite{Chiodaroli:2014xia},
we discussed in detail the ${\cal N}=2$ generic Jordan family YMESG theories with spontaneously-broken gauge symmetry and showed that they
continue to exhibit a double-copy structure on the Coulomb branch. By computing three-point 
and four-point scattering amplitudes we identified the map relating the double-copy asymptotic states and the asymptotic states of the supergravity Lagrangian.
Similar to the orbifold constructions of ref.~\cite{Carrasco:2012ca}, the supergravity fields are related to bilinears of the gauge theory fields which are neutral under an appropriately-identified global symmetry. The double-copy construction of the asymptotic states also follows closely the approach taken in ref.~\cite{Johansson:2014zca}.
Similar to the unbroken case~\cite{Chiodaroli:2014xia}, upon comparing the scattering amplitudes we identify 
the parameters of the supergravity Lagrangian in terms of the parameters of the two gauge-theory factors. 
This gives non-trivial relations between the dimensionful and dimensionless couplings of the various theories. 

The details of the double-copy construction extend to YMESG theories with ${\cal N}\leq 4$ supersymmetry with little change. 
In this paper, the ${\cal N} < 2$ YMESG theories have only 
been considered as obtained through the double copy, without detailing their Lagrangian formulation. 
Nevertheless, as pointed out in ref.~\cite{Chiodaroli:2014xia}, the formalism 
described there should extend to unbroken ${\cal N}=1$ supersymmetric 
and non-supersymmetric theories. Since ${\cal N}=1$ YMESG theories have no adjoint scalars they cannot be considered on the Coulomb branch.
For non-supersymmetric YME theories with adjoint scalars, the spontaneously-broken phase is straightforwardly obtained through the double copy. 

We addressed with more details the case of ${\cal N}=4$ MESG and YMESG theories. 
The ${\cal N}=4$ MESG theories are obtained as a double-copy of ${\cal N}=4$ SYM theory with the dimensional reduction of 
some higher-dimensional pure YM theory~\cite{Chiodaroli:2014xia}. The corresponding ${\cal N}=4$ YMESG theories are 
obtained by gauging a subgroup of the global symmetry group, which in terms of the double-copy construction amounts to adding a $\phi^3$ term 
to the non-supersymmetric gauge-theory factor~\cite{Chiodaroli:2014xia}. 
In analogy with the double-copy construction for ${\cal N}= 2$ theories, amplitudes in the spontaneously-broken $\cN=4$ 
theory should be obtained from a double copy 
between spontaneously-broken ${\cal N}=4$ SYM and explicitly broken YM~+~$\phi^3$ theory.
We leave for future work a thorough understanding of the $\cN=4$ YMESG theories from a Lagrangian perspective, as well as a comparison of the 
resulting scattering amplitudes with the results of the double-copy construction outlined here. 

To illustrate the power of the double-copy construction we presented several one-loop
four-point amplitudes. For the broken YM~+~$\phi^3$ theory, we considered one-loop diagrams with external massless scalars, 
and internal massless vectors and massive scalars. After double-copying this theory with the corresponding spontaneously-broken ${\cal N}=4,2$ 
SYM numerators, we obtained amplitudes in spontaneously broken ${\cal N}=4,2$ YMESG theories. Corresponding one-loop amplitudes with no supersymmetry were also presented. 

A future relevant study would be the case of ${\cal N}=2$ YMESG theories with hypermultiplets in the fundamental representation. 
A Higgs mechanism with fields in representations different from the adjoint gives distinct scenarios for breaking the gauge group. 
It would be interesting to explore whether gauge theories with a fundamental scalar VEV exhibit color/kinematics duality, 
and similarly to check the result of the double-copy construction against scattering amplitudes evaluated with the corresponding supergravity Lagrangian 
as a starting point.

\section*{Acknowledgements}

We would like to thank Marco Zagermann for for helpful discussions on $N=4$ supergravity theories. We also thank John Joseph Carrasco and Alexander Ochirov for helpful conversations on topics related to this work. 
The research of M.G. was supported in part by the US Department of Energy under DOE Grant No: de-sc0010534. 
The research of R.R. was also supported in part by the US Department of Energy under DOE Grants No: de-sc0008745 and de-sc0013699 and by the National Science Foundation under Grant No. PHY11-25915 while at KITP. 
The research of H.J. is supported in part by the Swedish Research Council under grant 621--2014--5722, the Knut and Alice Wallenberg Foundation under grant KAW~2013.0235 (Wallenberg Academy Fellow), and the CERN-COFUND Fellowship program (co-funded via Marie Curie Actions grant PCOFUND--GA--2010--267194 under the European Union's Seventh Framework Programme).


\newpage
\appendix

\section{Summary of index notation \label{appendixA}}
\renewcommand{\theequation}{A.\arabic{equation}}
\setcounter{equation}{0}

Here we give a brief summary of the various indices used in the paper, with the exception of \sec{Neq4}, 
which follows a different notation  consistent with the relevant supergravity literature. The types of indices are:
\begin{center}
\begin{tabular}{ll}
 $ A,B,C=-1,0, \ldots, \tilde  n$  & index running over all sugra vectors in $4D$, \\
 & global gauge-theory index (before symmetry breaking),
 \\[3pt]
 $ I,J,K=0, \ldots, \tilde  n$  & index running over matter vectors in $4D$; \\
 &index running over all vector fields in $5D$, \\[3pt]
 $ x,y = 1,2, \ldots  , \tilde n $ & curved target space indices in $5D$,   \\[3pt]
 $ a,b,c  $ & index running over massless vectors, \\
 & in the Higgsed supergravity; \\ 
 & global index in gauge-theory, \\[3pt]
 $ i,j,k $ & index running over massless scalars, \\
 & in the Higgsed supergravity ; \\
 & fundamental global indices in gauge theory, \\[3pt]
 $ \hat \imath, \hat \jmath, \hat k $ & fundamental rep. indices in gauge theory, \\[3pt]
 $ \alpha, \beta, \gamma $ & index running over massive fields, \\[3pt]   
 $ \ha , \hb, \hc $   &  gauge-theory adjoint indices,  \\[3pt]
 $ \haa , \hbb, \hgg $   &  gauge-theory matter-representation indices,  \\[3pt]
 $ \ffm, \ffn, \ffp $ & flavor indices. \\[3pt]
\end{tabular}\\
\end{center}
With this notation we have 
\be A =\big( -1, \ I \big) = \big(a, \alpha, \bar \alpha \big) = \big(-1, i, \alpha, \bar \alpha \big)  \ .  \ee

\section{Symmetry breaking vs. dimensional compactification}
\renewcommand{\theequation}{B.\arabic{equation}}
\setcounter{equation}{0}

\subsection{Spontaneously broken SYM \label{SYMdredSect}}

In this appendix we show that SYM spontaneously-broken by an adjoint scalar VEV is equivalent to a dimensional compactification $(D+1)\rightarrow D$ of SYM such that for each field the extra-dimensional momentum becomes a mass that is proportional to a U(1) charge. 

Consider that the gluons and scalars fields in the higher-dimensional theory satisfy the following differential equation with respect to a derivative in the internal direction $(D+1)$:
\be
\partial_{D+1} 
\left( \begin{array}{c} A^{\mu \hat A} \\  \phi^{a \hat A} \end{array}  \right)
 =  -g V f^{0 \hat A\hat B} \left( \begin{array}{c} A^{\mu \hat B} \\  \phi^{a \hat B} \end{array}  \right) \equiv  i \,m^{\hat A \hat B} \left( \begin{array}{c} A^{\mu \hat B} \\  \phi^{a \hat B} \end{array}  \right)\,.
\label{diffeqn}
\ee
This means that some fields have a momentum turned on in the internal direction $(D+1)$, corresponding to the eigenvalues of $-g V f^{0 \hat A\hat B} \equiv  i \,m^{\hat A \hat B}$. Fields that commute with the generator $t^0$ will not have a mass since that implies that $f^{0 \hat A\hat B}$ vanish. If needed one can decompose this equation into massive and massless field, with the corresponding renaming and complexification as in \sec{Higgs}, giving
\bea
\partial_{D+1} 
\left( \begin{array}{c} A^{\mu \ha} \\  \phi^{a \ha} \end{array}  \right)
&=&0 \,, \nn \\
\partial_{D+1}\left( \begin{array}{c} W^{\mu}_\haa \\  \varphi^{a}_\haa \end{array}  \right)
&=& -g V f^{0 \ \hbb}_{\ \haa} \left( \begin{array}{c} W^{\mu}_ \hbb \\  \varphi^{a}_\hbb \end{array}  \right) \equiv  i \, m_\haa^{\ \hbb} \left( \begin{array}{c} W^{\mu}_\hbb \\  \varphi^{a}_{\hbb} \end{array}  \right)\,.
\eea
However, for the exercise in this appendix it is more convenient to work with the real fields and mass matrix in \eqn{diffeqn}. 

The kinetic term of the scalars $\phi^{a>0}$ in $(D+1)$ dimensions can now be shown to be identical to a kinetic term in $D$ dimensions plus a $\phi^4$-term containing a VEV:
\bea
&&\frac{1}{2}\big( {\cal D}_\mu \phi^{a \hat A} \big)^2-\frac{1}{2}\big( \partial_{D+1} \phi^{a \hat A}+gf^{A \hat B \hat C} A^{\hat B}_{D+1} \phi^{a \hat C} \big)^2 \nn \\
&&=\frac{1}{2}\big(  {\cal D}_\mu \phi^{a \hat A} \big)^2-\frac{1}{2}\big(i \, m^{\hat A \hat B} \phi^{a \hat B}+ g f^{ABC}\phi^{0 \hat B} \phi^{a \hat C} \big)^2 \nn \\ 
&&= \frac{1}{2}\big(  {\cal D}_\mu \phi^{a \hat A} \big)^2+\frac{g^2}{2}{\rm tr}\big( [V t^0 + \phi^0, \phi^a]^2 \big)\,,
\eea
where the second term on the first row corresponds to the $(D+1)$ component of the kinetic term, similarly $A^{\hat A}_{D+1}$ is the gauge field in that direction. The latter is renamed to $\phi^{0 \hat A}$ on the second line. The full expression for ${\cal D}_\mu \phi^{a \hat A}$ can be found in \eqn{FAmn}, remembering that the global index $a$ does not yet include $a=0$. 

To get the kinetic term for the $\phi^{0}$ field we need to look at the $(D+1)$-dimensional vector-field kinetic term. It is straightforward to see that it is identical to the $D$-dimensional vector-field kinetic term plus the kinetic term of $\phi^{0}$, including a VEV for the latter,
\bea
&&-\frac{1}{4}\big({\cal F}^{\hat A}_{\mu \nu}\big)^2+\frac{1}{2}\big( \partial_{\mu}A_{D+1}^{\hat A}-\partial_{D+1}A_\mu^{\hat A}+g f^{\hat A \hat B \hat C}A_\mu^{\hat B}A_{D+1}^{\hat C}\big)^2 \nn \\
&&=-\frac{1}{4}\big({\cal F}^{\hat A}_{\mu \nu}\big)^2+\frac{1}{2}\big(\partial_{\mu}\phi^{0 \hat A}- i\,m^{\hat A \hat B} A_\mu^{\hat B}+g f^{\hat A \hat B \hat C} A_\mu^{\hat B}\phi^{0\hat C}\big)^2 \nn \\
&&=-\frac{1}{4}\big({\cal F}^{\hat A}_{\mu \nu}\big)^2+\frac{1}{2}\big(({\cal D}_{\mu}\phi^{0 })^{\hat A}- i\, m^{\hat A \hat B} A_\mu^{\hat B}\big)^2 \nn \\
&&=-\frac{1}{4}\big({\cal F}^{\hat A}_{\mu \nu}\big)^2+\frac{1}{2}\big(({\cal D}_{\mu}\phi^{0}+{\cal D}_{\mu}\langle \phi^0 \rangle)^{\hat A}\big)^2\,,
\eea
where $\langle \phi^0 \rangle =V t^0$. Similar to before, the second term on the first line is the contribution of the field-strength in the $ \mu \otimes (D+1)$ direction,  on the second line $A^{\hat A}_{D+1}$ is renamed to $\phi^{0 \hat A}$, and on the third and fourth lines terms are reassembled into covariant derivatives. Again, the full expressions for ${\cal F}^{\hat A}_{\mu \nu}$ and ${\cal D}_\mu \phi^{0 \hat A}$ can be found in \eqn{FAmn}.

Finally, including the quartic terms for the $\phi^{a>0}$ scalars, the $(D+1)$-dimensional (massless/unbroken) SYM Lagrangian has become a spontaneously-broken $D$-dimensional SYM Lagrangian,
\be 
{\cal L}_{\rm SYM} =  -{1\over 4} \big({\cal F}^{\hat A}_{\mu \nu}\big)^2 +\frac{1}{2}\big(({\cal D}_{\mu}\phi^{a}+{\cal D}_{\mu}\langle \phi^a \rangle)^{\hat A}\big)^2+\frac{g^2}{4}{\rm tr}\big( [\phi^a+\langle \phi^a \rangle, \phi^b+\langle \phi^b \rangle]^2 \big)+{\rm fermions}\,,
\label{unbroken2}
\ee
where $\langle \phi^a \rangle=\delta^{a0}V t^0$. 

A practical implication of this identification is that scattering amplitudes for SYM spontaneously-broken by a adjoint scalar VEV can be computed from unbroken $(D+1)$-dimensional SYM given that for each field there is a relation between the internal-space momentum and a U(1) charge. This relation can be stated as an operator equation, simply by rewriting the differential equation \eqref{diffeqn} as follows:
\be
i(p_{D+1}- g V q )
\left( \begin{array}{c} A^{\mu \hat A} \\  \phi^{a \hat A} \end{array}  \right)=0\,,
\label{operatoreqn}
\ee
where $p_{D+1}$ is the momentum operator pointing in the $(D+1)$ direction, and $q$ is a U(1)-charge operator that acts as $ q \Phi = [t^0,\Phi]$, for some field $\Phi$. Similarly, we have that $ p_{D+1} \Phi = m \Phi$, where $m$ is the mass of $\Phi$, implying that a (massive) field carries the U(1) charge $m/(gV)$.

For example, for tree amplitudes it is sufficient to impose the constraint \eqref{operatoreqn} on the external states, then the internal states will automatically have this satisfied by virtue of charge/momentum conservation. Similarly, for loop amplitudes, it is sufficient to have this constraint imposed once for each independent loop momenta.

\subsection{Explicitly broken YM~+~$\phi^3$ \label{YMphi3dredSect}}

Here we re-derive the Lagrangian \eqref{YMscalarGlobalBroken} for explicitly broken YM~+~$\phi^3$, without explicitly using color/kinematics duality. Similar to the derivation in \sec{SYMdredSect}, it is given by a dimensional compactification $(D + 1) \rightarrow D$ of the corresponding unbroken theory, after a proper identification of the extra-dimensional momentum and the U(1) charge of each field. Although, the details are strikingly different compared to the SYM case.

Consider that the scalars fields in the higher-dimensional theory satisfy the following differential equation with respect to a derivative in the internal direction $(D + 1)$:
\be
\partial_{D+1}\phi^{A \ha}=-\frac{1}{2}\rho\lambda F^{0AB} \phi^{B \ha} \equiv  \, i\, m^{AB} \phi^{B \ha}\,.
\ee
Similar to before we let $\widetilde{\phi}^{0 \ha}=A^\ha_{D+1}$ represent the gluon that is converted to a scalar upon dimensional reduction. The covariant derivative, applied in the $(D+1)$ direction, of the other scalars is then given by
\begin{align}
\!-\frac{1}{2}\big(({\cal D}_{D+1}\phi^A)^{\ha}\big)^2 \!= &-\frac{1}{2}\big(\partial_{D+1}\phi^{A \ha}+ g f^{\ha \hb \hc}A_{D+1}^{\hb} \phi^{A \hc} \big)^2 \nn \\ =&  -\frac{1}{2}\Big(\!-\frac{1}{2}\rho\lambda F^{0AB} \phi^{B \ha}+g f^{\ha \hb \hc}\widetilde{\phi}^{0 \hb} \phi^{A \hc}\Big)^2 \nn \\ =& -  \frac{1}{2} m^{AC}\!m^{CB} \phi^{A \ha}\phi^{B \ha}\!+\!\frac{1}{2} g\rho\lambda F^{0AB} f^{\ha \hb \hc} \widetilde{\phi}^{0 \ha} \phi^{A \hb} \phi^{B \hc}\!-\! \frac{g^2}{2} f^{\ha \hb \he} f^{\he \hc \hd}\widetilde{\phi}^{0 \ha} \phi^{A \hb}\widetilde{\phi}^{0 \hc} \phi^{A \hd} \nn \\ 
=&  - (m^2)^{\ \beta}_\alpha \overline{\varphi}^{\alpha \ha} \varphi^{\ \ha}_\beta +g \rho\lambda F^{0\ \beta}_{\ \alpha} f^{\ha \hb \hc} \widetilde{\phi}^{0 \ha} \overline{\varphi}^{\alpha \hb} \varphi^{\ \hc}_\beta- g^2 f^{\ha \hb \he} f^{\he \hc \hd}\widetilde{\phi}^{0 \ha} \widetilde{\phi}^{0 \hc} \overline{\varphi}^{\alpha \hb} \varphi^{\ \hd}_\alpha \nn \\
&   \null  - \frac{g^2}{2} f^{\ha \hb \he} f^{\he \hc \hd}\widetilde{\phi}^{0 \ha} \phi^{a \hb}\widetilde{\phi}^{0 \hc} \phi^{a \hd}\,.
\label{dimredtilde}
\end{align}
On the last line the proper massive fields have been identified (and complexified), and $m^{\ \beta}_\alpha$ is the proper mass matrix corresponding to these fields, similar to the presentation in \sec{YMBreakScalar}. 

An important difference from the derivation in \sec{SYMdredSect} is that the extra-dimensional gluon is also charged under the global group, since $F^{0AB}\neq 0$ is assumed in order to have a mass term. However, in the derivation in \eqn{dimredtilde} this field is a U(1) singlet in the $\rho \rightarrow 0$ limit, which appears to be inconsistent with this assumption. To ensure the existence of a non-singlet scalar in this limit we demand that the true $\phi^0$ scalar is a linear combination of $\widetilde{\phi}^0$ and a scalar $\widehat{\phi}^0$ that was present already before the dimensional compactification. The non-kinetic terms in the Lagrangian containing $\widehat{\phi}^0$ is then,
\be
g \lambda F^{0\ \beta}_{\ \alpha} f^{\ha \hb \hc} \widehat{\phi}^{0 \ha} \overline{\varphi}^{\alpha \hb} \varphi^{\ \hc}_\beta- g^2 f^{\ha \hb \he} f^{\he \hc \hd}\widehat{\phi}^{0 \ha} \widehat{\phi}^{0 \hc} \overline{\varphi}^{\alpha \hb} \varphi^{\ \hd}_\alpha - \frac{1}{2}g^2 f^{\ha \hb \he} f^{\he \hc \hd}\widehat{\phi}^{0 \ha} \phi^{a \hb}\widehat{\phi}^{0 \hc} \phi^{a \hd}\,.
\label{phiwithhat}
\ee
Indeed, if we add the terms in \eqn{dimredtilde} and \eqn{phiwithhat} and do the unitary rotation 
\be
\left( \begin{array}{c} \widehat{\phi}^{0}\\  \widetilde{\phi}^{0}  \end{array}  \right) = \frac{1}{\sqrt{1 + \rho^2}} \left( \begin{array}{cc} 1& - \rho \\  \rho  & 1  \end{array}  \right) \left( \begin{array}{c} \phi^{0} \\  \phi'^{0} \end{array}  \right)\,,
\ee
of the two scalars, then only the field ${\phi}^{0}$ has a cubic interaction, and $\phi'^{0}$ becomes a U(1) singlet of the global group. We may drop the latter field since it can be absorbed into the freedom of redefining the global group, e.g. $G_k\times U(1) \rightarrow G_k$. We then get the following modification of the covariant derivative considered in \eqn{dimredtilde}:
\bea
-\frac{1}{2}\big(({\cal D}_{D+1}\phi^A)^{\ha}\big)^2 \!\!\!\!&\rightarrow& \!\!\! - (m^2)^{\ \beta}_\alpha \overline{\varphi}^{\alpha \ha} \varphi^{\ \ha}_\beta +g \lambda \sqrt{1\! + \! \rho^2}  F^{0\ \beta}_{\ \alpha} f^{\ha \hb \hc} \phi^{0 \ha} \overline{\varphi}^{\alpha \hb} \varphi^{\ \hc}_\beta- g^2 f^{\ha \hb \he} f^{\he \hc \hd}\phi^{0 \ha} \phi^{0 \hc} \overline{\varphi}^{\alpha \hb} \varphi^{\ \hd}_\alpha \nn \\
&&\null - \frac{1}{2}g^2 f^{\ha \hb \he} f^{\he \hc \hd}\phi^{0 \ha} \phi^{a \hb}\phi^{0 \hc} \phi^{a \hd}\,.
\eea
Compared to a massless unbroken theory, the only new terms in this expression are the two first ones. It is not surprising that a quadratic mass term appears, but that the cubic term corresponding to the global-group coupling gets modified by a square-root function is striking. 
The remaining two terms are simply a group decomposition of certain quartic scalar terms already present in the original unbroken Lagrangian (\ref{YMscalarNf0}). Ignoring these, we get the explicitly broken YM~+~$\phi^3$ by adding the above mass-term to the Lagrangian (\ref{YMscalarNf0}) and at the same time swapping the cubic $\lambda$-dependent term as
\bea
g \lambda F^{a\ \beta}_{\ \alpha} f^{\ha \hb \hc} \phi^{a \ha} \overline{\varphi}^{\alpha \hb} \varphi^{\ \hc}_\beta \rightarrow g \lambda \Delta^{ab} F^{b\ \beta}_{\ \alpha} f^{\ha \hb \hc} \phi^{a \ha} \overline{\varphi}^{\alpha \hb} \varphi^{\ \hc}_\beta\,,
\eea
with $\Delta^{ab} = \delta^{ab} + \big(\sqrt{1+\rho^2} -1\big)  \delta^{a0} \delta^{0b}$. Finally, carrying out the full decomposition of $\phi^{A \ha}$ into real and complex massive fields, we obtain precisely the Lagrangian in \eqn{YMscalarGlobalBroken}.

Even though the terms in the Lagrangian can be obtained as a dimensional compactification of the unbroken $(D+1)$ dimensional theory, the amplitudes of this theory have to be treated with some care. The reason is that in the $(D+1)$-dimensional theory the (massive) scalars can potentially source $W$-bosons that are not part of the explicitly broken YM~+~$\phi^3$ theory. Without careful treatment of amplitudes in the $(D+1)$ theory, such ``illegal'' particles will appear as intermediate states. An example of such a treatment would be to impose the operator equation (\ref{operatoreqn}), with the gauge-group generator replaced by the global-group generator $t^0\rightarrow T^0$, on all external states in the tree amplitude. Because of $T^0$ charge conservation and extra-dimensional momentum conservation, the internal states will automatically have the correct mass, including gluons which are singlets of the global group.

In fact, it is no surprise that explicitly broken YM~+~$\phi^3$ theory cannot be a straightforward dimensional compactification. If it were then, through the double-copy construction, the spontaneously broken YMESG would inherit this property. This is impossible, spontaneously broken YMESG is clearly not a straightforward dimensional compactification of a $(D+1)$-dimensional theory; for example, it does not have massive modes of gravitons.

\section{Expansions for the supergravity Lagrangian \label{appex}}
\renewcommand{\theequation}{C.\arabic{equation}}
\setcounter{equation}{0}

In this appendix we list expansions for the period matrix and scalar metric entering the supergravity Higgs Lagrangian after the field redefinition (\ref{redefinition}) 
and up to terms linear in the physical scalar fields. 
The non-zero entries of the period matrix are the following,
\bea \cN_{-1-1} = - i + { O}(\phi^2) \ , \quad &&  \cN_{-1 a } = 2 z^a  + { O}(\phi^2) \ ,      \no \\
\cN_{-1 0 } = 2 z^0  + { O}(\phi^2) \ , \quad && \cN_{0a} =  2 \bar z^a  + { O}(\phi^2) \ ,   \no \\
\cN_{-1 1 } = 2 z^1  + { O}(\phi^2) \ , \quad && \cN_{-1}^{\ \beta} = \sqrt{2} ( \overline{\varphi}^\beta_x + i \overline{\varphi}^\beta_y )  + { O}(\phi^2) \ ,     \no \\
\cN_{00} = - i + {O}(\phi^2) \ , \quad && \cN_{-1 \beta } = \sqrt{2} ( \varphi_{x \beta} + i \varphi_{y \beta} )   + { O}(\phi^2) \ ,   \no \\
\cN_{01} = - 2 \bar z^1  + { O}(\phi^2) \ , \quad && \cN_{0}^{\ \beta} = \sqrt{2} ( \overline{\varphi}^\beta_x - i \overline{\varphi}^\beta_y )  + { O}(\phi^2) \ ,   \no \\
\cN_{11} = - i + 2 \bar z^0 + {O}(\phi^2) \ , \quad && \cN_{0 \beta} = \sqrt{2} ( \varphi_{x \beta} - i \varphi_{y \beta} )  + { O}(\phi^2) \ ,   \no \\
&& \cN_{ab} = (- i + 2 \bar z^0) \delta^{ab} + { O}(\phi^2) \ , \no \\
&& \cN^{\alpha}_{\ \beta} = (- i + 2 \bar z^0) \delta^{\alpha}_{\beta} + { O}(\phi^2) \ , 
\eea
where the indices $a,b$ run over fields transforming in the adjoint of the unbroken gauge group and additional (non universal) spectators. 
Similarly, the scalar metric has non-zero entries,
\bea g_{00} = 1 -2 \sqrt{2} y^0 + { O}(\phi^2) \ , \quad &  g_{1 a } = -y^a  + { O}(\phi^2) \ ,    & g_{ab} = (1 -  2 y^1) \delta^{ab} + { O}(\phi^2) \ ,   \no \\
g_{11} = 1 - 2 y^1  + { O}(\phi^2) \ , \quad & g_{1}^{\ \beta} =  - \overline{\varphi}_y^\beta  + { O}(\phi^2) \ ,  & g^{\alpha}_{\ \beta} = ( 1 -2 y^1)
\delta^{\alpha}_{\beta} + { O}(\phi^2) \ .  \no \\
 \quad & g_{1 \beta} =- \overline{\varphi}_{y\beta}   + { O}(\phi^2) \ ,  \quad &  \no \\
 \eea
Note that the differences between these expansions and 
the ones in the appendix of \cite{Chiodaroli:2014xia} arise because the vector field $A^1_\mu$ has not been dualized. 
These expansions are sufficient for calculating the three-point amplitudes presented in section \ref{secsugraamp}. 

\section{Lagrangians of  $\cN=4$ MESG and YMESG theories in five dimensions \label{N4Lagrangian}}
\renewcommand{\theequation}{D.\arabic{equation}}
\setcounter{equation}{0}
The   Lagrangian of the five-dimensional  $\cN=4$ MESG theory  is given by  \cite{AT85,Dall'Agata:2001vb}:
\bea
e^{-1}\,\cL &=& -\frac12 R -\frac12\,
\bar\psi_\mu^i\,\Gamma^{\mu\nu\rho}\,D_\nu\psi_{\rho
  i}-\frac14\Sigma^2\,a_{\tI\tJ}\,F_{\mu\nu}^\tI
F^{\mu\nu\tJ}-\frac14\Sigma^{-4}\,G_{\mu\nu}G^{\mu\nu}\nnu\\
&&\null-\frac12\,(\prt_\mu\si)^2-\frac12\,\bar\chi^i
\Dsl\,\chi_i-\frac12\bar\la^{ia}\Dsl\,\la_i^a-\frac12\,P_\mu^{aij}P^\mu_{aij}\nnu\\
&&\null-\frac{i}{2}\bar\chi^i\Gamma^\mu\Gamma^\nu\psi_{\mu
  i}\,\prt_\nu\si
+i\bar\la^{ia}\,\Gamma^\mu\Gamma^\nu\psi_\mu^j\,P_{\nu ij}{}^a\nnu\\
&&{}\null+\frac{\sqrt 3}{6}\Sigma
L_\tI^{ij}\,F_{\rho\si}^\tI\bar\chi_i\,\Gamma^\mu\Gamma^{\rho\si}\psi_{\mu
  j}-\frac14\Sigma L_\tI^a\,F_{\rho\si}^\tI\bar\la^{ai}\,\Gamma^\mu
\Gamma^{\rho\si}\psi_{\mu i}\nnu\\
&&\null-\frac{1}{2\sqrt6}\Sigma^{-2}\,\bar\chi^i\,\Gamma^\mu\Gamma^{\rho\si}\psi_{\mu
  i}G_{\rho\si}+\frac{5i}{24\sqrt
  2}\Sigma^{-2}\bar\chi^i\Gamma^{\rho\si}\chi_i G_{\rho\si}\nnu\\
&&\null-\frac{i}{12}\Sigma L_\tI^{ij}
F_{\rho\si}^\tI\bar\chi_i\,\Gamma^{\rho\si}\chi_j -\frac{i}{2\sqrt
3}\Sigma
L_\tI^a\,F_{\rho\si}^\tI\bar\la^{ia}\Gamma^{\rho\si}\chi_i-\frac{i}{8\sqrt
  2}\Sigma^{-2} G_{\rho\si}\bar\la^{ia}\Gamma^{\rho\si}\la_i^a\nnu\\
&&\null+\frac{i}{4}\Sigma L_\tI^{ij}
F_{\rho\si}^\tI\bar\la^a_i\Gamma^{\rho\si}\la_j^a-\frac{i}{4}\Sigma
L_\tI^{ij} F_{\rho\si}^\tI\,[\bar\psi_{\mu
i}\Gamma^{\mu\nu\rho\si}\psi_{\nu
  j}+2\,\bar\psi^\rho_i\,\psi^\si_j]\nnu\\
&&\null-\frac{i}{8\sqrt2}\,\Sigma^{-2} G_{\rho\si}\,[\bar\psi^i_{\mu}
\Gamma^{\mu\nu\rho\si}\psi_{\nu
  i}+2\bar\psi^{\rho i}\,\psi^\si_i]\nnu\\
&&{}\null+\frac{\sqrt2}{8}e^{-1}\,C_{\tI\tJ}\,\epsilon^{\mu\nu\rho\si\la}\,
F_{\mu\nu}^\tI F_{\rho\si}^\tJ\,a_\la +e^{-1}\cL_{4\!f}\,,
\label{Lagrange1}
\eea
where $\cL_{4\!f}$ denotes the four fermion terms in the Lagrangian. The supersymmetry transformation laws 
are given by\footnote{All symmetrizations $(ij)$ and anti-symmetrizations $[ij]$ are of weight one.}
\bea
\delta e_{\mu}^{m}&=&\frac{1}{2}\bar\varepsilon^{i}
\Gamma^{m}\psi_{\mu i}\,, \nonumber  
\eea 
\bea 
\delta \psi_{\mu i}&=& D_{\mu}\varepsilon_{i} +
\frac{i}{6}\Sigma L_{\tI ij} F_{\rho \sigma}^{\tI}
(\Gamma_{\mu}^{\,\,\rho \sigma}
-4 \delta^{\rho}_{\mu}\Gamma^{\sigma}) \varepsilon^{j} \,,\nonumber\\
&&\null +\frac{i}{12\sqrt{2}} \Sigma^{-2}G_{\rho \sigma}
(\Gamma_{\mu}^{\,\,\rho \sigma}
-4 \delta^{\rho}_{\mu}\Gamma^{\sigma})\varepsilon_{i} +
\textrm{3-fermion terms} \,, \nonumber\nnu\\
\delta \chi_{i}&=&-\frac{i}{2}\prtsl \sigma
\varepsilon_{i}+\frac{\sqrt{3}}{6}
\Sigma L_{\tI ij} F_{\rho \sigma}^{\tI} \Gamma^{\rho
\sigma}\varepsilon^{j}
-\frac{1}{2\sqrt{6}}\Sigma^{-2}G_{\rho \sigma} \Gamma^{\rho \sigma}
\varepsilon_{i} \,, \nonumber\\
\delta \lambda^{a}_{i}&=&iP_{\mu ij}^{a}
\Gamma^{\mu} \varepsilon^{j} -\frac{1}{4}\,\Sigma\,
L_{\tI}^{a} F_{\rho \sigma}^{\tI} \Gamma^{\rho \sigma}\varepsilon_{i}
  +
\textrm{3-fermion terms} \,, \nonumber\\
\delta A_{\mu}^{\tI}&=&\vartheta_{\mu}^{\tI}\,, \nonumber\\
\delta a_{\mu}&=&\frac{1}{\sqrt{6}}\Sigma^{2}{\bar{\varepsilon}}^{i}
\Gamma_{\mu}\chi_{i}-\frac{i}{2\sqrt{2}}\Sigma^{2}{\bar{\varepsilon}}^{i}
\psi_{\mu i}\,, \nonumber\\
\delta \sigma&=&\frac{i}{2}{\bar{\varepsilon}}^{i}\chi_{i}\,,\nonumber\\
\delta
L_{\tI}^{ij}&=&-iL_{\tI}^{a}(\delta^{[i}_{k}\delta^{j]}_{l}-\frac{1}{4}
\Omega^{ij}\Omega_{kl}){\bar{\varepsilon}}^{k}\lambda^{la}\,,\nonumber\\
\delta L_{\tI}^{a}&=&-iL_{\tI ij}{\bar{\varepsilon}}^{i}\lambda^{ja}\,,
\label{trafo1}
\eea
where
\begin{equation}\label{theta}
\vartheta^{\tI}_{\mu}\equiv  -\frac{1}{\sqrt{3}}\Sigma^{-1}L^{\tI}_{ij}
{\bar{\varepsilon}}^{i}\Gamma_{\mu}\chi^{j}-i\Sigma^{-1}L^{\tI}_{ij}
{\bar{\varepsilon}}^{i}\psi^{j}_{\mu}+\frac{1}{2}L^{\tI}_{a}\Sigma^{-1}
{\bar{\varepsilon}}^{i}\Gamma_{\mu}\lambda^{a}_{i}\,.
\end{equation}
and \begin{equation}
\Sigma \ = \ e^{\frac{1}{\sqrt 3}\si}.
\end{equation}
The abelian field strengths of vector fields are defined as
\begin{equation}
F_{\mu\nu}^\tI\ = (\prt_\mu A_\nu^\tI-\prt_\nu A_\mu^\tI)~,\quad
G_{\mu\nu}\ = (\prt_\mu a_\nu-\prt_\nu a_\mu)\,,
\end{equation}
and the covariant derivative, $D_{\mu}$ involves the composite connections:
\begin{equation}
D_\mu\la_{i}^a \ = \ \nabla_\mu\la_{i}^a+Q_{\mu i}{}^j\la_{j}^a+Q_{\mu}^{ab}
\la_{i}^b\, ,
\end{equation}
where $\nabla_\mu$ is  the Lorentz- and spacetime covariant
derivative.

To construct an $\cN=4$ YMESG theory with a semisimple subgroup $K_S$ of the global symmetry group $SO(5,n)$ as the non-abelian gauge symmetry one replaces  all derivatives acting on fields that transform non-trivially under $K_S$  with $K_S$ gauge covariant derivatives \cite{AT85,Dall'Agata:2001vb}.
As explained in \sec{Neq4} this is implemented by the following substitutions in the Lagrangian:
\bea F^\tI_{\mu\nu}& \lra& \cF^\tI_{\mu\nu} \ = \ F^\tI_{\mu\nu}
+\gs\,A_\mu^\tJ
f_{\tJ\tK}^\tI A_\nu^\tK\nnu \,, \\
\partial_\mu L^\tI_A & \lra& \fD_\mu L^\tI_A \ = \ \prt_\mu L^\tI_A  +\gs  A_\mu^\tJ f_{\tJ\tK}^\tI L^\tK_A.
\eea
where $f_{\tI\tJ}^{~~\tK}$ are non-vanishing only when the indices take values in the adjoint representation of the semisimple gauge group $K_S$  and vanish whenever any one of the indices  labels the spectator vector fields.
The $USp(4)$ and $SO(n)$ connections, as well as
the vielbein $\,P_{\mu ij}^a\,$ are also modified by
 the new $\gs$ dependent contributions, i.e.,
\bea
\cQ_{\mu i}^{\ \ \ j} &=& Q_{\mu i}^{\ \ \ j}
+\gs A_\mu^\tJ L^\tK_{ik} f_{\tJ\tK}^\tI L_{\tI}^{kj}\,,\label{compco1}\\
\cQ_{\mu a}^{\ \ \ b} &=& Q_{\mu a}^{\ \ \ b}
-\gs A_\mu^\tJ L^\tK_a f_{\tJ\tK}^\tI L_{\tI}^b\,,\label{compco2}\\
\cP_{\mu ij}^a &=& P_{\mu ij}^a -\gs
A_\mu^\tJ L^{\tK}_{ij}f_{\tJ\tK}^\tI L_{\tI}^a\,.\label{compco3}
\eea
The derivatives acting on  the fermions get modified accordingly
\begin{equation}
\mathcal{D}_{\mu}\lambda^{a}_{i}=\nabla_{\mu}\lambda^{a}_{i}
+\mathcal{Q}_{\mu
i}^{\,\,\,\,\,\,j}\lambda_{j}^{a}+\mathcal{Q}_{\mu}^{ab}
\lambda^{b}_{i}\,,
\end{equation}
where $\cQ_{\mu i}^{\ \ \,  j}$ and $\cQ_{\mu a}^{\ \ \, b}$ now include   the $\gs$ dependent terms.

To restore supersymmetry with the above covariantizations one adds to the Lagrangian following  Yukawa couplings  as well as potential terms\cite{AT85,Dall'Agata:2001vb}
\be
\Delta{\cL} = \cL_{\textrm{Yukawa}} + \cL_{\textrm{Potential}}\,,
\ee
where
\bea
\cL_{\textrm{Yukawa}} &=& \frac{3i}{2} \gs
S_{ij}\,\bar\psi_\mu^i\Gamma^{\mu\nu}\psi_\nu^j+i\gs
I_{ijab}\,\bar\lambda^{ia}\lambda^{jb}
+\frac{i}{2}\gs S_{ij}\,\bar\chi^i\chi^j+\gs
T_{ij}^a\,\bar\psi_\mu^i\Gamma^\mu
\lambda^{ja}\,,\nnu\\
&&{}+\sqrt3 \gs S_{ij}\,\bar\psi_\mu^i\Gamma^\mu\chi^j
-\frac{2i}{\sqrt3} \gs T^a_{ij}\,\bar\chi^i\lambda^{ja}\,, \eea
and
\begin{equation}
\cL_{\textrm{Potential}}= -\gs^2 \cV^\uS \ = -\gs^2 \left(
-\frac92\,S_{ij}\Delta^{ij}+\frac12\,T_{ij}^a\,T^{aij} \right).
\end{equation}
Various  scalar field dependent quantities  above  are defined as follows:
\bea
S_{ij} &=& -\frac29 \Sigma^{-1} L^{\tJ}_{(i|k|} f_{\tJ\tI}^\tK L_{\tK}^{kl}
L^{\tI}_{|l|j)}\,,\\
T_{ij}^a &=& -\Sigma^{-1} L^{\tJ a}L^{\tK ~k}_{(i} f_{\tJ\tK}^\tI L_{\tI|k|j)}\,,\\
I_{ijab}&=& -\frac32
S_{ij}\delta_{ab}-\Sigma^{-1}L^{\tJ a}L^\tK_{ij}f_{\tJ\tK}^\tI L_{\tI}^b\,.
\eea
Furthermore one needs to modify  the transformation rules of the fermions  as follows:
\bea
\delta_{\textrm{new}} \psi_{\mu i} &=& i\gs
S_{ij}\Gamma_\mu\varepsilon^j,\\
\delta_{\textrm{new}} \lambda^a_i &=& \gs T_{ij}^a\varepsilon^j,\\
\delta_{\textrm{new}} \chi_i &=& \gs \sqrt3\,S_{ij} \varepsilon^j.
\eea

\end{document}